\documentclass[a4paper,11pt]{article}
\usepackage[top=2.5cm,bottom=3cm,left=2.5cm,right=2.5cm]{geometry}
\usepackage{amssymb}
\usepackage{amsmath}
\usepackage{graphicx}
\usepackage{color}
\usepackage{subcaption}
\usepackage[usenames,dvipsnames]{xcolor}
\usepackage{hyperref}
\usepackage{multirow}
\usepackage{soul}

\numberwithin{equation}{section}

\newcommand{\e}{\eta}

\newcommand{\sgn}{\mathrm{sgn}}

\newcommand{\rmd}{\mathrm{d}}
\newcommand{\la}{\lambda}
\newcommand{\ud}{\mathrm{d}}

\def\R{{\mathbb R}}

\def\be{\begin{equation}}
\def\ee{\end{equation}}
\def\bea{\begin{eqnarray}}
\def\eea{\end{eqnarray}}
\newcommand{\dos}{\rho_{\rm p}}
\newcommand{\iddr}{\text{id}^{\text{dr}}}
\newcommand{\veff}{v^{\rm eff}}

\newcommand{\bc}{\begin{center}}
\newcommand{\ec}{\end{center}}
\def\ba#1{\begin{array}{#1}\displaystyle}
\newcommand{\ea}{\end{array}}

\newcommand{\beq}{\begin{equation}}
\newcommand{\eeq}{\end{equation}}
\newcommand{\beqa}{\begin{eqnarray}}
\newcommand{\eeqa}{\end{eqnarray}}

\newcommand{\n}{\nonumber\\}
\newcommand{\bi}{\begin{itemize}}
\newcommand{\ei}{\end{itemize}}

\def\lt#1{\left#1}
\def\rt#1{\right#1}
\def\t#1{\tilde{#1}}
\def\h#1{\hat{#1}}

\def\frc#1#2{\frac{#1}{#2}}

\newcommand{\p}{\partial}

\newcommand{\bra}{\langle}
\newcommand{\ket}{\rangle}

\newcommand{\Tr}{{\rm Tr}}

\newcommand{\dd}{{\rm d}}

\newcommand{\Or}{{\cal O}}

\newcommand{\ep}{\epsilon}
\newcommand{\varep}{\varepsilon}
\newcommand{\ri}{{\rm i}}

\newcommand{\s}{\sigma}

\newcommand{\orcid}[1]{\href{https://orcid.org/#1}{\includegraphics[width=10pt]{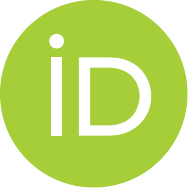}}}

\title{Generalized hydrodynamics of the KdV soliton gas}

\author{Thibault Bonnemain\orcid{0000-0003-0969-2413}$^1$, Benjamin Doyon\orcid{0000-0002-5258-5544 }$^2$  and Gennady El\orcid{0000-0003-1962-5388}$^1$\thanks{Email address for correspondence: gennady.el@northumbria.ac.uk}}
\date{$^1$Department of Mathematics, Physics and Electrical
Engineering, Northumbria University, Newcastle upon Tyne, United
Kingdom, \\
$^2$Department of Mathematics, King's College, London, United Kingdom.
}

\begin{document}

\maketitle

\begin{abstract}
We establish the explicit correspondence between the theory of soliton gases in classical integrable dispersive hydrodynamics, and generalized hydrodynamics (GHD), the hydrodynamic theory for many-body quantum and classical integrable systems. This is done by constructing the GHD description of the soliton gas for the Korteweg-de Vries (KdV) equation. We further predict the exact form of the free energy density and flux, and of the static correlation matrices of conserved charges and currents, for the soliton gas. For this purpose, we identify the solitons' statistics with that of classical particles, and confirm the resulting GHD static correlation matrices by numerical simulations of the soliton gas. Finally, we express conjectured dynamical correlation functions for the soliton gas by simply borrowing the GHD results. In principle, other conjectures are also immediately available, such as diffusion and large-deviation functions for fluctuations of soliton transport.
\end{abstract}

\section{Introduction} 
\label{sec:intro}

Long wavelength, hydrodynamic theories play a prominent role in physics, from fluids to optics,
condensed matter to quantum mechanics, and beyond. Recently, two new hydrodynamic theories have come to the fore, which present a large amount of conceptual and formal similarities, and which have been extremely successful in their respective fields of study: the theory of soliton gases \cite{el2021soliton}, and generalised hydrodynamics \cite{doyon_lecture_2020}.

On the one hand, the theory of soliton gases is concerned with the emergent, large-scale behaviours of integrable partial differential equations of dispersive hydrodynamics, such as the Korteweg-de Vries (KdV) and the nonlinear Schr\"odinger (NLS) equations. Dispersive hydrodynamics theories are of spectacularly different character than their dissipative counterparts. For instance, dispersive shock waves consist of coherent, rank-ordered, nonlinear oscillations that
continually expand. Remarkably, the multi-scale nonlinear dynamics of dispersive shock waves are  asymptotically described by a {\em new, emergent hydrodynamic theory}, arising as a result of averaging over fast oscillations---the famous Whitham  modulation theory \cite{whitham_linear_1974}. The Whitham equations are a system of first-order quasilinear PDEs describing slow evolution of the wave's parameters such as  amplitude, wavelength, mean etc. For integrable equations such as KdV or NLS, the Whitham modulations are expressed in terms of slow deformations of hyperelliptic Riemann surfaces  associated with  spectral finite-gap solutions that locally approximate  solutions to initial-value problems for the original dispersive hydrodynamics \cite{lax_small_1983,flaschka_multiphase_1980}. 

This new hydrodynamics is in fact of a much more general character. Many physically relevant wave phenomena exhibit complex spatiotemporal behaviours that cannot be restricted to macroscopically coherent  dispersive shock waves. 
In spite of integrability, the inherent randomness of many real-life systems (due to initial and boundary conditions or to complex mechanisms \cite{lax_zero_1991, gurevich_development_1999}) ultimately means results can only be obtained in statistical terms. This kind of random wave motion in nonlinear infinitely-dimensional Hamiltonian systems has been  dubbed   `integrable turbulence'   \cite{zakharov_turbulence_2009} and recently attracted significant attention. At a microscopic scale associated with the system's coherence length, integrable dispersive hydrodynamics feature solitons---the localised nonlinear waves that exhibit particle-like properties such as elastic, pairwise interactions. The soliton-dominated integrable turbulence emerging at the macroscopic scale is often called {\it soliton gas}, a moniker that can be traced back to V. Zakharov's seminal paper in 1971 \cite{zakharov_kinetic_1971}. It is in this work that he derived the kinetic equation for a ``rarefied" gas of KdV solitons, by considering the modification of the soliton velocity due to the position shifts in
its pairwise collisions with other solitons in the gas. This kinetic picture was then generalized in 2003, taking the thermodynamic limit of spectral finite-gap solutions and their modulations, by G. El to KdV soliton gases of arbitrary density \cite{el2003thermodynamic}, and further in \cite{el_spectral_2020} to soliton and breather gases of the focusing NLS equation. The spectral thermodynamic limit derivation in \cite{el2003thermodynamic, el_spectral_2020} has enabled in \cite{el_kinetic_2005, congy_soliton_2021} a formal construction of kinetic equations for isotropic and anisotropic bidirectional soliton gases such as e.g., shallow-water soliton gas described by the Kaup-Boussinesq equation. Integrability of the spectral  kinetic equation for soliton gas has been studied in \cite{el_kinetic_2011, ferapontov2021kinetic} at the level of various hydrodynamic reductions.  Beyond these important results in the mathematics of integrability, the physical relevance of soliton gases has been demonstrated in  recent ocean observations \cite{costa_soliton_2014} and laboratory experiments  \cite{redor_experimental_2019, suret_nonlinear_2020}. 

On the other hand, generalised hydrodynamics (GHD) is concerned with the emergent, large-scale behaviours of integrable, quantum and classical, many-body systems out of equilibrium \cite{bertini2016transport,castro2016emergent, doyon_lecture_2020}. Many-body systems in states that are inhomogeneous at large distance scales, and that evolve on long time scales, may be described using hydrodynamic principles, in terms of ``fluid cells'' and their emergent dynamics \cite{spohn2012large}. Before the advent of GHD, however, a common paradigm was that the applicability of the hydrodynamic principles -- including the key notion of local thermalisation -- required the system to be chaotic. Integrable systems admit an infinite number of conserved quantities which restrict their dynamics, and are usually considered to be non-chaotic. Integrability profoundly affects non-equilibrium physics as the presence of infinitely many conservation laws impedes thermalisation, while the lack of macroscopic homogeneity makes implementing standard methods of many-body integrability based on the inverse scattering method, difficult. 

GHD was initially developed simultaneously by two independent teams in 2016 \cite{bertini2016transport,castro2016emergent}, in order to discuss non-equilibrium states of integrable quantum many-body systems. In \cite{castro2016emergent} it is proposed to adapt the hydrodynamic principles to ``generalised thermalisation": the local entropy maximisation with respect to the infinitely many conserved quantities. That is, the hydrodynamic principles still apply to non-chaotic, integrable systems, but local fluid cells are in so-called Generalised Gibbs Ensembles (GGE) \cite{eisert2015quantum}, instead of Gibbs ensembles. This, in conjunction with an extension to GGE of the celebrated Thermodynamic Bethe Ansatz (TBA) \cite{yang1969thermodynamics,zamolodchikov1990thermodynamic,mossel2012generalized}, constitute the main building blocks of GHD in the hydrodynamic picture. In \cite{bertini2016transport}, by contrast, a kinetic picture was taken, where a kinetic equation for stable quasi-particles is derived based on the quasi-particle velocity first proposed in \cite{bonnes2014light}, with the same results. These works gave rise to an understanding of the non-trivial, large-scale, emergent (hydro)dynamics of quantum gases, chains and field theories \cite{ilievski2017microscopic,doyon2017large,piroli2017transport} from microscopic properties. Ultimately this approach also has proven well-suited to describe classical integrable systems \cite{bastianello2018generalized,doyon2019generalized,SpoToda,BCM19,spohn2020collision}, and the GHD equation correctly specialises to the hydrodynamic equation for the classical gas of hard rods, proven rigorously earlier \cite{boldrighini_one-dimensional_1983}. The hydrodynamic picture advocated in \cite{castro2016emergent}, in particular, is very powerful, and has provided a wealth of exact results, including dynamical correlation functions \cite{SciPostPhys.3.6.039,doyon2018exact,de_nardis_hydrodynamic_2018} and fluctuations \cite{doyon2020fluctuations,myers2020transport}. GHD has been extended to include external forces \cite{doyon2017note} and other type of space-time varying parameters \cite{bastianello2019Integrability,BasGeneralised2019} as well as diffusion \cite{de_nardis_hydrodynamic_2018,denardis2019diffusion,gopalakrishnan2020hydrodynamics}, and to probe integrability breaking \cite{cao2018incomplete,caux2019hydrodynamics,friedman2020diffusive,DurninTherma2020,bulchandani2021quasi-particle}. Importantly, the Euler-scale GHD equations with external force have been confirmed in experiments on cold atomic gases \cite{schemmer2019generalised,moller2020extension,malvania2020ghd}. GHD is now an extremely active field of research.

The kinetic equation for soliton gases, and the (Euler-scale) GHD equation for many-body integrable models, are strikingly similar, something already pointed out both in the contexts of GHD and dispersive hydrodynamics, see e.g. \cite{bulchandani_classical_2017,bulchandani2018bethe,doyon_soliton_2018, doyon_lecture_2020, el2021soliton}. Much like for soliton gases,  in GHD it can be seen as an effective propagation equation for the stable quasi-particles of many-body integrable systems, as advocated in \cite{bertini2016transport}. Semi-classical arguments made in \cite{doyon_soliton_2018} and \cite{bettelheim2020whitham} further argue that the GHD equation can be recovered using Zakharov's 1971 idea that scattering shifts lead to modified velocities. This idea is also at the root of the early derivation in the 1970's of the hydrodynamic equations for the hard rod gas \cite{aizenman1975ergodic}, which GHD generalises. But besides these remarks, the precise mathematical relation between the kinetic theory of soliton gases in dispersive hydrodynamics, and GHD, has not been established so far.

The main purpose of this paper is to establish an explicit connection between the spectral kinetic theory of soliton gases in dispersive hydrodynamics, and the generalised hydrodynamics of integrable many-body systems. We achieve this in two steps. 

First, we develop the GHD description of the KdV soliton gas. This involves making the explicit relation between quantities defined for soliton gases, and those within the context of GHD. In order to make this relation as clear as possible, the respective conventional notations for soliton gases and GHD are kept troughout, and we provide a dictionary between them.

Second, and perhaps most importantly, we give evidence to the fact that the (generalised) thermodynamics of the soliton gas can be described in terms of the classical Thermodynamic Bethe Ansatz (TBA). That is, the general form expressed in \cite{doyon_lecture_2020} for the thermodynamics of integrable many-body systems holds for soliton gases as well. In this, a piece of information about the gas is required, that was not discussed yet in the literature on soliton gases: the ``statistics" of the solitons. We find that solitons are associated to the Maxwell-Boltzmann statistics, like classical indistinguishable particles. As a result, we identify the key thermodynamic quantities associated with the KdV soliton gas (including the entropy, the free energy, the temperature). Using the general results of GHD, we then infer the static covariances, the space-integrated two-point correlation functions involving densities and currents. {The general statement that ensembles of solitons should admit the Maxwell-Boltzmann statistics was first proposed in \cite{bastianello2018generalized}, from previous results on the sine-Gordon model obtained by taking appropriate classical limits of the quantum TBA. However, here we provide the first (non-rigorous) derivation directly for the classical soliton gas. We also confirm for the fist time the conjecture of the Maxwell-Boltzmann statistics, by comparing static GHD covariances against numerical simulations of KdV soliton gases.}

The dictionary between the KdV soliton gas and GHD, and the uncovering of the thermodynamics of the KdV soliton gas, allows us to use results in one field in order to strengthen the other. For instance, the derivation of the KdV soliton gas kinetic equations, from the spectral theory of the KdV equation, gives a mathematically stronger basis for the GHD equation than the principle of local entropy maximisation used up to now to justify it in many-body integrable systems. Also, the various GHD results can now be applied to soliton gases, such as the exact form of dynamical correlation functions at the Euler scale.

The paper is organised as follows: In Section \ref{sec:solgas} we provide a brief introduction to the theory of soliton gases from the point of view of dispersive hydrodynamics, putting the emphasis on the KdV equation. Section \ref{sec:GHDLL} introduces GHD by way of a paradigmatic case study, through its direct application to the Lieb-Liniger model. Section \ref{sec:dico} draws explicit parallels between the two aforementioned theories, making the relations between the relevant quantities and notations in both fields. Section \ref{sec:conj} builds on the previous one by straightforwardly exploiting results from GHD to construct the thermodynamics of the KdV soliton gas. Section \ref{sec:concl} contains a summary of our results and some concluding remarks. The  paper is complemented by three appendices providing  details of our numerical simulations.

\section{Soliton gas in dispersive hydrodynamics}\label{sec:solgas}

We start our discussion by introducing the theory of soliton gases in the context of integrable dispersive hydrodynamics. More specifically, we will focus on how it relates to the KdV equation through its $N$-soliton solutions and further, the thermodynamic limit of finite-gap potentials.

\subsection{Basic construction}
\label{sec:basic_constr}

Let us consider the KdV equation in the   form
\begin{equation}\label{KdV}
 \partial_t u\ +\ 6u\partial_x u\ +\ \partial_{xxx}^3u\ =\ 0 \, .
\end{equation}
This equation \eqref{KdV} belongs to the family of  completely  integrable equations and, for a  broad class of initial conditions, its integrability  is realised via the inverse scattering transform (IST) method \cite{gardner_method_1967}.
The inverse scattering theory  associates a  soliton  of the KdV equation  with a point of discrete spectrum $\la=\la_n$,  of the Schr\"odinger operator
\be\label{schr}
\mathcal{L}= -\partial_{xx}^2  - u(x,t) \, . 
\ee
 Assuming $u \to 0$ as $x \to \pm \infty$, the KdV soliton solution corresponding to $\la_i = -\eta^2_i$, $\eta_i>0$, is given by
\be\label{kdv1sol}
u_s(x,t; \eta_i)=2\eta_i^2 \hbox{sech}^2 [\eta_i(x-4\eta_i^2 t - x_i^{0})],
\ee
where we denote by $a_i= 2\eta_i^2$ the soliton amplitude, by $s_i=4\eta_i^2$ its speed, and by $x_i^0$ its initial position or  ``phase''. Note that solitons have finite width $\sim 1/\eta_{i}$, which affects the notion of interaction range, particularly for small-amplitude solitons. In what follows we will be referring to  $\eta$ as a spectral parameter  with the understanding that $\eta = \sqrt{-\la}$. Along with the simplest single-soliton solution, the KdV equation supports $N$-soliton solutions $u_N(x,t)$ characterised by $N$ discrete spectral parameters $\eta_1 < \eta_2 < \dots <\eta_N$ and the set of initial positions  $\{x_i^0 | i=1, \dots, N\}$.

The integrable structure of the KdV equation has profound implications for the dynamics of soliton interactions.

\begin{enumerate}
\item  The KdV evolution preserves the IST spectrum,  $\frac{\rmd}{\rmd t}\eta_j=0$, implying that solitons retain their ``identity'' (amplitude, speed) upon  interactions.

\item The  collision of two solitons with spectral parameters $\eta_i$ and $\eta_j$, $i \ne j$ results in their phase (position) shifts given by
\be \label{shift_kdv}
\Delta_{ij} \equiv \Delta (\eta_i, \eta_j)= \frac{\sigma_{ij}}{\eta_i}   \log\left|\frac{\eta_i + \eta_j}{\eta_i-\eta_j} \right|, \quad
\sigma_{ij}= \sgn(\eta_i-\eta_j),
\ee
so that the taller soliton acquires shift forward and the smaller one -- shift backwards.

\item Solitons interact pairwise, i.e. the resulting phase shift $\Delta_i$ of a given  soliton with spectral parameter $\eta_i$  after its interaction with  $M$ solitons with parameters $\eta_j$, $j \ne i$, is equal to the sum of the individual phase shifts,
\be\label{delta_total}
\Delta_i= \sum \limits_{j =1, j\ne i}^M \Delta_{ij}.
\ee
\end{enumerate}
It is important to stress that  the collision phase shifts are far-field effects. Mathematically they are artefacts of the asymptotic representation of the exact two-soliton solution of the KdV equation  in the form of a sum of two individual solitons: $u_2(x,t; \eta_1, \eta_2) \simeq u_s(x+\Delta_{12},t; \eta_1)+u_s(x+\Delta_{21},t; \eta_2)$, which
is only valid if solitons are sufficiently separated (the long-time asymptotics). The interaction of solitons is a complex nonlinear process  \cite{lax_integrals_1968} and the resulting wave field in the interaction region  cannot be represented as a superposition of the phase-shifted one-soliton solutions.

We first introduce soliton gas phenomenologically, as an infinite random ensemble of solitons distributed on $\mathbb{R}$ with some non-zero spatial density $\beta$, characterised by certain distributions over the spectral parameter $\eta_i \in \Gamma=[\eta_{\min}, \eta_{\max}] \subset \mathbb{R}^+$ and the phase $x_j^{0} \in \mathbb{R}$. Without loss of generality one can assume  $\Gamma =[0,1]$, but we shall keep the general notation for the time being. Assuming macroscopic equilibrium (spatial homogeneity), the {\it spectral density of states} (spectral DOS)  $f(\eta)$ of  soliton gas is  introduced in such a way that $f(\eta_0)\dd\eta$ gives the number of solitons with the spectral parameter $ \eta \in [\eta_0; \eta_0 + \dd \eta]$  contained  in the  portion of soliton gas over a unit interval of $x \in \mathbb{R}$ (the individual solitons can be counted by ``cutting out'' the relevant portion of the gas and letting them separate). The corresponding spectral flux density $v(\eta)$ represents the temporal counterpart of the spectral DOS, i.e. $v(\eta_0) \dd \eta$ is the number of solitons with the spectral parameter $ \eta \in [\eta_0; \eta_0+ \dd \eta]$ crossing any given  point $x=x_0$ per unit interval of time. These definitions are physically suggestive in the context of rarefied soliton gas where solitons are identifiable as individual localised wave structures. 
The  total spatial density of the soliton gas 
\begin{equation}\label{kappa}
\beta = \int_\Gamma f(\eta) \rmd \eta \, .
\end{equation}
For a rarefied gas, $\beta \ll 1$, the phases $x_i^0$  can be assumed to follow Poisson distribution on $\mathbb{R}$ with density $\beta$. In such a gas, the  total spatial shift of a ``tracer'' soliton with spectral parameter $\eta=\eta_1$ (we shall call it $\eta_1$-soliton) over the time interval  $\rmd t$, due to the interactions with $\mu$-solitons, $\mu \in \Gamma$, is given by $\Delta_1 \approx \int_\Gamma [\Delta(\eta_1, \mu) |s_0(\eta_1) - s_0(\mu)| f(\mu)  \rmd \mu ] \rmd t$, where $s_0(\eta)=4\eta^2$ is the speed of a free, non-interacting soliton. This simple argument was used by Zakharov in 1971 \cite{zakharov_kinetic_1971}.

It turns out that a straightforward generalisation of Zakharov's construction to the case of a dense gas can be made. This can be formulated as the {\it collision rate ansatz},  whereby the total position shift of the $\eta_1$-soliton, due to soliton collisions in a gas with DOS $f(\mu)$ over the time interval $\rmd t$,  is given by 
\be \label{collision_rate}
\Delta_1 = [\int_\Gamma \Delta(\eta_1, \mu)|s(\eta_1) - s(\mu)| f(\mu)\rmd \mu ] \rmd t,
\ee 
where  $s(\eta)$ is the effective velocity of the `tracer' soliton with spectral parameter $\eta$. In simple terms \eqref{collision_rate} represents an extrapolation of the rarefied gas properties to a dense gas, realised by replacing $s_0(\eta) \to s(\eta)$ in the collision rate expression.

It is important to stress  that the validity of \eqref{collision_rate}  for  a dense soliton gas is far from being obvious.  Indeed, as we mentioned,  the very notion of the phase shift in classical soliton theory is only applicable in the context of the long-time asymptotics, i.e. when solitons have sufficient time to separate from each other  after the interaction, which can only happen in a rarefied gas. In other words, the interactions between solitons can be treated as short-range only  if the typical distance between solitons in a gas is much greater than soliton's width. The assumption that, in a dense gas, where solitons experience significant overlap and continual interaction, the net effect of soliton collisions on the mean velocity is expressed in the same way as in the rarefied gas requires justification.  
Such a justification has been provided in \cite{el2003thermodynamic} in the framework of the  finite-gap spectral theory \cite{belokolos_algebro-geometric_1994}.

The collision rate ansatz \eqref{collision_rate} implies the integral equation for the effective velocity of a tracer soliton: 
\begin{equation}\label{eq_state_kdv} 
s(\eta)=4\eta^2+\frac{1}{\eta}\int _\Gamma \log
\left|\frac{\eta + \mu}{\eta-\mu}\right|f(\mu)[s(\eta)-s(\mu)]\dd\mu\, .
\end{equation}
Equation \eqref{eq_state_kdv} can be viewed as the {\it equation of state} of a dense homogeneous (equilibrium) KdV soliton gas. For a weakly non-homogeneous  (out of equilibrium) gas  we have $f(\eta) \rightarrow f(\eta; x, t)$, $s(\eta) \rightarrow s(\eta; x, t)$, where the $(x,t)$-variations of $f$ and $s$ occur on macroscopic, hydrodynamic scales, much larger than the typical scales associated with variations of the wave field $u(x,t)$ in individual solitons.
Now, isospectrality of the KdV evolution within the IST framework  implies the conservation equation
\be \label{kin_eq0}
\partial_tf+\partial_x(sf) = 0,
\ee
which, together with the equation of state \eqref{eq_state_kdv}, provides a spectral hydrodynamic/kinetic description of  the KdV soliton gas. {As we will now discuss, those last two equations can be rigorously recovered even in the context of a dense gas, in which the collision rate ansatz should a priori not apply.}

\subsection{Nonlinear dispersion relations for soliton gas and spectral kinetic equation}\label{sec:nonline_disp_rel}

As shown in \cite{el2003thermodynamic} (see also \cite{el2021soliton}), equation \eqref{eq_state_kdv} can be derived in the framework of
the multiphase  modulation (Whitham) theory \cite{whitham_linear_1974}, \cite{flaschka_multiphase_1980}  without making any assumptions about the gas' density or the collision rate. The derivation in \cite{el2003thermodynamic}  makes an extensive use of the spectral properties of the so-called  finite-gap potentials \cite{belokolos_algebro-geometric_1994}, the quasi-periodic generalisations of $N$-soliton solutions of the KdV equation exhibiting the band Lax spectrum, $\la \in  \mathcal{S}_N \equiv \cup _{i=1}^{N+1} \gamma_i , \quad \gamma_i \cap \gamma_j= \emptyset, \ \ i \ne j$, where  the spectral bands $\gamma_j \subset \mathbb{R}, j=1, 2, \dots, N$ are finite intervals and $\gamma_{N+1}$ -- a semi-infinite interval. The $N$-soliton limit of an $N$-gap solution is achieved by collapsing all the finite bands $\gamma_j$ into double points corresponding to the solitonic spectral values $\la_j$.

It was proposed in \cite{el2003thermodynamic} that the soliton gas can be described by the thermodynamic limit of the spectral $N$-gap KdV solutions,  achieved by assuming a special band-gap distribution (scaling) of the  spectral set $\mathcal{S}_N$ for $N \to  \infty$ on a fixed  interval (say $\la \in [-1, 0]$) whereby the finite spectral bands are required to be exponentially narrow compared to the gaps $\sim 1/N$ between them. Such spectral scaling enables the  soliton limit  of finite-gap potentials as $N \to \infty$ while preserving finite spectral DOS. 

Within the spectral thermodynamic limit approach, the spectral DOS $f(\eta)$ and the spectral flux density $v(\eta)$ of an equilibrium (uniform) soliton gas are  defined via the {\it nonlinear dispersion relations}. The nonlinear dispersion relations for soliton gas are 
 derived by applying the thermodynamic spectral limit to the wavenumber-frequency relations for the finite-gap KdV solutions  \cite{flaschka_multiphase_1980} (see \cite{el2003thermodynamic, el2021soliton} for details). Let $\eta =\sqrt{-\la}$, then  the spectral DOS $f(\eta)$ and the corresponding spectral flux density $v(\eta)$  satisfy 
 \be \label{inta}
\int_\Gamma \log \left|\frac{\eta + \mu}
{\eta-\mu}\right| f(\mu)\rmd\mu  + \sigma(\eta) f(\eta)
=  \eta\, ,
\ee
\be \label{intat}
\int_\Gamma \log \left|\frac{\eta + \mu}
{\eta-\mu}\right|v(\mu) \rmd \mu  + \sigma(\eta)v(\eta)
=4\eta^3 \, .
\ee
Here $\sigma(\eta)\geq 0$ is the {\it spectral scaling} function characterising the Lax spectrum of the soliton gas while $\Gamma$ is the support of $f(\eta)$ and $v(\eta)$. 
Eliminating $\sigma$  one obtains the equation of state
\eqref{eq_state_kdv} for the effective velocity of a ``tracer'' soliton in the gas, $s(\eta)=v(\eta)/f(\eta)$. The  derivation of the equation \eqref{eq_state_kdv}  via the thermodynamic limit thus provides the justification of the collision rate assumption \eqref{collision_rate} for the KdV soliton gas. 
The  continuity  (kinetic) equation \eqref{kin_eq0} for slowly varying spectral DOS $f(\eta; x, t)$ in a non-equilibrium (weakly non-unifiorm) soliton gas is obtained as a thermodynamic limit of  kinematic modulation equations expressing the ``conservation of waves'' \cite{el2003thermodynamic, el2021soliton}. A similar derivation has been recently performed in \cite{el_spectral_2020} for the focusing NLS equation where the spectral parameter $\la$ is complex. We note that the spectral support $\Gamma$ should not necessarily be a fixed simply connected set: the general case when $\Gamma$ is given by a union of disjoint intervals whose endpoints are allowed to vary in space-time, has been considered for the KdV gas in \cite{congy2022dispersive}.

The equation for the evolution of $\sigma(\eta; x,t)$ in a non-equilibrium soliton gas follows from \eqref{inta}, \eqref{eq_state_kdv} and \eqref{kin_eq0}. A direct computation   yields for a fixed $\Gamma$
\be \label{cont_riemann}
\partial_t\sigma + s \partial_x\sigma=0,
\ee
which  is equivalent to the spectral kinetic equation.
Thus the function $\sigma(\eta, x, t)$ plays the role of the Riemann invariant in the soliton gas kinetics/hydrodynamics.

Furthermore, it can be shown that the following conditions are satisfied \cite{bulchandani2017solvable}:
\begin{equation}\label{semi-h}
	\int \dd\nu \left[\frac{\delta}{\delta\sigma(\nu)}\left(\frac{\delta s(\eta)/\delta\s(\mu)}{s(\mu)-s(\eta)}\right)\right]=\int \dd\mu \left[\frac{\delta}{\delta\sigma(\mu)}\left(\frac{\delta s(\eta)/\delta\s(\nu)}{s(\nu)-s(\eta)}\right)\right], \quad \mu \ne \nu \ne \eta \, ,
\end{equation}
\begin{equation}\label{lin_degen}
	\frac{\delta s(\eta)}{\delta\sigma(\eta)}=0, \quad \forall \eta \; .
\end{equation}
Relations \eqref{semi-h} and \eqref{lin_degen} are the  continuum limit counterparts of the semi-Hamiltonian (integrability) and linear degeneracy properties  respectively of the  hydrodynamic reductions of the  kinetic equation \eqref{kin_eq0}, \eqref{eq_state_kdv}, established in \cite{el_kinetic_2011}, see also \cite{bulchandani_classical_2017, ferapontov2021kinetic} for further developments on integrability of hydrodynamic reductions.

Finally we note that the  limits $\sigma \to \infty$ and $\sigma \to 0$  correspond to the special cases of soliton gas: the ideal (noninteracting) soliton gas ($\sigma \to \infty$) and the {\it soliton condensate}  ($\sigma \to 0$). In the latter case, equations \eqref{inta} and \eqref{intat} can be integrated explicitly yielding the critical DOS $f_c(\eta)$ and the corresponding spectral flux $v_c(\eta)$ \cite{congy2022dispersive}-- see Appendix~\ref{app:solclass}.

\subsection{Conserved quantities}
\label{sec:cons_quan}

One of the fundamental properties of integrable dispersive hydrodynamics  is the availability of 
an infinite set of local conservation laws 
\be\label{cons_law}
\partial_tP_n+\partial_xQ_n=0, \ i=0, 1, 2, \dots\; , 
\ee
where the  $P_i$ and $Q_i$ are functions of the field variable $u$ and its derivatives (the so-called local polynomial  functionals). Of particular interest are the first three conserved KdV densities
\be \label{kruskal_norm}
P_0=u, \quad P_1=u^2, \quad P_2 = \frac{u_x^2}{2} -  u^3 \, , 
\ee
typically associated with  conservation of ``mass'', ``momentum'' and ``energy''.
Let $u(x,t)$ describe the random KdV wave field in a soliton gas -- the ``integrable turbulence'' \cite{zakharov_turbulence_2009}. In the spirit of the Whitham modulation theory \cite{whitham_linear_1974}  we invoke  scale separation and, assuming ergodicity of the soliton gas,  apply the ensemble averaging to the KdV conservation laws  \eqref{cons_law} to obtain the  modulation system (note however that the original Whitham theory uses spatial averaging)
\begin{equation}\label{stoch_Whitham}
\partial_t \langle P_n[u ] \rangle + \partial_x \langle Q_n[u]  \rangle = 0, \quad
i=0, 1, 2, \dots
\end{equation}
Naturally, system \eqref{stoch_Whitham} must be consistent with the kinetic equation. Indeed, the kinetic equation \eqref{kin_eq0} implies that for any $h(\eta) \ne 0$, $\int_\Gamma h(\eta)f (\eta; x, t) \rmd \eta$ is a conserved quantity (density) with  $\int_\Gamma h(\eta)f (\eta; x, t) s(\eta; x, t)\rmd \eta$ being the corresponding flux density. The specific ``Kruskal'' series  in \eqref{stoch_Whitham} is obtained as \cite{el2003thermodynamic}
\be \label{kdv_integrals}
\langle P_n[u ] \rangle = C_n\int_\Gamma \eta^{2n+1} f(\eta) \rmd \eta\; , \quad \langle Q_n[u ] \rangle = C_n\int_\Gamma \eta^{2n+1} f(\eta) s(\eta) \rmd \eta\; ,
\ee
where, for the first three conserved densities,  we have 
\begin{equation}\label{moments_KdV}
\langle P_0 \rangle \ =\ 4\int_\Gamma \eta f(\eta )\,\ud\eta \,, \quad
\langle P_1 \rangle \ =\ \dfrac{16}{3}\int_\Gamma \eta ^{3}f(\eta)\,\ud\eta \, , \quad \langle P_2 \rangle =\dfrac{32}{5}\int_\Gamma \eta ^{5}f(\eta)\,\ud\eta \; .
\end{equation}
We should remark that the specific coefficients $C_n$ in \eqref{moments_KdV} are relevant in the context of the particular normalisation \eqref{kruskal_norm} of the polynomial KdV integrals. However, the resulting  conservation system 
\begin{equation}
\frac{\partial}{\partial t} \int_{\Gamma} \eta^{2n+1}f(\eta;x,t) \ud\eta + \frac{\partial}{\partial x} \int_{\Gamma} \eta^{2n+1} f(\eta;x,t)s(\eta; x,t) \ud\eta =0,
\end{equation}
equivalent to \eqref{stoch_Whitham}, does not depend on the normalisation choice.

We conclude this section on KdV soliton gas by mentioning two fundamental restrictions imposed on the admissible spectral DOS  $f(\eta)$. The first one follows naturally from non-negativity of the variance  
\begin{equation}\label{var}
  \mathcal{A}\ =\ \sqrt{\langle {u^2} \rangle \ -\ \langle {u} \rangle ^2}\ \geqslant\ 0,
\end{equation}
or equivalently, recalling equations \eqref{kruskal_norm} and \eqref{kdv_integrals},
\begin{equation}
    \int_\Gamma \eta ^{3}f(\eta)\,\ud\eta - 3\left(\int_\Gamma \eta f(\eta )\,\ud\eta\right)^2 \geqslant\ 0\; .
\end{equation}
The second constraint, of purely spectral nature, arises due to the requirement of nonnegativity of the spectral scaling function $\sigma(\eta)$ in the nonlinear dispersion relations \eqref{inta}, \eqref{intat}, hence
\begin{equation}
    \frac{1}{\e}\int_\Gamma \log \left|\frac{\eta + \mu}
{\eta-\mu}\right| f(\mu)\rmd\mu \geqslant\ 1 \; .
\end{equation}

{\subsection{Thermodynamics and large-scale correlation functions}}

{The main results of this paper, besides making the dictionary between the solition gas and generalised hydrodynamics, is to provide new conjectures for exact results in the soliton gas. These results are concerned with the thermodynamics and large-scale correlation functions, which have not been studied yet in soliton gases. We therefore conclude this section by providing an early heads up for the dispersive hydrodynamics readership to these main results, without the need for a full understanding of GHD. Namely, we present expressions for the temperature, free energy and correlation functions in the KdV solilton gas in terms of the DOS or, equivalently, the spectral scaling function $\sigma(\eta)$. The full discussion detailing the GHD conjectures necessary to explicitly derive the expressions below can be found in Section \ref{sec:conj}. The relevant GHD tools are introduced in Sections \ref{sec:GHDLL} and \ref{sec:dico}.}

{For the sake of simplicity we shall focus here on expressing the spectral density for thermal states, and on expressing correlations of the KdV field $u(x,t)$ in generic states; generalised Gibbs ensembles and correlations of other densities $P_n$ can also be accessed.}

{The first result is the spectral scaling function for a thermal state. The definition of a thermal state for the KdV equation, or for a soliton gas, is a subtle question. In this paper we take a natural definition based on the soliton gas viewed as the limit of large soliton numbers, and large lengths, of ensembles of solitons. An ensemble of $N$ soliton on an interval of length $L$ is a ensemble of configurations of the KdV field, supported on the interval (and quickly tending to zero beyond it), that decomposes asymptotically at large times into a train of $N$ separate solitons. We distribute these solitons according to a Boltzmann weight, with the temperature being the parameter associated to the Hamiltonian, $\int P_2(x)\dd x$. See Subsection \ref{ssectGGEkdv}. With this definition, the spectral scaling function for a thermal state in terms of the inverse temperature $\beta$ is}
\begin{equation}
    \sigma(\e) = \frac{\pi}{4}\exp\left[\frac{32}{5}\beta\eta^5 + \int_{\Gamma}\frac{\dd\mu}{\s(\mu)}\log\left|\frac{\e+\mu}{\e-\mu}\right|\right] \; .
\end{equation}

{In fact for a general spectral function (not necessarily that of a thermal state), the inverse temperature is defined more generally as}
\begin{equation}
\begin{aligned}
   \beta 
   & = \frac{1}{768}\frac{\partial^{5}}{\partial\e^5}\left\{\log \left[\frac{4\sigma(\e)}{\pi}\right] - \int_\Gamma \frac{\dd\mu}{\s(\mu)}\log\left|\frac{\e+\mu}{\e-\mu}\right|\right\}\; .  
\end{aligned}
\end{equation}
{At this level of generality, one may also evaluate the entropy (this is the analogue of the Yang-Yang entropy of quantum integrability \cite{yang1969thermodynamics}), which takes the simple form}
\be
\mathcal S  = \int_\Gamma\left(\log\left[\frac{4\sigma(\e)}{\pi}\right]+1\right)f(\e)\dd \e \;,
\ee
{reaching its minimal value in the condensate limit $\s \to 0$ alluded to in Section \ref{sec:nonline_disp_rel}. 

Finally we can evaluate correlations and fluctuations of the field: the static covariance is given by}
\be
\int \dd x \left[\bra u(x,t)u(0,t)\ket-\bra u(x,t)\ket\bra u(0,t)\ket\right] = 16\int_\Gamma\dd\eta\,\sigma(\e)^2\,f(\eta)^3 \; ,
\ee
{and the dynamical correlation function of the KdV field at the Euler scale of large position and time separation is}
\be
\begin{aligned}
	 \frc1{2\lambda} \int_{-\lambda}^{\lambda} \dd x\,\big[\bra u(\xi t+x, t)u(0,0)\ket&-\bra u(\xi t+x, t)\ket\bra u(0,0)\ket\big]\\
	 &\sim
	\frc1t \sum_{\eta\in \eta_*(\xi)} \frc{16}{|s'(\eta)|} \,\sigma(\e)^2\,f(\eta)^3\quad (t\gg \lambda\to\infty) \; ,
\end{aligned} 
\ee
{where $\eta_*(\xi) = \{\eta:s(\eta) = \xi\}$.}

\section{Generalised hydrodynamics}\label{sec:GHDLL}

\subsection{Introduction}

In the theory of Generalised Hydrodynamics (GHD), a rather different viewpoint is taken from that described in the previous section. The idea is to directly apply the general principles of hydrodynamics (we concentrate on Euler hydrodynamics here) to many-body integrable systems, in order to predict their dynamics away from equilibrium.

In the simplest expression of GHD, we consider a system, composed of many particles in interaction, which has the property of integrability. The system can be either quantum or classical, but the theory was initially developed to discuss the quantum case in the original works \cite{castro2016emergent} and \cite{bertini2016transport}. This includes the Lieb-Liniger model, considered in \cite{castro2016emergent}, which will be our main example. Additionally, there is no need to restrict to systems of particles, and the Heisenberg spin chain was considered in \cite{bertini2016transport}.

In GHD, we assume that the system is initially in a fluctuating state which is, at every point in space, very well described by a state of Gibbs form, where entropy is maximised with respect to all available conservation laws. In integrable models these are the generalised Gibbs ensembles (GGEs), widely studied \cite{eisert2015quantum,essler2016quench,ilievski2015complete,langen2015experimental}. For instance, the initial state could be a profile of temperature, where at each point the system is thermal; but it could be something more general, where many conserved charges are involved. The system is taken to be infinite, although GHD with external force fields, which may constrain the system to a finite length, has also been developed. The fluctuations of the full system may be with respect to any ensemble (micro-canonical, canonical, grand-canonical, and more), as this does not affect the description of local states.

Then, the evolution of this state is obtained by applying the hydrodynamic principles: the state is assumed to be, throughout time, well described by local Gibbs-like states, and the dynamics is determined by the continuity equations for all conservation laws admitted by the model.

As usual, the above program requires knowledge of the thermodynamics -- all the ``equilibrium" states to which the system locally relaxes, in integrable models these are the GGEs -- and of the equations of state -- how the average ``currents" are related to the average ``densities", these being the objects involved in the local conservation laws. Solving the latter question was the main technical achievement of the original works \cite{castro2016emergent,bertini2016transport}.

In this program, no consideration is made of the explicit microscopic dynamics: one only needs thermodynamic quantities, and one uses hydrodynamic principles to conjecture an emergent evolution equation. The result is supposed to be valid whenever changes of all local probes occur on large enough scales in space and time.

In order to illustrate these ideas, we will consider the  Lieb-Liniger model \cite{lieb1963exact} (which can be viewed as a quantum nonlinear Schr\"odinger equation \cite{rosenzweig2019mean}). This is a system of $N$ bosons with repulsive delta interactions described by the Hamiltonian
\begin{equation}\label{HLL}
	\hat H_\text{LL} = -\frac{1}{2}\sum_{i=1}^{N} \frac{\partial^2}{\partial x_i^2} + 2c\sum_{i>j} \delta(x_i-x_j) \; ,
\end{equation}
where $c$ is a positive constant parametrising the strength of interactions. This system is integrable in the sense that its Hamiltonian is part of a family of conserved quantities (also called charges) in involution
\begin{equation}\label{LLinv}
	\hat H_\text{LL} \equiv \hat Q_2 \; , \quad \quad [\hat Q_n,\hat Q_m]=0 \; .
\end{equation} 
This includes the number of particles and momentum,
\beq
    \h Q_0 = N{\bf 1}\, ,\quad
    \h Q_1 = \h P = \sum_{i=1}^N (-\ri) \frc{\p}{\p x_i}\;.
\eeq
Higher conserved charges have a more complicated form \cite{davies1990higher,davies2011higher}. Their explicit expressions are not important, except for noting that they are deformations of the natural conserved charges of free-particle models, the higher-power of momenta:
\beq\label{hatQn}
    \hat Q_n = \frc1{n!}\sum_{i=1}^{N} \lt(-\ri \frac{\partial}{\partial x_i}\rt)^n + \sum_{i=1}^N \h V_n(x_i-x_1,\ldots,x_i-x_N)\;.
\eeq
Here $\h V_n(x_i-x_1,\ldots,x_i-x_N)$ is a differential operator with derivatives up to a maximal order $n-1$, and with coefficients that are functions of $x_i-x_1,\ldots,x_i-x_N$. It does not depend on $x_j$ whenever $|x_i-x_j|$ is large enough, and tends to zero when $|x_i-x_j|$ tend to infinity for all $j\neq i$. The normalisation in \eqref{hatQn} is purely conventional. The existence of this set of conserved charges ensures that many-particle scattering is elastic (it preserves all momenta), and factorizes into separate two-body scattering events  \cite{arutyunov2019factorised}.

A crucial notion is that of extensivity of the charges $\h Q_n$: locality or quasi-locality of their densities  \cite{ilievski2016quasilocal}. That is, we may write
\begin{equation}\label{GHDcharges}
	\h Q_n=\int \dd x\,\h q_n(x,t) \; ,
\end{equation}
where the density $\h q_n(x,0)$ is an operator ``supported" at, or on some finite region around, the point $x$. A full and accurate notion of support would need more care, but, as a part of this notion, $\h q_n(x,0)$ should commute with $\h q_0(x',0) = \sum_i \delta(x'-x_i)$ for $|x-x'|$ large enough. By conservation of the charges $\p_t \h Q_n=0$, these satisfy the microscopic conservation laws
\begin{equation}\label{microcons}
	\partial_t  \h q_n(x,t) + \partial_x  \h j_n(x,t) =0 \; ,
\end{equation} 
for some local or quasi-local currents $\h j_n(x,t)$.

We will start by discussing the thermodynamic limit of such a model and how integrability impacts the properties of its stationary states. To that end we will rely on the celebrated Thermodynamic Bethe Ansatz (TBA). This was initially developed to specifically deal with the Lieb-Liniger model \cite{yang1969thermodynamics}, but it turns out that its applications are significantly more far-reaching \cite{franchini2017introduction}. We will then use these results to construct a hydrodynamic theory of integrable systems according to \cite{doyon_lecture_2020}. As we shall see in the next section, parallels are to be drawn between quantum particles and solitons.

Note that, in the following, we consider systems featuring only one type of particle but all of this can easily be extended as discussed for instance in \cite{doyon2018exact}.

\subsection{Stationary states}\label{sec:stationary}

One of the most important principles of statistical mechanics is that the set of all states to which a many-body system may relax, as seen by local probes, is expected to be the set of all stationary, clustering states. A clustering state is one where local observables have vanishing correlations at large distances; clustering state are ``ergodic" (in some sense). In typical, non-integrable gases, a so-called ``ergodic principle" says that these are the Gibbs states and their Galilean (or relativistic) boosts, with density matrix
\beq
    \h\rho \propto e^{-\beta (\h H-\mu \h N + \nu \h P)} \, ,\quad \Tr \h\rho = 1\; .
\eeq
Thus, in the inertial frame, these are equilibrium states. In integrable models, they may involve any combination of the local charges
\beq\label{GGE}
    \h\rho \propto e^{-\h W} \; ,
\eeq
for any extensive conserved quantity $\h W$, typically of the form
\beq
    \h W = \sum_n \beta_n \h Q_n\;.
\eeq
These are the GGEs. This in fact implies that stationary, clustering states include those which carry nonzero currents which cannot be made to vanish by simple boosts, and thus are non-equilibrium steady states. Extensivity of $\h W$ guarantees that the resulting state is clustering. As the Hamiltonian is homogeneous, the stationary, clustering states are invariant under space-time translations.

An equivalent description of the stationary, clustering states \eqref{GGE} is as those which maximise entropy with respect to the conserved charges. The $\beta_n$'s are the generalised ``inverse temperatures" associated to the conserved charges $Q_n$ (the Lagrange multipliers for the constraints of conserved charges). 

The GGEs are most efficiently described via the TBA.

\subsubsection{Bethe Ansatz}\label{sec:BA}

Consider a $N$-particle eigenstate of the Hamiltonian \eqref{HLL} on a circle of length $L$. As the Hamiltonian is free whenever coordinates are disjoint, in each region of a given ordering, the wave-function is a solution of the free Schr\"odinger equation. Thus it can simply be written as a superposition of plane waves
\begin{equation}\label{psill}
	\Psi_\text{LL}(x_1,...,x_N ) = \sum_{\mathcal{P}} \mathcal{A}_{\mathcal{P}}(\mathcal Q)\exp\left[\ri\sum_{j=1}^{N}p_{\mathcal P(j)}x_j\right] \; ,
\end{equation}
where the set of ``quasi-momenta" ${p_i}$, no two of which are identical \cite{yang1969thermodynamics}, will be determined by the periodicity conditions. {The sum is over the group of all permutations $\mathcal{P}$ of $N$ elements; $\mathcal{P}$ represents permutations of the quasi-momenta. $\mathcal{Q}$ is also a permutation of $N$ elements; it is not summed over, but rather is fixed by the positions $x_1,\ldots,x_N$ (thus it is a function of the positions) and represents the ``particle order". By definition, it is the permutation satisfying $x_{\mathcal{Q}^{-1}(i+1)} > x_{\mathcal{Q}^{-1}(i)}$ (where $\mathcal{Q}^{-1}$ is the inverse permutation). Amplitudes are permutation invariant, in the sense that $\mathcal{A}_\mathcal{P}(\mathcal Q)= \mathcal{A}_{\mathcal{P}\circ \mathcal{P}'}(\mathcal Q\circ \mathcal{P}')$ for all permutations $\mathcal{P}'$; this in fact guarantees that the resulting wave-function is indeed bosonic, i.e.~invariant under permutations of its arguments (the particles' positions).} Amplitudes are fixed by the delta-function interaction, and their ratios give rise to scattering phases, as the scattering problem can be obtained from the limit $L\to\infty$. With $\mathcal{P'}$ differing from $\mathcal{P}$ only through permutation of quasi-particles $i$ and $j$, the ratio involves the two-particle phase shift $\phi(p_i,p_j)$,
\begin{equation}\label{ALL}
	\frac{\mathcal{A}_\mathcal{P}}{\mathcal{A}_\mathcal{P '}} =\exp\left[\ri \phi(p_i,p_j)\right] =\frac{p_i-p_j -\ri c}{p_i-p_j +\ri c} \; .
\end{equation}
The above form of the wave function implies that three-body and higher processes factorise into two-body processes \cite{arutyunov2019factorised}. This is the blueprint for every Bethe-ansatz integrable model.

If we have $N$ particles on a circle, the wave function must be periodic.  Assuming the particles did not interact, the wave function would simply acquire a factor $e^{ip_iL}$ if we were to take one particle of momentum $p_i$ around the circle and back to its position, leading to the usual quantization conditions 
\begin{equation}\label{FQC}
	p_i = \frac{2\pi I_i}{L} \; , \quad \quad I_i\in \mathbb{Z} \; .
\end{equation}
Taking into account the interactions, however, the same particle will have to scatter through all the others, accumulating phase-shifts and yielding the Bethe equations \cite{levkovich2016bethe}
\begin{equation}\label{BA}
	p_i L = 2\pi I_i + \sum_{j\neq i}\phi(p_i,p_j) \; , \quad i=1,...,N \; ,
\end{equation}
where $I_i \in \mathbb{Z}$. Indeed, if $\phi = 0$, without interactions, equation \eqref{BA} becomes equivalent to equation \eqref{FQC}, of which it can be considered an extension.

As mentioned, the Lieb-Liniger model admits an infinite number of conserved charges. Although these are difficult to construct in generality, as they are local deformations of higher powers of momenta, their eigenvalues can be evaluated by acting on asymptotic states, where particles are well-separated and do not interact; it turns out that the finite-$L$ effects on the wave-function \eqref{psill} identically vanish, hence these are the correct eigenvalues in general. Eigenvalues of charges $\hat Q_n$ on eigenstates $|\{p_i\}\ket$ take the form
\begin{equation}
	Q_n\equiv \sum_i h_n(p_i) \; , \quad \quad \hat Q_n|\{p_i\}\rangle = Q_n|\{p_i\}\rangle  \; ,
\end{equation}
where $h_n(p_i)$ is the amount of charge $Q_n$ carried by particle $i$. In the case of the Lieb-Liniger model, we have the total number of particles $\hat Q_0 = N{\bf 1}$, the total momentum $\hat Q_1 =  -\ri\sum_i \frc{\p}{\p x_i}$, and the total energy  $\hat Q_2 = \hat H_{\rm LL}$, with
\beq
    h_0(p) = 1\, ,\quad h_1(p) = p\, ,\quad
    h_2(p) = p^2/2\;.
\eeq
More generally eigenvalues take the form $Q_n\propto\sum_i p_i^n$.

As the momentum operator $\h Q_1 = \h P$ and the energy $\h Q_2 = \h H_{\rm LL}$ play an important role, we give their one-particle eigenvalues special names:
\beq\label{LLPE}
    P(p) = h_1(p)\,,\quad E(p) = h_2(p)\;.
\eeq
TBA can be written in universally applicable forms using these objects.

\subsubsection{Thermodynamics}\label{sec:TBA}

We are looking to take the thermodynamic limit of the states \eqref{GGE}, in the sense that both $L\rightarrow \infty$ and $N\rightarrow \infty$ while $N/L$ is kept constant,
\beq
    \bra \cdots\ket = \lim_{L\to\infty}
    \Tr \Big(\h\rho \cdots\Big).
\eeq
In fact it will be simpler to allow $N$ to fluctuate and keep $\bra N\ket/L$ constant; thus $\h Q_0$ is included in the exponential in \eqref{GGE}. We can characterise $\h\rho$ by its eigenvalues on the eigenstates $|\{p_i\}\ket$:
\beq
    \rho \propto e^{-\sum_i w(p_i)}.
\eeq
The function $w(p)$ replaces the Lagrange multipliers, as
\beq
    w(p) = \sum_n \beta_n h_n(p)\;.
\eeq
There is no explicit need for the Lagrange multipliers $\{\beta_n\}$, the GGE is fully fixed by the spectral function $w(p)$. This defines a good GGE under the condition that it grows fast enough as $|p|\to\infty$, and is bounded from below. The set $\{\beta_n\}$, and the function $w(p)$, form systems of coordinates in the manifold of maximum entropy states.

The thermodynamic limit of the Lieb-Liniger model was first investigated in \cite{yang1969thermodynamics}. For a generic state, we can approximate the distribution of momenta by a spectral, or phase space, density $\rho_{\rm p}(p)$, which is such that $\rho_{\rm p}(p_i) \approx (L(p_{i+1}-p_i))^{-1}$. Defining a density of holes by $\rho_{\rm h}(p_i)= (I_{i+1}-I_i)\rho_{\rm p}(p_i)$, the Bethe equations \eqref{BA} become \cite{van2016introduction}
\begin{equation}\label{TBA}
	2\pi \rho_{\rm s}(p)= P'(p) - \int \dd p'\,\varphi(p,p')\rho_{\rm p}(p') \; ,
\end{equation}
where
\beq
    \rho_{\rm s}(p) = \rho_{\rm p}(p) + \rho_{\rm h}(p) \; ,
\eeq
is the total density or ``asymptotic space density" (a notion that will become clear later), where
$P'= \dd P/\dd p$ is the derivative of the momentum eigenvalue ($=1$ in the Lieb-Liniger model), and $\varphi(p,p')=\partial_p\phi(p,p')$ is the differential scattering phase. Eigenvalues of conserved charges can now be computed as
\begin{equation}\label{QTBA}
	\lim_{L\to\infty} \frc{Q_n}L = \int \dd p \,h_n(p)\rho_{\rm p}(p) \; .
\end{equation}

In a ``micro-canonical ensemble", there are several states with almost identical charges $\{Q_n\}$, i.e. with the same charge density, as averaged densities are not sensitive to the redistribution of particles within infinitesimal regions of quasi-momenta. Equivalently, these are the states that are consistent with a given pair $(\rho_{\rm p},\rho_{\rm h})$. In an interval $\dd p$ the number of different configurations $\mathcal N$ yielding the same densities $\rho_{\rm p}$ and $\rho_{\rm h}$ is given by
\begin{equation}
	\mathcal N = \frac{\left[L(\rho_{\rm p}+\rho_{\rm h})\dd p\right]!}{\left[L\rho_{\rm p} \dd p\right]!\left[L\rho_{\rm h} \dd p\right]} \; .
\end{equation}
We may then define an entropy  from $\log\mathcal N(\rho_{\rm p},\rho_{\rm h})$ which can be evaluated in the thermodynamic limit ($L\rightarrow \infty$) using Stirling's formula
\begin{equation}
	\begin{aligned}
		&\mathcal S[\rho_{\rm p},\rho_{\rm h}] = L\int \dd p\,\big( (\rho_{\rm p}+\rho_{\rm h})\log(\rho_{\rm p}+\rho_{\rm h})-\rho_{\rm p}\log\rho_{\rm p}-\rho_{\rm h}\log\rho_{\rm h}\big)
	\end{aligned}\; .
\end{equation}

Following Yang and Yang's argument \cite{yang1969thermodynamics} and its generalization \cite{mossel2012generalized}, the GGE distribution can be described using a partition function
\begin{equation}
	\mathcal{Z}=\int\mathcal{D}[\rho]\exp\big(\mathcal S\left[\rho_{\rm p},\rho_{\rm h}[\rho_{\rm p}]\right]-w[\rho_{\rm p}]\big) \; ,
\end{equation}
where, in the entropy $\mathcal S$, $\rho_{\rm h}$ has been expressed as a functional of $\rho_{\rm p}$ using the thermodynamic Bethe equation \eqref{TBA}, and $w[\rho] = \int \dd p\,w(p)\rho_{\rm p}(p)$. We evaluate this partition function under the saddle point approximation, with saddle point condition
\begin{equation}
	\begin{aligned}\label{YY}
		& \epsilon(p) = w(p) - \int \frc{\dd p'}{2\pi}\,\varphi(p,p')\log\left(1+e^{-\epsilon(p')}\right)  \; ,
	\end{aligned}
\end{equation}
where $\epsilon=\log[\rho_{\rm h}/\rho_{\rm p}]$. This function is referred to as the ``pseudo-energy". This is because it allows us to define an occupation function 
\beq\label{defn}
    n=\frc{\rho_{\rm p}}{\rho_{\rm s}} \; ,
\eeq
and recover the form it takes for the Fermi-Dirac statistics
\begin{equation}\label{occupation}
	\begin{aligned}
		& n=\frac{e^{-\epsilon}}{1+e^{-\epsilon}} 
	\end{aligned} \;.
\end{equation}
In a similar fashion, the free energy contribution of a Fermi particle of energy $\epsilon$ \cite{zamolodchikov1990thermodynamic}
\begin{equation}\label{fEf}
	F(\ep) = -\log\left(1+ e^{-\epsilon(p)}\right)\; , \quad \quad n(p) = \left.\frac{\dd F(\epsilon)}{\dd\epsilon} \right|_{\epsilon=\epsilon(p)} \; ,
\end{equation}
yields the free energy density $\mathcal{F} = -L^{-1}\log \mathcal Z\big|_{\text{saddle point}}$ of the system, in the form
\begin{equation}\label{fEd}
	\mathcal{F} = \int \frac{\dd P(p)}{2\pi} F\left(\epsilon(p)\right) \; .
\end{equation}
The Fermi-Dirac statistics is natural for the Lieb-Liniger model, even though it is originally a bosonic theory, because of the delta-function interaction, which forbids particles from being at the same point.

For a given function $w(p)$, or set of Lagrange multipliers $\{\beta_n\}$, equation  \eqref{YY} completely determines the stationary state. Although the state is fluctuating, the fluctuations accounted for do not affect the charge densities (they are fixed to the saddle-point value), hence their averages still take the form \eqref{QTBA},
\beq\label{avq}
    \bra \h q_n\ket = \int \dd p \,h_n(p)\rho_{\rm p}(p)\;,
\eeq
where $\rho_{\rm p}(p)$ is evaluated at the saddle point. The value of $\rho_{\rm p}(p)$ can be established by expressing averages as derivatives of the free energy density
\begin{equation}
	\bra \hat q_n(0,0)\ket = \frac{\partial \mathcal{F}}{\partial \beta_n}\;.
\end{equation}
This gives us the relation between phase-space densities and Lagrange multipliers $\beta_n$, or spectral weight $w(p)$, which is the main expression of the thermodynamics of the model. We obtain it in the form of integral equations,
\beq\label{rhopdress}
    2\pi \rho_{\rm p} = n ({P'})^{\rm dr} \; ,
\eeq
where the ``dressing operation" is a linear operation defined by the integral equation
\beq\label{dressing}
    h^{\rm dr}(p)
    = h(p) + \int \frac{\dd p'}{2\pi}\,
    \varphi(p,p') n(p') h^{\rm dr}(p') \; .
\eeq
Recall that in the Lieb-Liniger model $P'(p) = 1$ is the constant unit function.

Conversely, if $\rho_{\rm p}$ is known, then $\rho_{\rm s}$ is evaluated by \eqref{TBA} and therefore the pseudo-energy is known, from which the Lagrange multipliers can be computed by differentiating the saddle-point condition \eqref{YY},
\begin{equation}\label{LmLL}
	\beta_n = \frac{\partial^n}{\partial p^n}\left.\left[\epsilon(p)-\int \frc{\dd p'}{2\pi}\,\varphi(p,p')F\left(\epsilon(p')\right)\right]\right|_{p=0} \; .
\end{equation}

\subsubsection{Equation of state}

In order to construct a hydrodynamic theory we need a way to relate averages of currents to that of charges: we need an equation of state. In statistical mechanics, this information is fully encoded within the ``free energy flux", first considered in \cite{castro2016emergent}. This is a function $\mathcal G$ of the state whose derivatives give rise to average currents
\beq
    \bra \h j_n\ket = \frc{\p \mathcal G}{\p \beta_n}\;.
\eeq
It is shown in \cite{castro2016emergent} that such a generating function always exist. In free particle models, one obtains a form similar to \eqref{fEd}, where the free energy $F\left(\epsilon(p)\right)$ controls the contribution of each mode, but with a measure determined by the energy $E(p)$ instead of the momentum. It is then natural to expect that this generalises to integrable models as
\begin{equation}\label{G}
	\mathcal{G} = \int \frac{\dd E(p)}{2\pi} F\left(\epsilon(p)\right)\;.
\end{equation}
This expression was first derived in the Lieb-Liniger model in \cite{castro2016emergent} using a property of relativistic quantum field theory called crossing symmetry, and taking the non-relativistic limit. Simple calculations then give the rather intuitive expression
\beq\label{Jvrho}
 \bra \h j_n\ket=\int \dd p\, h_n(p)v^{\rm eff}(p)\rho_{\rm p}(p) \; ,
\eeq
where
\begin{equation}\label{veff_LL}
	\begin{aligned}
		v^\text{eff}(p) &=  v^\text{gr}(p) + \int \dd p'\, \frac{\varphi(p,p')}{P'(p)}\rho_{\rm p}(p')\left[v^\text{eff}(p')-v^\text{eff}(p)\right] 
	\end{aligned} \; .
\end{equation}
Here $v^\text{gr}= E'/P'$ ($=p$ in the Lieb-Liniger model) is the group velocity (and recall that $P'(p) = 1$ in the Lieb-Liniger model). This effective velocity can be interpreted as the large-scale modification of the group velocity of a particle going through a gas as it accumulates scattering shifts when colliding with other particles. The expression \eqref{Jvrho} for the currents was given an alternative proof in quantum field theory \cite{cubero2021form}, and a full Bethe ansatz derivation in quantum spin chains in \cite{pozsgay2020algebraic}. Recently a general proof was obtained, valid in quantum and classical systems, using self-conserved currents in \cite{spohn2020collision,yoshimura2020collision}.

\subsection{Hydrodynamics}
\subsubsection{Hydrodynamic approximation}

As in regular hydrodynamics, GHD relies heavily on the separation of scales: systems are divided in fluid cells considered thermodynamically large, so that we assume entropy to be locally maximised, but small compared to the laboratory scales over which variations of local averages occur. That is, we assume that the averages of local observables can be well approximated, at large times, by averages evaluated in local GGEs
\begin{equation}\label{hydas}
	\langle \h o(x,t\rangle) \approx \langle \h o\rangle_{\{\beta_n(x,t)\}} \equiv \bar o_n(x,t)\; ,
\end{equation} 
where the Lagrange multipliers $\{\beta_n(x,t)\}$ now depend on $(x,t)$. Integrating the conservation laws \eqref{microcons} over a contour $[0,X]\times[0,T]$ and taking the average under approximation \eqref{hydas} we may write \cite{doyon_lecture_2020}
\begin{equation}\label{intcons}
	\int_{0}^{X} \dd x\,\left[\bar q_n(x,T)-\bar q_n(x,0)\right] + \int_{0}^{T}\dd t\,\left[\bar j_n(X,t)-\bar j_n(0,t)\right]=0	\; .
\end{equation}
If $\bar q_n$ and $\bar j_n$ are differentiable, macroscopic conservation laws \eqref{intcons} can be re-written in their differential form
\begin{equation}\label{avcons}
	\partial_t \bar q_n(x,t) + \partial_x \bar j_n(x,t) =0 \; ,
\end{equation}
where, this time, $\partial_x$ and $\partial_t$ represent large-scale derivatives encoding variations between fluid cells. Note that hydrodynamics is a derivative expansion; we will here only focus on the Euler scale, giving the lowest order approximation, and neglect for example diffusive,  dispersive and higher order terms.

\subsubsection{Fundamental equations}\label{sec:FunEq}
With the above, the fundamental equations of GHD are easy to derive. To that end we shall now characterize the states of locally maximised entropy by using results from section \eqref{sec:TBA}. Now that the Lagrange multipliers depend on $(x,t)$, we must introduce the space-time dependence
\begin{equation}\label{key}
	\rho_{\rm p}(p) \rightarrow \rho_{\rm p}(p;x,t) \; , \quad \quad v^\text{eff}\rightarrow v^\text{eff}(p;x,t) \; ,
\end{equation}
and may write the large-scale continuity equations \eqref{avcons} using TBA quantities \eqref{avq} and \eqref{Jvrho}
\begin{equation}\label{GHDint}
	\int \dd p\, h_n(p)\left(\partial_t\rho_{\rm p}(p;x,t) + \partial_x\left[v^\text{eff}(p;x,t)\rho_{\rm p}(p;x,t)\right]\right)=0 \; .
\end{equation}
The space of pseudo-local charges being complete \cite{doyon2017thermalization} we can finally obtain the fundamental equation of (Euler scale) generalised hydrodynamics
\begin{equation}\label{eulerGHD}
	\partial_t\rho_{\rm p}(p;x,t) + \partial_x\left[v^\text{eff}(p;x,t)\rho_{\rm p}(p;x,t)\right] = 0 \; .
\end{equation}
It is important to note that we can diagonalise this continuity equation using the occupation function defined in \eqref{defn}
\begin{equation}\label{neq}
	\partial_tn(p;x,t) + v^\text{eff}(p;x,t)\partial_xn(p;x,t) = 0 \; .
\end{equation}
Thus, the TBA occupation function plays the role of a continuous Riemann invariant for the GHD equation.

\section{Relation between KdV soliton gas and GHD}\label{sec:dico}

One can see some striking parallels between the spectral soliton gas and the GHD theories. In particular, the equation of state \eqref{veff_LL} and the continuity equation \eqref{eulerGHD} in GHD have the same structure as the spectral kinetic equations \eqref{eq_state_kdv}, \eqref{kin_eq0} in the soliton gas theory (up to the form of the phase shift kernel). Furthermore, the thermodynamic Bethe equation \eqref{TBA}  and the equation \eqref{neq} for the occupation function have their counterpart equations \eqref{inta}  and  \eqref{cont_riemann} respectively in the spectral soliton gas theory.

We shall now make explicit connections between the formalisms of soliton gases and GHD. We will re-contextualise the relevant quantities of one theory in terms of the other, methodically identifying the various notations, to make apparent what each approach can bring to the table. As direct by-product of this consideration, in the next section we shall then develop the thermodynamic characterisation of the KdV soliton gas in terms of entropy, free energy, temperature etc., and from this obtain and verify new conjectures for static covariances in the soliton gas.

\subsection{``Classical Bethe Ansatz" and KdV}\label{sec:TBAKdV}

We start this discussion by a heuristic, TBA-inspired, approach to soliton gases.  More explicitly, we follow the classical TBA approach, already discussed in the context of Toda gas in \cite{doyon2019generalized}, as a way to heuristically recover the dispersion relation \eqref{inta}, analogous to the thermodynamic Bethe ansatz equation \eqref{TBA}, albeit in the context of a  soliton gas. The aim of this section is to build intuition and further reinforce the aforementioned parallels between soliton gas theory and GHD. It will also serve as a vector to naturally introduce notions such as the ``asymptotic space'', and a geometric interpretation of the classical TBA which will provide much insight into the hydrodynamics and thermodynamics of the soliton gas.

As we previously discussed, the KdV equation supports multi-soliton solutions which, at sufficiently large time, can be asymptotically represented as a superposition of well-separated single-soliton solutions \eqref{kdv1sol},
\begin{equation}\label{Nsol}
 u_N \sim \sum_{i=1}^{N} 2\eta_i^2\text{sech}^2\left[\eta_i\left(x-4\eta_i^2t-x_i^{\pm}\right)\right]  \ \ \hbox{as} \ \ t \to \pm \infty \; .
\end{equation}
{Here $x_i^-$ and $x_i^+$ refer to the phases of soliton $i$ as $t \to - \infty$ and $t \to + \infty$, respectively. The ``phase" of soliton $i$ is the shift of the soliton's center with respect to the straight trajectory of velocity $4\eta_i^2$ passing by the origin. Then, $x_i^-$ is the phase before the soliton interacted with any other soliton in the gas, and $x_i^+$ is the phase after it accumulated shifts through interactions with every other soliton. This distinction is important as will be made clear below.} These solitons will serve as the quasi-particles of our ``gas", making for a rather literal interpretation of the ``soliton gas" appellation.

In order to describe this gas, we consider two different sets of dynamical variables that can be extracted from the multi-soliton solution; their relations will allow us to derive the TBA.

One set is simply composed of the $\{\eta_i\}$'s and the $\{x_i^-\}$'s. That is, we consider the positions at time $t=0$ obtained by extending the linear trajectories of asymptotic solitons, and their spectral parameters. These asymptotic coordinates can be seen as dynamical variables under the KdV equation: under the KdV evolution
\beq
	\mathfrak U_t u_N(\cdot,0) = u_N(\cdot,t) \; ,
\eeq
we have
\beq\label{evolveatxminus}
	\mathfrak U_t \eta_i = \eta_i,\quad\mathfrak U_t x_i^- = x_i^-(t) = x_i^- + 4 \eta_i^2 t.
\eeq
They are the action-angle variables of the integrable system, and the KdV equation makes $(\eta_i,x_i^-)$ into a set of phase-space coordinates for free particles with velocities
\beq\label{baregroup}
    v^{\rm gr}(\eta_i) = 4\eta_i^2\;.
\eeq

But the $\{x_i^-\}$'s do not help in measuring averages of physical observables in the KdV gas, or in establishing its thermodynamics. The second way is to assume that, even at intermediate times $t$, solitons can be considered as quasi-particles indexed by $i$, of position $x_i^t$ and spectral parameter $\eta_i$. The KdV soliton physics is implemented, on these quasi-particles, by assuming that quasi-particles undergo 2-soliton scattering shifts \eqref{shift_kdv},
\beq\label{defdelta}
	\delta(\eta,\mu) \equiv {{\rm sgn}(\mu-\eta)}\Delta(\eta,\mu) = -\frc1\eta\log
	\lt|\frc{\eta+\mu}{\eta-\mu}\rt|\; ,
\eeq
at all scattering events. Note that, by convention, in the GHD literature $\delta(\eta,\mu)>0$ means that if the object $\eta$ is to the left (right) of the object $\mu$ before the collision, then it will be to the right (left) after collision, and its trajectory has been shifted with respect to the non-interacting trajectory towards the left (right). In other words, for $\delta(\eta,\mu)>0$ the shift is ``backward" with respect to the direction of the collision. For the KdV soliton gas, $\delta(\eta,\mu)<0$.

With this prescription, the KdV evolution now induces a nontrivial dynamics $\mathfrak U_t x_i^0 = x_i^t$. This dynamics has an explicit algorithmic implementation as the ``flea gas" \cite{doyon_soliton_2018}, but its microscopic realisation is not important here. It satisfies the asymptotic condition $x_i^t \sim x_i^- + 4\eta_i^2 t$ ($t\to-\infty$), thus the quasi-particle describes correctly the trajectory of the soliton $i$ at large negative times. Using this, we may relate the coordinates $x_i^0$ of the quasi-particles at time 0, to the asymptotic coordinates $x_i^-$, taking into account all 2-soliton scattering events that occurred at negative times. As such, the quasi-particle gas, with coordinates $x_i^0$, is akin to the Bethe wave function \eqref{psill}.

\begin{figure}[ht]
	\includegraphics[width=\textwidth]{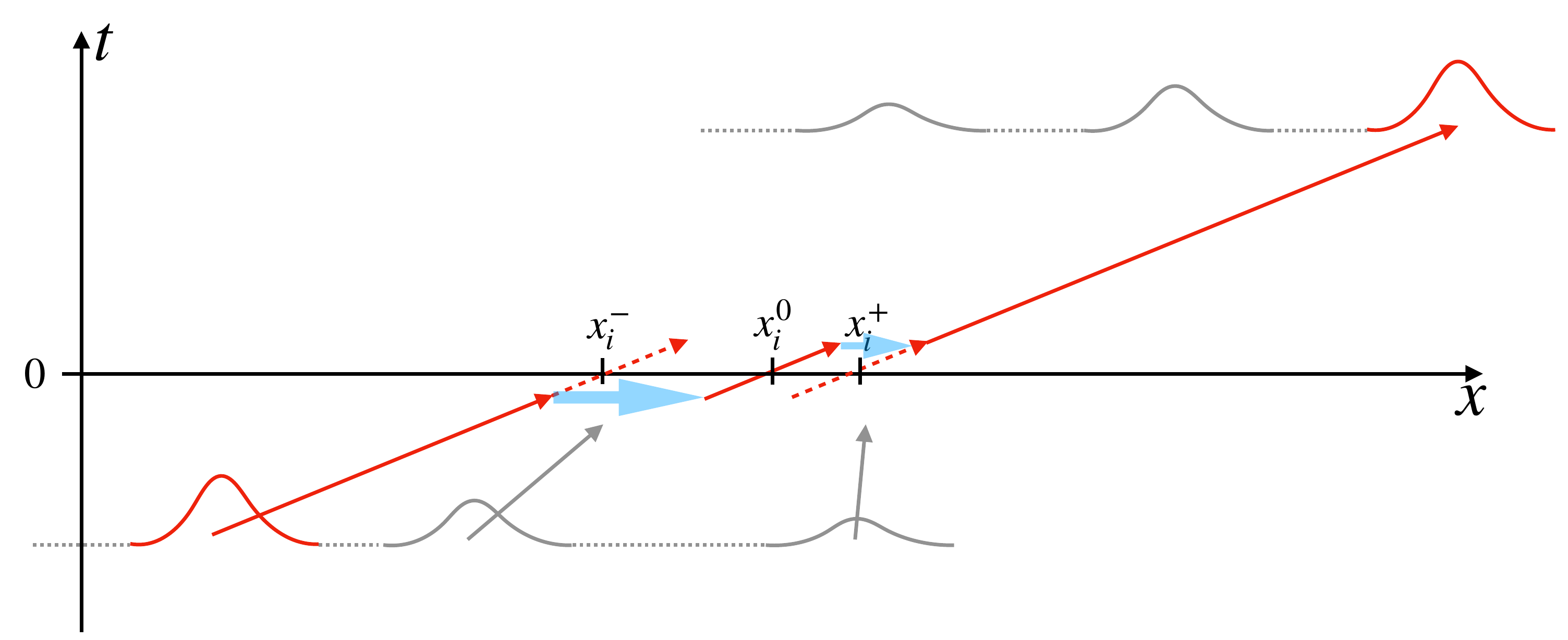} 
\caption{Schematic illustration of the asymptotic coordinates $x_i^\pm$ and the quasi-particle coordinate $x_i^0$. The red soliton (quasi-particle) from the initial soliton train moves ballisticaly at speed $4\e_i^2$, and its extended straight trajectory crosses the zero-time line at $x_i^-$. As it interacts with other solitons, the quasi-particle jumps, and after accumulating jumps at negative times, it crosses the zero-time line at $x_i^0$. It further accumulates jumps at positive times and agrees with the straight trajectory of the soliton from the large-time soliton train. The extended large-time trajectory crosses the zero-time line at $x_i^+$}
\label{figx}
\end{figure}

The exact total scattering shift  $x_i^{+}-x_i^{-}$ of the $i$-th soliton, after full scattering has occurred, is given by the sum of individual phase shifts from the collisions with other solitons in the gas, as per the well known factorised scattering theory of KdV. Therefore, by the definition of the dynamics of quasi-particles, $x_i^t \sim x_i^+ + 4\eta_i^2 t$ ($t\to+\infty$), hence the quasi-particle also describes correctly the trajectory of the soliton $i$ /at large positive times. The main assumption is that the quasi-particle description is in fact good enough also for intermediate times, even if actual solitons are not well separated. In other words, the effect of the $i^{\rm th}$ soliton at time $t$ is restricted to an area around $x_i^t$, even if the KdV field is of a rather complicated form and does not look at all like separate solitons; from the viewpoint of the large-scale physics and thermodynamics, this knowledge is sufficient. See Figure \ref{figx} for an illustration of the meaning of $x_i^\pm$ and $x_i^0$

\begin{figure}[ht]
	\includegraphics[width=\textwidth]{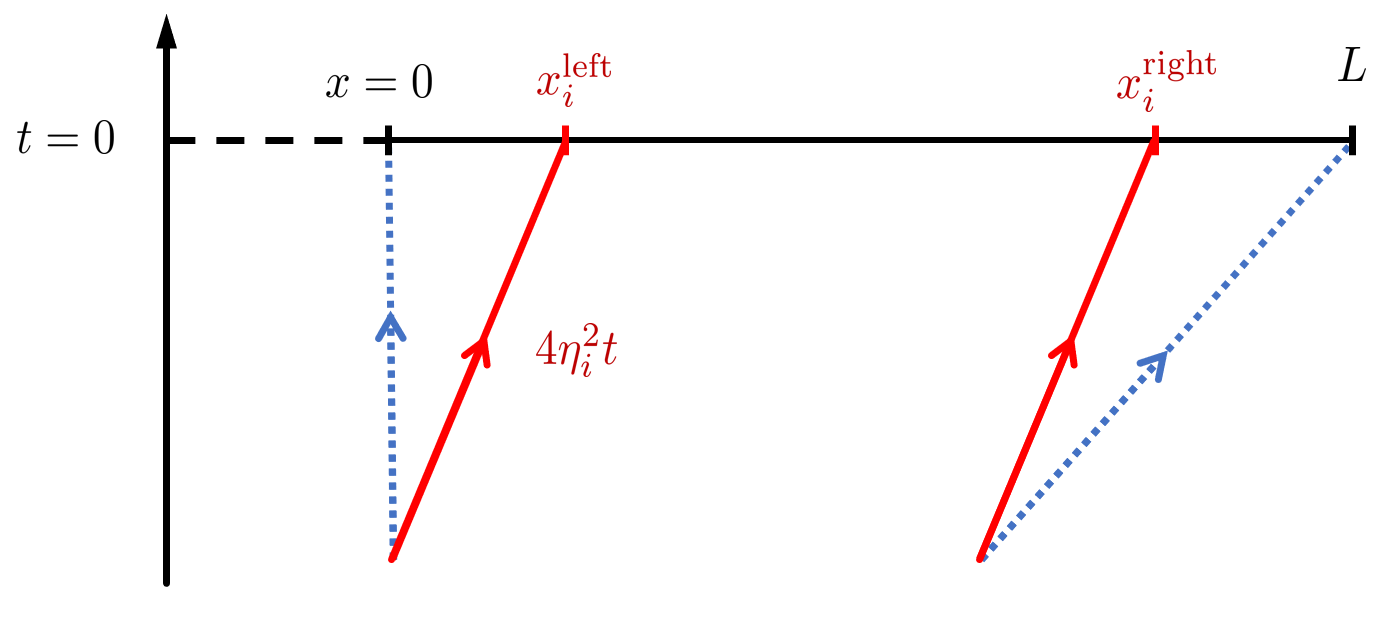}  
\caption{Schematic illustration of the asymptotic change of metric. Repeated interactions within the gas affect the effective (shifted) trajectory of any given soliton $i$ (dotted blue line), deviating from its ballistic trajectory at velocity $4\e_i^2$ (solid red line). Given an interval $[0,L]$ on which the gas is supported at $t=0$, the space of length $L_i = x_i^\text{right} - x_i^\text{left}$ perceived asymptotically by soliton $i$ is made smaller by the accumulated scattering shifts, either from the left by $\sum_{j\in T_i^\text{left} }\Delta_{ij}$, or the right by $\sum_{j\in T_i^\text{right} }\Delta_{ij}$.}
\label{fig:TBA}
\end{figure}

Now, consider $x_i^0$, the position of quasi-particle $i$ at $t=0$. In a gas of finite density lying on the interval $[0,L]$, we take $N$ solitons and require for the gas at time 0 to be confined within this interval, i.e. $\{\forall i=1\ldots N, \  x_i^0\in[0,L]\}$, for large $N$ and $L$ along with fixed $N/L$. We would like to ``measure" the resulting spatial extent on the asymptotic, free-particle coordinates $x_i^-$ given the constraint $x_i^0\in[0,L]$. The dynamics being 2-body reducible, to obtain $x_i^0$ from $x_i^-$ we need to add all spatial shifts incurred by quasi-particle $i$ as it crosses other quasi-particles in the gas from time $t\to-\infty$ to reach $x_i^0$ at time $t=0$. Now, consider a choice of value of $x_i^-$ such that quasi-particle $i$, at time 0, is all the way to the left, $x_i^0=0$. We denote $x_i^-=x_i^\text{left}$. Since for $t\to -\infty$ quasi-particles are ordered from fastest to slowest, and since at $t=0$ they all lie within $[0,L]$, collisions have almost surely all come from the left and therefore involve all faster quasi-particles. Denoting by $T_i^\text{left}$ the set of all quasi-particles that are faster than $i$, we find
\begin{equation}\label{left}
	0 = x_i^\text{left} - \frac{1}{\eta_i}\sum_{j\in T_i^\text{left}}\log\left|\frac{\eta_i+\eta_j}{\eta_i-\eta_j}\right| \; ,
\end{equation}
By a similar argument, choosing instead a value of $x_i^-$, denoted $x_i^-=x_i^\text{right}$, which is such that the same quasi-particle $i$ is all the way to the right at time 0, $x_i^0=L$, collisions have come from the right and involve all slower quasi-particles, from the set $T_i^\text{right}$:
\begin{equation}\label{right}
	L = x_i^\text{right} + \frac{1}{\eta_i}\sum_{j\in T_i^\text{right}}\log\left|\frac{\eta_i+\eta_j}{\eta_i-\eta_j}\right| \; .
\end{equation}
Writing both equations \eqref{left} and \eqref{right} allows us to compute the length of the asymptotic space $L_i \equiv x_i^\text{right} - x_i^\text{left}$ (the ``free length" as perceived by quasi-particle $i$). Noting that the set $T_i^\text{left} \cup T_i^\text{right}$ contains all quasi-particles except for $i$ (recall from Section \ref{sec:basic_constr} that all solitons have different velocities since $\e_i\neq\e_j$ for $i\neq j$), we have
\begin{equation}\label{aL}
	L_i = L - \frac{1}{\eta_i}\sum_{j\neq i}\log\left|\frac{\eta_i+\eta_j}{\eta_i-\eta_j}\right| \; .
\end{equation} 
This is illustrated in Figure \ref{fig:TBA}.

Let $L_N(\eta)$ be a smooth function interpolating $L_i$'s  for $i=1,\ldots N$ and define an ``asymptotic space density" ${\mathcal K}_{N}$ as
\begin{equation}\label{fs}
{\mathcal K}_{N}(\eta) := \frac{L_N(\eta)}{L} = 1 - \frac{1}{L\eta}\sum_{j}\log\left|\frac{\eta+\eta_j}{\eta-\eta_j}\right| \; .
\end{equation}
Taking the thermodynamic limit of \eqref{fs} we obtain 
\be \label{TBA_dr}
{\mathcal K}(\eta) = 1- \frac{1}{\eta} \int\dd \mu\, f(\mu) \log \left|\frac{\eta + \mu}{\eta - \mu} \right| \; ,
\ee
in which the DOS  $f(\eta)$ is the density of the spectral points $\eta_i$ per unit distance and per unit spectral interval (see Section \ref{sec:basic_constr}).   Here and elsewhere in this section the spectral integration  is performed over the support $\Gamma$ of the DOS $f(\mu)$.

The classical thermodynamic Bethe ansatz equation for KdV, equation~\eqref{TBA_dr}, is the classical version of the equation \eqref{TBA} in the quantum context, under the identifications
\beq\label{equiv1}
    \eta\equiv p,\quad f(\eta) \equiv \rho_{\rm p}(\eta),\quad
    \mathcal K(\eta) \equiv \frc{2\pi\rho_{\rm s}(\eta)}{P'(\eta)},\quad
    \delta(\eta,\mu) \equiv \frc{\varphi(\eta,\mu)}{P'(\eta)}
\eeq
(recall the definition \eqref{defdelta} for $\delta(\eta,\mu)$). The first two equations follow by identifying Bethe roots with quasi-particle (soliton) spectral parameters; the last two follow from comparing \eqref{TBA} with \eqref{TBA_dr}. Recall that for the Lieb-Liniger model $P'(\eta) =1$, however we keep it arbitrary here as this is a more universal expression -- we provide an in-depth discussion of $P(\e)$ in the context of the KdV soliton gas in Section \ref{sec:momfun}. The classical TBA equation is also equivalent to one half of the dispersion relations, equation~\eqref{inta}, from the soliton gas theory, upon the identification
\be 
{\mathcal K}(\eta) \equiv \frac{\sigma(\eta)}{\eta} f(\eta)\; .
\ee
In particular, with the occupation function $n = \rho_{\rm p}/\rho_{\rm s}$, this leads to
\beq\label{sigman}
    \sigma(\eta) \equiv \frc{2\pi \eta}{n(\eta)P'(\eta)}.
\eeq

A few important comments are in order. 
\begin{enumerate}
\item {We assumed that  spatial density of solitons does not depend on their spectral parameter. This is appropriate for the thermodynamics, and is what happens in the spectral theory of ``periodic'' soliton gases \cite{tovbis2022recent}. In the more general, quasi-periodic setting of \cite{el2003thermodynamic}, \cite{el_spectral_2020} the DOS is given by $f(\eta)= \phi(\eta)\varkappa(\eta)$, where $\phi(\eta)$ is the spectral measure  on $\Gamma$ (the distribution of solitons over the spectrum) and $\varkappa(\eta)$ is the function  responsible for the variable spatial density of solitons depending on their spectral parameter (in the periodic setting $\varkappa(\eta)={\alpha}$ -- the inverse period).}

\item The formula \eqref{TBA_dr}, {\it as derived}, is, strictly speaking, valid only for a rarefied gas. However, as mentioned, its validity in the context of a dense gas may be argued for by the idea that the effect of the $i^{\rm th}$ soliton is indeed restricted to where the $i^{\rm th}$ quasi-particle is located. Full justification is obtained by the thermodynamic finite-gap derivation of \cite{el2003thermodynamic} (see also \cite{el2021soliton} and the discussion in Section~\ref{sec:basic_constr}).

\item The classical interpretation of $\mathcal K(\eta)$ in \eqref{equiv1} as a density of asymptotic space was first proposed in \cite{doyon2018geometric} in the context of a geometric, semi-classical interpretation of the GHD equations. {According to the definition \eqref{fs} taken in the thermodynamic limit ($L,N\to\infty$, $L/N$ fixed), the quantity $\mathcal K(\eta)$ is the finite ratio of the perceived length $L_N$ (which grows to infinity) to the real length $L$ (which also grows to infinity). As we assume the gas to be homogeneous, this translates into the ratio of a length element $\dd x^-(\eta)$ in the asymptotic space of a soliton with spectral parameter $\eta$, to the length element $\dd x$ in real space (in the thermodynamic limit). Thus, there is a change of coordinates $\dd x^-(\eta) = \mathcal K(\eta) \dd x$ as viewed by quasi-particle $\eta$, and we may identify the metric as $g(\eta) = \mathcal K(\eta)^2$. Because the KdV equation translates into the free evolution in asymptotic coordinates, the fluid equation in terms of asymptotic coordinates is simple, and the change of coordinates to real space gives rise to the GHD equation. This is reviewed in Subsection \ref{ssectmetric}.} Interestingly, it relates the classical concept of a metric in asymptotic space, with the quantum Bethe ansatz concept of the total density $\rho_{\rm s}(\eta)$. In the soliton gas theory, it gives a geometric interpretation of the spectral scaling function $\sigma(\eta)$.
\end{enumerate}

\subsection{Conserved quantities in KdV}

It is important to identify the charges of GHD with the conserved quantities of KdV soliton gases (c.f. section~\ref{sec:cons_quan}). It is often convenient to work with their local densities. Here, the notations in the KdV and GHD contexts collide. In the GHD notation, for instance for the quantum Lieb-Liniger model (see Section \ref{sec:GHDLL}), the densities are denoted $q_i(x,t)$, the total charges $Q_i$, and the index usually starts at $i=0$ with an ``ultra local" charge (if the model admit such a charge), whose Hamiltonian flow is trivial.

With the GHD convention, the first few conserved densities are defined, according to \eqref{kruskal_norm}, as 
\beq
	q_0 = u,\quad q_1 = u^2,\quad q_2 = -u^3 + \frc12 u_x^2,\quad\ldots \; .
\eeq
The latter two, when integrated, reproduce the momentum and energy (Hamiltonian in one canonical formulation), so we have
\beq
	Q_0 = N = \int \, \dd x \, u \; , \quad
	Q_1 = P = \int \, \dd x \, {u^2} \; , \quad
	Q_2 = H = \int  \,  \dd x \,\Big(\frc{u_x^2}2 - u^3\Big)\; .
\eeq
While $Q_1$ and $Q_2$ respectively generate space and time translations using the canonical Poisson bracket, $Q_0$ generates a trivial Hamiltonian flow: it is, in the GHD language, an ``ultra-local charge", much like $\h Q_0$ for the Lieb-Liniger model
\beq
	\{u(x),u(x')\} = \delta'(x-x')\; ,\quad
	\{u,N\} = 0\; ,\quad
	\{u,P\} = \p_x u\; ,\quad
	\{u,H\} = \p_t u\; .
\eeq
We may express those charges in terms of the spectral parameter of the elementary solution that is the soliton $u_\eta(\cdot,\cdot) = u_s(\cdot,\cdot;\eta)$, given by equation~\eqref{kdv1sol},
\beq\label{solitonquantities}
	N|_{u_\eta} = h_0(\eta) = 4\eta\; , \quad P|_{u_{\eta}} = h_1(\eta) = \frc{16}{3} \eta^3\; ,\quad
	H|_{u_\eta} = h_2(\eta) =  \frc{32}{5} \eta^5\; .\
\eeq

Given any uniform density of solitons $\dos(\eta)$ per unit distance and per unit spectral parameter -- recall that this is equivalent to $f(\eta)$ within the soliton gas notations -- then the average of conserved densities can be expressed as
\beq\label{avcharge}
	\bra q_n\ket = \int \dd\eta \,\dos(\eta)h_n(\eta)\; ,
\eeq
just as it was in the quantum case \eqref{avq}; the $h$-functions are model dependent but the equations take the same form. This follows from the invariance of $Q_n$ under the KdV time evolution, and the asymptotically separated form \eqref{Nsol} of the $N$-soliton solution. This is perfectly equivalent to equation~\eqref{kdv_integrals}.

As the densities are conserved, they have associated local currents, which can be evaluated from the KdV equation since each $q_n$ only depends on $u$ and its derivatives, hence
\beq
	\p_t q_n + \p_x j_n = 0\; .
\eeq
This is just equation~\eqref{cons_law}, the classical version of \eqref{microcons}. In particular, the KdV equation itself directly leads to the relation
\beq\label{j1q3}
	j_0 = 3 \left(q_1 + \frc13 u_{xx}\right)\; .
\eeq

\subsection{GHD equation for the gas of solitons from classical TBA}

Following the logic of the quantum TBA, we would now construct the free energy. However, this has no direct equivalent in the theory of soliton gas as developed until now. But within the classical TBA, it is in fact possible to directly justify the GHD equations, without the need for the full thermodynamics as was done in the quantum context. Hence, we keep the development of the KdV thermodynamics for Section \ref{sec:conj}, and here complete the parallels between TBA and the KdV soliton gas theory of Section \ref{sec:solgas}.

We would like to derive a hydrodynamic equation for the gas of quasi-particles of the classical TBA developed in Section \ref{sec:TBAKdV}. We present two separate arguments.

\subsubsection{Collision rate ansatz} 
First, the soliton collision rate ansatz \eqref{collision_rate} is immediately applicable to the quasi-particles of the classical TBA. Within the gas every quasi-particle has ``bare" group velocity \eqref{baregroup}, $v^{\rm gr}(\eta) = 4\eta^2$, while the two-body scattering is characterised by the scattering shift \eqref{shift_kdv}, or equivalently (up to a sign convention) $\delta(\eta,\mu)$ \eqref{defdelta}. The shift is then used to calculate the effective velocity of a quasi-particle with given spectral parameter as a modification to its group velocity to account for collisions with other quasi-particles within the gas,
\beq\label{veff}
	v^{\rm eff}(\eta) = v^{\rm gr}(\eta) + \int \dd\mu\, \dos(\mu) \delta(\eta,\mu) (v^{\rm eff}(\mu) - v^{\rm eff}(\eta)) \; .
\eeq
This is the classical version of the quantum TBA result \eqref{veff_LL} under the identification for $\delta(\eta,\mu)$ in \eqref{equiv1}. Further, it agrees with \eqref{eq_state_kdv} from the soliton gas theory if we identify
\beq\label{sveff}
	s(\eta) \equiv v^{\rm eff}(\eta)\; .
\eeq

The collision rate ansatz is then used 
to express the average currents within the gas,
\beq\label{avcurr}
	\bra j_n\ket = \int \dd\eta\,v^{\rm eff}(\eta)\dos(\eta) h_n(\eta)\; ,
\eeq
which, given equation~\eqref{avcharge}, has a very intuitive interpretation: the average current is obtained by multiplying the amount $h_n(\eta)$ of charge carried by a soliton of parameter $\eta$ by the density of solitons of the same parameter and by the velocity at which they move within the gas. Note that from \eqref{j1q3} we have
\beq\label{j1q3average}
	\bra j_0\ket = 3\bra q_1\ket\; ,
\eeq
such that KdV admits a self-conserved current \cite{spohn2020collision,yoshimura2020collision}.

Finally, following from Section \ref{sec:FunEq}, in a long-wavelength inhomogeneous gas, the GHD Euler equation
\beq
	\p_t \dos + \p_x (v^{\rm eff} \dos) = 0 \; ,
\eeq
agrees with the KdV soliton gas kinetic equation \eqref{kin_eq0}. The relation \eqref{sigman} between the soliton-gas quantity $\sigma(\eta)$ and the TBA quantity $n(\eta)$ makes it clear that the ``continuum'' of Riemann invariants found in soliton gas, equation~\eqref{cont_riemann}, and in GHD, equation~\eqref{neq}, are essentially just inverse of each other.

\subsubsection{Change of metric}\label{ssectmetric}

The second way does not involve the collision rate ansatz, but rather the metric interpretation of $\mathcal K$ that arose from the classical TBA.

For a weakly inhomogeneous gas confined to a large volume $[0,L]$, we may repeat the classical TBA arguments of Section \ref{sec:TBAKdV}, and obtain a relation between the asymptotic coordinates of quasi-particles $x_i^-$ of spectral parameter $\eta_i$, and the real coordinates $x_i^0$. Recall from \eqref{evolveatxminus} that under KdV evolution,
\beq
    x^-_i(t) = x^-_i(0) +  v^\text{gr}(\eta_i)t\;,
\eeq
and $x_i^0$ evolves to $x_i^t$. Denoting by $x^-(\eta;x,t)$ the asymptotic coordinate for spectral parameter $\eta$ corresponding to real coordinates $x$ at time $t$, we have, according to the results of Section \ref{sec:TBAKdV},
\beq\label{xmk}
    \dd x^-(\eta;x,t) = \mathcal K(\eta;x,t) \dd x\; .
\eeq
Integrating,
\beq\label{xmkint}
    x^-(\eta;x,t) - x^-(\eta;x',t) = \int_{x'}^x \dd y\,\mathcal K(\eta;y,t)\; ,
\eeq
and far enough on the left, $x'\to-\infty$, where the gas is absent, $\mathcal K(\eta;y,t) = 1$, which gives a full characterisation of $x^-(\eta;,x,t)$ up to an overall constant shift.

As the evolution of the asymptotic coordinates is trivial,
the density $\rho^-_{\rm p}(\eta;x^-,t)$ of quasi-particles in the asymptotic space satisfies the  Liouville equation,
\beq\label{hydroasymp}
    \p_t \rho^-_{\rm p}(\eta;x^-,t) + v^{\rm gr}(\eta)
    \p_{x^-} \rho^-_{\rm p}(\eta;x^-,t) = 0.
\eeq
Clearly, taking into account the transformation property of a density, we have
\beq
    \rho^-_{\rm p}(\eta;x^-,t) = \frc{\dd x}{\dd x^-(\eta;x,t)} \rho_{\rm p}(\eta;x,t)
    = \frc{\rho_{\rm p}(\eta;x,t)}{\mathcal K(\eta;x,t)}
\eeq
under the relation \eqref{xmk}. Using the results established above, this gives
\beq\label{asympn}
    \rho^-_{\rm p}(\eta;x^-,t) = \frc{P'(\eta) }{2\pi}n(\eta;x,t)
    \equiv \frc{\eta}{\sigma(\eta;x,t)}.
\eeq
The combination of \eqref{xmk}, \eqref{hydroasymp} and \eqref{asympn} leads to equation~\eqref{cont_riemann}, equivalently equation~\eqref{neq}; the derivation is provided in \cite{doyon2018geometric}. Therefore, the soliton gas kinetic equation, equivalently the GHD equation, is nothing else but the Liouville equation for asymptotic coordinates, under the transformation of metric that accounts for the relation between the asymptotic and real spaces induced by the two-body scattering shifts.

A by-product of this viewpoint is an integral-equation ``solution" to the soliton gas kinetic equation, a.k.a.~GHD equation. Indeed, \eqref{hydroasymp} is solved as
\beq
    \rho^-_{\rm p}(\eta;x^-,t)
    = \rho^-_{\rm p}(\e;x^--v^{\rm gr}(\eta) t,0).
\eeq
Therefore, 
\beq\label{ncharact}
    n(\eta;x,t) = n(\e;U(\eta;x,t),0),\quad
    \sigma(\eta;x,t) = \sigma(\e;U(\eta;x,t),0)
\eeq
where $U(\eta;x,t)$ is such that $x^-(\e;U(\eta;x,t),0) = x^-(\eta;x,t)-v^{\rm gr}(\eta)t$, giving
\beq\label{charact}
    \int_{x'}^{U(\eta;x,t)}
    \dd y\,\mathcal K(\e;y,0)-
    \int_{x'}^{x}
    \dd y\,\mathcal K(\eta;y,t)
    + v^{\rm gr}(\eta) t = 0.
\eeq
The combination of equations~\eqref{ncharact} and \eqref{charact}, along with the TBA equation, gives a system of integral equations where time $t$ enters explicitly. Thus, it is not necessary to solve the hydrodynamic equation with finite-element analysis on the time coordinate; time can be set directly to the desired value, and the system of equations solved by recursion. The function $u(\eta;,x,t)$ has the interpretation as the coordinate in real space such that the characteristic curve for the fluid mode $\eta$ passes by both $(U(\eta;x,t),0)$ and $(x,t)$. See \cite[Sec 4.3]{doyon_lecture_2020}.

\subsection{Nonlinear dispersion relations}

Let us recall the relations obtained until now: from \eqref{equiv1} and \eqref{sigman}
\beq
    \eta\equiv p,\quad f(\eta) \equiv \rho_{\rm p}(\eta),\quad
    \sigma(\eta) \equiv \frc{2\pi \eta}{n(\eta)P'(\eta)},\quad
    \delta(\eta,\mu) \equiv \frc{\varphi(\eta,\mu)}{P'(\eta)}
    \;,
\eeq
and from \eqref{sveff},
\beq
    s(\eta) \equiv v^{\rm eff}(\eta)\; .
\eeq
The left-hand sides are quantities from soliton gas theory: the spectral parameter $\eta$, the DOS $f(\eta)$, the spectral scaling function $\sigma(\eta)$, the two-body soliton scattering shift $\delta(\eta,\mu) = {\rm sgn}(\alpha-\eta)\delta(\eta,\mu)$ (up a sign convention), and the tracer velocity $s(\eta)$. The right-hand sides are those from TBA and GHD: the Bethe root, or quasi-momentum, $p$, the quasi-particle phase-space density $\rho_{\rm p}(\eta)$, the occupation function $n(\eta)$, the differential scattering phase $\varphi(\eta,\mu)$, and the effective velocity $v^{\rm eff}(\eta)$. The right-hand side also involves a momentum function $P(\eta)$, which has no equivalent in the soliton gas theory and which we discuss below.

The last missing element of the soliton gas theory that needs comparison to GHD are the nonlinear dispersion relations, equations~\eqref{inta} and \eqref{intat}. We have seen that \eqref{inta} is in fact the TBA equation \eqref{TBA_dr} for the density of asymptotic space in the classical TBA, or \eqref{TBA} for the total density of quasi-particles and holes in the quantum TBA. Using the relation \eqref{defn}, $\rho_{\rm p} = n\rho_{\rm s}$, the total density takes the form, in terms of soliton gas objects,
\beq
    2\pi\rho_{\rm s}(\eta) \equiv \frc{P'(\eta)}{\eta}\,\sigma(\eta)f(\eta)\; .
\eeq
Then, one can interpret \eqref{inta} simply as the {\em dressing equation} \eqref{dressing} for $2\pi\rho_{\rm s}(\eta)$ expressed as the dressed momentum derivative via the thermodynamic relation \eqref{rhopdress},
\be\label{intadress}
\int\dd \mu\,\eta \delta(\eta,\mu) f(\mu) + \sigma (\eta) f(\eta) = \eta  \quad \Leftrightarrow\quad  2\pi\rho_{\rm s}(\eta) = ({P'})^{\rm dr}(\eta)\;.
\ee
The soliton gas nonlinear dispersion relation also includes equation~\eqref{intat}, involving the spectral flux density $v(\eta) = s(\eta) f(\eta)$. This can again be seen as a dressing equation, this time for the quantity
\beq
    2\pi v^{\rm eff}(\eta)\rho_{\rm s}(\eta)
    \equiv \frc{P'(\eta)}{\eta}\,\sigma(\eta)v(\eta)\;.
\eeq
It is, using $v^{\rm gr}(\eta) = 4\eta^2$, the dressing equation
\be\label{intatdress}
\int\dd \mu\,\eta \delta(\eta,\mu) v(\mu) + \sigma (\eta) v(\eta) = 4\eta^3  \quad \Leftrightarrow\quad  2\pi v^{\rm eff}(\eta)\rho_{\rm s}(\eta) = (v^{\rm gr} P')^{\rm dr}(\eta)\;.
\ee
This justifies the introduction of the {\em energy function} $E(\eta)$, whose derivative is
\beq
    E'(\eta) = v^{\rm gr}(\eta) P'(\eta).
\eeq
In this sense, \eqref{intadress} and \eqref{intatdress} really are ``dressed" dispersion relations. They lead to the well-known GHD relation for the effective velocity
\beq\label{veffratio}
    v^{\rm eff} = \frc{({E'})^{\rm dr}}{({P'})^{\rm dr}}\;,
\eeq
which is a dressed version of the usual definition of the group velocity in terms of the dispersion relation.

In the Lieb-Liniger model, the energy function as defined here agrees with the energy one-particle eigenvalue $E(p) = p^2/2$ in \eqref{LLPE}. In the soliton gas theory, it is an object that depends on our choice of the momentum function $P(\eta)$, which we now discuss.

Note that the dressing operation has no explicit equivalent in soliton gas theory, although the $y_\alpha$ functions introduced in \cite[Eq 49]{gurevich2000statistical} can be intrepreted as the dressed quantities $h_\alpha^{\rm dr}$.

\subsection{Momentum function}\label{sec:momfun}

As seen above, in many relations between objects from soliton gas theory and from GHD, the momentum function
\beq
    P(\eta)
\eeq
is involved. This function does not enter any of the physical equations of soliton gas theory: the kinetic equation \eqref{kin_eq0}, the dispersion relations \eqref{inta}, \eqref{intat} or the effective velocity relation \eqref{eq_state_kdv}. Therefore, within the soliton gas theory, it is superfluous: the choice of momentum function does not alter any of the physical quantities evaluated in the soliton gas. Nevertheless, it is a part of its relation to GHD.

This is because GHD was developed for quantum systems. In quantum systems, the momentum function $P(\eta)$ has an unambiguous physical meaning. Quantum scattering is described by a phase $S = e^{\ri \phi}$, and in order to obtain a scattering shift in space, we need to know ``what space is". This information is given by the one-particle eigenvalue $P(\eta)$ of the physical momentum operator $\h P$. {In continuous quantum systems with translation-invariant dynamics such as the Lieb-Liniger model -- the systems of interest here -- the momentum operator $\h P$ always exist and} defines the notion of space, as operators $\h\Or$ are displaced via $\h\Or(x) = e^{-\ri \h P x}\Or e^{\ri \h Px}$ (with $\hbar=1$). Here one-particle states are parametrised by some ``rapidity" $\eta$; the precise choice of parametrisation is not important, but the momentum eigenvalue is unambiguous. From it, the semi-classical scattering shift is
\beq
    \delta(\eta,\mu) = -\ri \,\frc{\p \log S(\eta,\mu)}{\p P(\eta)}\; .
\eeq
The choice of parametrisation $\eta$ is determined by the momentum function $P(\eta)$. Once a parametrisation is chosen, the conventional definition of the differential scattering shift is in terms of this parametrisation,
\beq
    \varphi(\eta,\mu) = -\ri \,\frc{\p \log S(\eta,\mu)}{\p \eta}\; .
\eeq
This gives the relation between differential scattering phase and scattering shift,
\beq\label{varphidelta}
	\varphi(\eta,\mu) = P'(\eta) \delta(\eta,\mu)\; .
\eeq

The parametrisation of quasi-particles in GHD in terms of a ``rapidity" $\eta$ is in fact an arbitrary choice. Indeed, all GHD equations are invariant under re-parametrisation, when written in universal ways in terms of the fundamental quantities; this justifies the writing in ``universal" forms used above. These quantities themselves transform, under re-parametrisation, either as vector fields (like derivatives $\p/\p\eta$, hence getting an additional $P'$ factor) or as scalar fields (without getting any additional factors). Under a change of parametrisation, all equations written in their universal form stay invariant if the following quantities transform as follows:
\beq
	\dos(\eta),\ \rho_{\rm s}(\eta),\ P'(\eta),\ E'(\eta)\quad\mbox{as vector fields}
\eeq
and
\beq
	v^{\rm gr}(\eta),\ v^{\rm eff}(\eta) ,\ n(\eta),\ \mathcal K(\eta),\ \delta(\eta,\mu)\; ,\ h_n(\eta) \quad\mbox{as scalar fields (in all variables);}
\eeq
recall that $h_n(\eta)$ are the one-particle eigenvalues used in evaluating average conserved densities. As a consequence of \eqref{varphidelta}, $\varphi(\eta,\mu)$ is a vector field in the first variable, and scalar field in the second. Reparametrisation invariance was discussed in the context of GHD in \cite{doyon2018exact}.

The relation to soliton gas, however, makes clear that the universal formulation of GHD is in fact {\em redundant}, something which had not been pointed out before. Indeed, once a parametrisation is chosen, all physical quantities (other than the momentum itself) in fact {\em do not involve the momentum function}, and can be recast in terms of momentum-independent, ``fundamental" quantities. Note how the following combinations only depend on the fundamental quantities of soliton gases $\dos(\eta)$, $\delta(\eta,\mu)$, $v^{\rm gr}(\eta)$, and not on $P(\eta)$:
\beq
	\frc{E'(\eta)}{P'(\eta)},\  n(\eta)P'(\eta),\ \sigma(\eta),\ 
	\frc{2\pi \rho_{\rm s}}{P'(\eta)},\ 
	\mathcal K(\eta),\ 
	v^{\rm eff}(\eta) = \frc{(E')^{\rm dr}(\eta)}{(P')^{\rm dr}(\eta)}\quad\mbox{are independent of $P(\eta)$.}
\eeq
This means that, when re-writing the soliton gas theory in terms of GHD, we may in fact choose the momentum function that we wish; there is no ``more meaningful" choice coming from physical scattering phases, as there are no scattering phases, just scattering shifts. As this is a structural observation, it is true for the GHD of any quantum system as well: it can first be written in terms of fundamental soliton-gas quantities, and then re-written in a universal form using any chosen momentum function. For quantum systems, however, choosing the incorrect momentum, although still giving correct hydrodynamic equations, would give the incorrect thermodynamics; this is not so in classical systems, as discussed in the next section.

The freedom in choosing the momentum function in the soliton gas theory can be used fruitfully, in order to simplify the relation with GHD. A ``physical" choice is the momentum carried by the soliton, from \eqref{solitonquantities}
\beq\label{physmom}
	P_{\rm phys}(\eta) = h_1(\eta)= \frc{16}3\eta^3\;.
\eeq
With this choice, the energy function defined as $E' = v^{\rm gr} P'$ is then the actual physical energy of the soliton,
\beq
	E_{\rm phys}(\eta) = h_2(\eta) = \frc{32}5\eta^5
\eeq
and the differential scattering phase is
\beq\label{diffscatphys}
	\varphi_{\rm phys}(\eta,\mu) = 16\eta \log\left|\frc{\eta-\mu}{\eta+\mu}\right|.
\eeq

In fact, this is {\em not} the most convenient. Instead, as in the classical TBA, we can see solitons as classical quasi-particles: it is natural to simply choose the momentum corresponding to unit-mass particles of velocities $v^{\rm gr}(\eta)$
\beq\label{solpart}
    P(\eta) = v^{\rm gr}(\eta) = 4\eta^2\;.
\eeq
In this way, the energy function  takes its natural form
\beq
    E(\eta) = \frc12 P(\eta)^2 = 8\eta^4\;.
\eeq
These, in the soliton gas theory, do not correspond to any conserved charge in the Kruskal series, expressed in terms of extensive functionals of the KdV field. Nevertheless, they are natural in the quasi-particle picture of soliton gases. With this choice, the differential scattering phase becomes
\beq
    \varphi(\eta,\mu) = 8\log\left|\frc{\eta-\mu}{\eta+\mu}\right| \; .
\eeq
This now satisfies the {\em symmetry relation}
\beq
	\varphi(\eta,\mu) = \varphi(\mu,\eta)\;,
\eeq
which simplifies many of the equations. This choice also streamlines the relation between two fundamental quantities of both theories: $\sigma$ and $n$
\beq\label{sig_n}
    \sigma(\eta) \equiv \frc{\pi}{4 n(\eta)} \; ,
\eeq
making one simply the inverse of the other, up to constant factors.

In the TBA literature, one typically chooses a parametrisation that guarantees symmetry of the differential scattering phase. As discussed in \cite{doyon2018exact}, the dressing operation in general is {\em different} for quantities $h_n(\eta)$ that are scalar fields, and those that are vector fields such as $P'(\eta)$: the differential scattering phase appears with opposite order of its arguments. This guarantees that the dressing operation preserves the transformation property. But with the symmetric choice, the dressing has the same form in both case.

Finally, we note that average densities and currents may be re-written using the dressing operation in various ways:
\beqa\label{averagesdressed}
	\bra q_n\ket &=& \int \dd \eta\,\dos(\eta)h_n(\eta)
	= \int \frc{\dd\eta}{2\pi}\,({P'})^{\rm dr}(\eta)n(\eta)h_n(\eta) = \int \frc{\dd\eta}{2\pi}\,P'(\eta)n(\eta)h_n^{\rm dr}(\eta) \; , \\
	\bra j_n\ket &=& \int \dd \eta\,v^{\rm eff}(\eta)\dos(\eta)h_n(\eta)
	= \int \frc{\dd\eta}{2\pi}\,({E'})^{\rm dr}(\eta)n(\eta)h_n(\eta) = \int \frc{\dd\eta}{2\pi}\,E'(\eta)n(\eta)h_n^{\rm dr}(\eta)\; .\n
	\label{averagesdressedcurrents}
\eeqa
{The last two equalities in equations \eqref{averagesdressed} and \eqref{averagesdressedcurrents} are ``symmetry relations" consequences of the dressing definition. One can deduce from {this symmetry}  that{\em, in average, the current of solitonic charge {$\bra j\ket$, with $h$-function} $h(\eta)=P'(\eta)$, is equal to the density of charge $\bra q\ket$, with  $h$-function} $\t h(\eta)=E'(\eta)$}. {Equality of the current of a certain conserved charge with the density of another conserved charge is in fact ``boost-invariance relation". For instance, in Galilean systems, the particle current equals the momentum density, and in relativistic systems, the energy current equals the momentum density. Many of the thermodynamic consequences of boost invariance can be deduced from such equality.  Given our current conventions  $P'(\eta) = 2h_0(\eta)$ and $E'(\eta) = 6h_1(\eta)$, we conclude from this symmetry that}
\beq\label{j0q1}
	\bra j_0\ket = 3\bra q_1\ket
\eeq
recovering \eqref{j1q3average}, a direct consequence of the KdV equation itself. 

We note that relation \eqref{j0q1} can be used, in combination with the exact expression of the free energy \eqref{F} derived in the next section, in order to obtain the exact expression of the currents \eqref{averagesdressedcurrents} using the methods of \cite{spohn2020collision,yoshimura2020collision}, instead of directly using a collision rate ansatz. However, here this does not increase the rigour of the derivation, as the classical TBA leading to the free energy of the soliton gases is not on fully rigorous grounds.

\subsection{Overview of the soliton gas - GHD relation}

The parallels drawn in this section are summarised in Table \ref{tab:recap} under the conventional choice of momentum $P(\e)=v^{\rm gr}(\e)$. With this formulation, we can now argue as in \cite{doyon2019generalized} in order to derive the thermodynamics of the soliton gas.

{\renewcommand{\arraystretch}{2}
\begin{table}
\centering
\begin{tabular}{ |c|c|c| }
 \hline
 Quantity & Soliton gas theory & Generalized Hydrodynamics \\
 \hline
 Conservation laws & $(P_n)_t+(Q_n)_x=0$ & $\p_t q_n + \p_x j_n = 0$\\
 Scattering shift & $\Delta (\e, \mu)= \frac{\sgn(\e-\mu)}{\e}   \log\left|\frac{\e + \mu}{\e-\mu}\right|$ & $\delta(\eta,\mu) = \frac{\varphi(\eta,\mu)}{8\e} = \frac{1}{\e}   \log\left|\frac{\e - \mu}{\e+\mu}\right|$ \\
 Spectral density of states & $f(\e)$ & $\dos(\e)$\\
 Effective velocity & $s(\e)$ & $\veff(\e)$\\
 Occupation function & $\frac{\pi}{4\s(\e)}$ & $n(\e)$\\
 Average density of charges & $\bra P_n[u ] \rangle = C_n\int_\Gamma \rmd \eta\, f(\eta)\eta^{2n+1} $ & $\bra q_n\ket = \int \dd \eta\,\dos(\eta)h_n(\eta)$\\
 Average density of currents & $\bra Q_n[u ] \rangle = C_n\int_\Gamma \rmd \eta\,s(\e)f(\eta)\eta^{2n+1}  $ & $\bra j_n\ket = \int \dd \eta\,\veff(\e)\dos(\eta)h_n(\eta)$\\
 \hline
\end{tabular}
\caption{Dictionary table for the various notations used in soliton gas theory and GHD under the conventional choice of the momentum function $P(\e)=4\e^2$.}
\label{tab:recap}
\end{table}}

\section{Thermodynamics of the KdV soliton gas: conjectures from GHD}\label{sec:conj}

In this section we shall use the previously established correspondence in order to construct the thermodynamics of soliton gases by directly applying results from GHD and TBA in this new context. As such, we will exclusively use notations associated with soliton gas theory, otherwise referring the reader to the dictionary Table \ref{tab:recap}.

\subsection{Generalised Gibbs Ensemble for KdV soliton gases}\label{ssectGGEkdv}

The thermodynamics for a soliton gas can be constructed by writing the generalised Gibbs measure for an ensemble of $N$-soliton solutions \eqref{Nsol}, using the asymptotic, action-angle coordinates $(\eta_n,x_n^-)$ discussed in Section \ref{sec:TBAKdV}. As these evolve as free particles, the a priori invariant measure for a single particle, with conventional normalisation coming from the choice $\hbar=1$, is $(2\pi)^{-1} \dd P(\eta_n)\dd x_n^-$, where the momentum function is the quasi-particle velocity, $P(\eta)= v^{\rm gr}(\eta) = 4\eta^2$. Therefore, the invariant measures -- the GGEs -- for a gas on a finite length $L$ are of the form
\beq\label{GGEKdV}
	\mu_w = \sum_{N=0}^{\infty} \frc1{N!} \prod_{i=1}^N \frac{\dd P(\eta_i)}{2\pi} \dd x^-_i\, \exp\Big[-\sum_{i=1}^N w(\eta_i)\Big]\,
	\chi\Big( |u_N(x,0)| < \varep_{|x|-L/2} \ \mbox{for}\ |x|>L/2\Big) \; ,
\eeq
for functions $w(\eta)$. Here $\varep_y\to0^+$ quickly enough as $y\to\infty$, say as $\varep_y = e^{-y/a}$ for some $a>0$. This measure is to be taken on $\eta_n>0,\  x_n^-\in\R$. The measure factorises into the scattering data except for a single non-trivial factor, which confines the gas (at time $t=0$) to lie on the interval $[-L/2,L/2]$, up to exponentially small corrections, and plays the role of an entropy term.

The measure may be seen as coming from the conserved charges in the GGE way, as was done in the quantum context in Section \ref{sec:stationary}. It involves some conserved charge $W$ such that
\beq
	\exp\Big[-\sum_{i=1}^N w(\eta_i)\Big] =
	\exp\Big[- W|_{u_N}\Big] \; ,
\eeq
where
\beq
	W = \sum_n\beta_n P_n \; ,
\eeq
if
\beq\label{w}
	w(\eta) = \sum_n \beta_n h_n(\eta) \; .
\eeq
As in the quantum case, there is no necessity for $w(\eta)$ to be expressible as a Taylor series for the measure to be a valid GGE. A valid GGE is a space-time translation invariant measure which is clustering at large distances (which is ergodic), and a large space of functions $w(\eta)$, or equivalently conserved charges $W$, will work, going beyond those which possess a Taylor expansion.

We are interested in taking the limit $L\to\infty$; note that for ``good" $w(\eta)$, the average number of solitons will increase in proportion to $L$, to give a finite density, in a way that is determined by the precise form of $w(\eta)$.

\subsection{Free energy and entropy of KdV soliton gases}

The GGEs now characterised, the next step in discussing the thermodynamics of our soliton gas would naturally be to construct its free energy and entropy.

For this purpose, we need to adapt \eqref{YY} and \eqref{fEd}, derived in the quantum context, to the classical TBA. As stressed in \cite{doyon_lecture_2020}, even if the TBA was originally developed to tackle quantum integrable systems, its general structure is not limited to those: classical limits have been investigated in \cite{de2016equilibration} or \cite{timonen1986exact}, and \cite{takayama1985extended} explored TBA approaches to soliton-bearing systems. In all cases, the TBA structure remains valid, with $F(\ep)$ being replaced by the free energy function representing the correct statistics of the asymptotic particles.

The universality of the TBA has, in fact, a geometric underpinning which follows from the geometric interpretion \cite{doyon2018geometric} of the classical TBA developed in Section \ref{sec:TBAKdV}. Recall that in the geometric viewpoint, the interaction can be implemented as a density-dependent change of metric on a system of free particles. This is because the scattering shifts $\delta(\eta,\mu)$ effectively change the perceived space. In this interpretation, each ``mode" $\mu$ induces an extra amount of physical space given by $\delta(\eta,\mu)$ as perceived by the mode $\eta$. This change of metric is exactly what we need to take into account when recasting the entropy factor in \eqref{GGEKdV} as a constraint on the asymptotoic coordinates $x_n^-$. 

The full calculation from this geometric idea was explained in \cite{doyon2019generalized}. It is easy however, from this perspective, to re-interpret the formulae obtained in the quantum context, in order to explain the classical result. In the quantum context, equation \eqref{fEd} is the free energy density for a system of free particles where each quasi-momentum $p$ is independently distributed in accordance with the Boltzmann weight determined by the ``energy" $\ep(p)$. But the Yang-Yang equation \eqref{YY} says that the interaction ``dresses" this energy: it modifies the probability for the level to be occupied by taking into account the free energy within the additional space introduced by mode $p$.

With this interpretation, for the soliton gas, the free energy density (per unit length) is
\beq
\mathcal{F} = \int_\Gamma \frac{\dd P(\e)}{2\pi} F\left(\epsilon(\e)\right)
\eeq
where, because of the measure \eqref{GGEKdV}, we must use the Maxwell-Boltzamann statistics of classical particles,
\beq\label{FMB}
F\left(\epsilon(\e)\right) = -\exp\left[ -\epsilon(\e)\right]\;,
\eeq
and
\beq\label{classicalYY}
	\underbrace{\ep(\eta)}_{\text{dressed weight}} =  \quad \underbrace{w(\eta)}_{\text{bare weight}} \quad +
	\underbrace{\int \frc{\dd \mu}{2\pi}\,\varphi(\mu,\eta)
	F(\ep(\mu))}_{\text{free energy in additional space introduced by quasi-particle $\eta$}} \; .
\eeq
Note that $\dd\mu\,\varphi(\mu,\eta) = \dd P(\mu) \delta(\mu,\eta)$ and, as such, the right-hand side actually involves the scattering shift. Hence the extra length perceived by quasi-particle (or soliton) $\mu$, as it should.

In thermodynamics the occupation function is fundamentally defined by \eqref{fEf}, and thus
\beq
n(\e) = \left.\frac{\dd F(\epsilon)}{\dd \epsilon}\right|_{\epsilon=\epsilon(\e)}
= e^{-\ep(\eta)}\;,
\eeq
as per the Maxwell-Boltzmann statistics. Recalling the equivalence \eqref{sig_n} we can straightforwardly obtain our conjectured expressions for the thermodynamics of the soliton gas. We have
\be\label{eps}
    \epsilon(\e) = \log \left[\frac{4\sigma(\e)}{\pi}\right] ,\quad
    F\left(\epsilon(\e)\right) = -\frac{\pi}{4\sigma(\e)}\; ,
\ee
which is directly related to the free energy density of the soliton gas
\be\label{F}
\mathcal{F} = -\int_\Gamma \frac{\e }{\sigma(\e)}\dd\e \; . 
\ee
The classical Yang-Yang equation \eqref{classicalYY} then allows for re-contextualisation of the notions of inverse temperature and chemical potential in terms of the soliton gas characteristic quantity $\s(\e)$,
\be
\begin{aligned}
w(\e) &= \epsilon(\e) - \int_\Gamma \frac{\dd\mu}{2\pi} \varphi(\e,\mu)F(\epsilon(\mu))\\
& = \log \left[\frac{4\sigma(\e)}{\pi}\right] + \int_\Gamma \frac{\dd\mu}{\s(\mu)}\log\left|\frac{\e-\mu}{\e+\mu}\right| \; .
\end{aligned}
\ee
Recalling the definition \eqref{w} of $w(\e)$, and that $h_k(\e) = C_k\e^{2k+1}$ for KdV -- given the normalisation \eqref{kdv_integrals} --  we may write
\be\label{betaksoliton}
\beta_k = \frac{1}{C_k(2k+1)!}\frac{\partial^{2k+1}}{\partial\e^{2k+1}}\left. \left\{\log \left[\frac{4\sigma(\e)}{\pi}\right] + \int_\Gamma \frac{\dd\mu}{\s(\mu)}\log\left|\frac{\e-\mu}{\e+\mu}\right|\right\}\right|_{\e=0} \; ,
\ee
and, in particular, the inverse temperature of a soliton gas (i.e. the Lagrange multiplier associated with its energy) is given by
\be
\beta_2 = \frac{1}{768}\frac{\partial^{5}}{\partial\e^5}\left. \left\{\log \left[\frac{4\sigma(\e)}{\pi}\right] + \int_\Gamma \frac{\dd\mu}{\s(\mu)}\log\left|\frac{\e-\mu}{\e+\mu}\right|\right\}\right|_{\e=0} \; .
\ee
Finally, the free energy density also provides a way to compute the entropy density $\mathcal S$
\be
\mathcal F = W - \mathcal S \; ,
\ee
hence we infer
\be
\begin{aligned}
    \mathcal S &= \int_\Gamma\left[\frac{\e}{\s(\e)}+w(\e)f(\e)\right]\dd \e \\
    & = \int_\Gamma\left\{\frac{\e}{\s(\e)} +\log\left[\frac{4\sigma(\e)}{\pi}\right]f(\e) + \int_\Gamma \frac{f(\mu)}{\s(\e)}\log\left|\frac{\e-\mu}{\e+\mu}\right|\dd \mu\right\}\dd \e \\
    & = \int_\Gamma\left(\log\left[\frac{4\sigma(\e)}{\pi}\right]+1\right)f(\e)\dd \e 
\end{aligned} \; ,\label{Ssoliton}
\ee
by successively making use of Yang-Yang equation \eqref{classicalYY} and then of the first dispersion relation \eqref{inta}, after commuting integrations on $\e$ and $\mu$. The soliton gas spectral entropy density $S(\e)=f(\eta) [1+\log\frac{4\s(\e)}{\pi}]$ is then consistent with both the classical entropy, that typically takes the from $S_\text{cl}(\e)=\dos(\e)[1-\log n(\e)]$, and the notion of condensate (examined in more details in Appendix \ref{app:solclass}), since it reaches its minimal value for $\s = 0$.

Above we have used the Maxwell-Boltzmann statistics: we are explicitly working under the assumption that solitons behave as classical particles. The validity of this assumption is examined in Appendix \ref{app:solclass}. A further justification is that the Maxwell-Boltzmann statistics \eqref{FMB} is the only choice which has the property that, when \eqref{classicalYY} is written in terms of the fundamental soliton gas quantity $\delta(\alpha,\eta)$, a change of momentum function $\dd P(\eta)\to r(\eta) \dd P(\eta)$ is equivalent to a change of bare weight $w(\eta)\to w(\eta)-\log r(\eta)$, in agreement with the GGE measure \eqref{GGEKdV}. This property is desirable, as the choice of the momentum function is irrelevant to the physics of the soliton gas, besides its effect on the chosen {\em a priori} measure.

Equations~\eqref{classicalYY}, \eqref{eps}, \eqref{F}, \eqref{betaksoliton} and \eqref{Ssoliton} are the main new results of this paper.

\subsection{Correlation functions}

With the above results, we can now borrow the GHD results from \cite{SciPostPhys.3.6.039,doyon2018exact}.

If we take \eqref{GGEKdV} as describing the correct soliton gas measure, assuming solitons behave as classical particles, then we are able to conjecture fluctuations (thermodynamics) and correlations (hydrodynamic linear response) in the soliton gas. These directly follow from the expression of the free energy \eqref{F} as per standard statistical mechanics. Recalling from definition \eqref{averagesdressed} and using the KdV notations $P_n$ and $Q_n$ for the charges and currents
\be
\begin{aligned}
    \langle P_n\rangle &= \int_\Gamma \frac{\dd P(\e)}{2\pi}n(\e) h_n^{\rm dr}(\e)\\
    &= \int_\Gamma \frac{\dd P(\e)}{2\pi}\left.\left[\frac{\dd F(\epsilon)}{\dd \epsilon}\frac{\partial \epsilon}{\partial \beta_n}\right]\right|_{\epsilon=\epsilon(\e)}\\
    &= \frac{\partial \mathcal F}{\partial\beta_n} \; ,
\end{aligned} 
\ee
space-integrated connected correlation functions take the form
\beqa\label{conncorr}
	\mathsf C_{ab} &:=& \int \dd x\,\bra P_a(x)P_b(0)\ket^{\rm c}
	:=\int \dd x\,\big(\bra P_a(x)P_b(0)\ket - \bra P_a(x)\ket\bra P_b(0)\ket\big)\n
	&=& -\frac{\partial^2 \mathcal F}{\partial\beta_a\partial\beta_b}= \int_\Gamma \dd\eta\,f(\eta)\theta(\eta)h_a^{\rm dr}(\eta)h_b^{\rm dr}(\eta) \; ,\n 
\eeqa
where $\theta(\eta)$ characterises the statistics, details of the computations being found in \cite{doyon2018exact}. In particular, for the Maxwell-Boltzmann statistics,
\beq
	\theta(\eta) = 1\;.
\eeq
Correlation functions, by way of the statistical factor $\theta(\e)$, thus provide an avenue for investigation into the statistics of solitons, which we perform in Appendix \ref{app:solclass}. Similarly, we can construct the field-current correlator by differentiating the free energy flux $\mathcal G$ such that
\be
\langle Q_n \rangle = \frac{\partial\mathcal G}{\partial\beta_n} \; ,
\ee
and hence
\beqa
	\mathsf B_{ab} &:=& \int \dd x\,\bra Q_a(x)P_b(0)\ket^{\rm c}\n
	&=& -\frac{\partial^2\mathcal G}{\partial\beta_a\partial\beta_b}= \int_\Gamma \dd\eta\,f(\eta)\theta(\eta)s(\eta) h_a^{\rm dr}(\eta)h_b^{\rm dr}(\eta) \; .
\eeqa
The spatial integrals are taken over any region that is large enough surrounding the point 0. We expect correlation functions to decay exponentially, so the integration should go over ``a few correlation lengths". Note that general principles of statistical mechanics imply \cite{castro2016emergent,denardis2019diffusion}
\beq
	\mathsf B_{ab} = \mathsf B_{ba}.
\eeq

The above correlations represent the thermodynamics, but more interesting are maybe the quantities that involve the dynamics. The Drude weight, for instance, is an important quantity for characterising ballistic transport. It can be expressed in terms of a time integral of correlation functions using the Kubo formula and takes the form
\beqa\label{drude}
	\mathsf D_{ab} &:=& \lim_{t\to\infty}\frc1{2t} \int_{-t}^t \dd \tau\int\dd x\
	\bra Q_a(x,\tau)Q_b(0,0)\ket^{\rm c}\n
	&=& \int_\Gamma \dd\eta\,f(\eta)\theta(\eta) s(\eta)^2 h_a^{\rm dr}(\eta)h_b^{\rm dr}(\eta) \; .
\eeqa
For this result to hold, the integration time $t$ typically needs to be larger than the relaxation time to pure ballistic effects, although it may be difficult to assess how large that is. Going further, we may evaluate the actual two-point functions of conserved densities and currents. The usual results from hydrodynamics apply at the Euler scale
\beq\label{2pcorr}
	\frc1{2\lambda} \int_{-\lambda}^{\lambda} \dd x\,\bra P_a(\xi t+x, t)P_b(0,0)\ket^{\rm c} \sim
	\frc1t \sum_{\eta\in \eta_*(\xi)} \frc1{|s'(\eta)|} f(\eta)\theta(\eta) h_a^{\rm dr}(\eta)h_b^{\rm dr}(\eta)\quad (t\gg \lambda\to\infty)\; ,
\eeq
where $\eta_*(\xi) = \{\eta:s(\eta) = \xi\}$. By convention, evaluation of two-point functions \eqref{2pcorr} often involves an averaging (usually either over space, time or space-time along the ray) to eliminate fluctuactions. We therefore predict a decaying behaviour in $1/t$.

To conclude with an example, we shall consider the static covariance $\mathsf C_{00}$ and the two-point auto-correlation function for the KdV wave field $P_0(\eta) = u(\eta)$, with $h_0(\eta) = 4\eta$. Explicitly, we start by evaluating the dressed identity in terms of the soliton gas notations
\beq\label{iddr}
	{\rm id}^{\rm dr}(\eta) = \eta + \int \frc{\dd\mu}{\sigma(\mu)} \log\left|\frc{\eta-\mu}{\eta+\mu}\right|\,{\rm id}^{\rm dr}(\mu) \; ,
\eeq
obtained from definition \eqref{dressing}. Comparison with the  dispersion relation \eqref{inta} allows for direct identification ${\rm id}^{\rm dr}(\eta)=\sigma(\eta)f(\eta)$. In the more general case one typically needs to evaluate dressed quantities numerically (c.f. Appendix \ref{app:numint}). As such the simplest correlations readily take the form
\be
\begin{aligned}
   	&\mathsf C_{00} = 16\int_\Gamma\dd\eta\,\sigma(\e)^2\,f(\eta)^3\\
	& \frc1{2\lambda} \int_{-\lambda}^{\lambda} \dd x\,\bra u(\xi t+x, t)u(0,0)\ket^{\rm c} \sim
	\frc1t \sum_{\eta\in \eta_*(\xi)} \frc{16}{|s'(\eta)|} \,\sigma(\e)^2\,f(\eta)^3\quad (t\gg \lambda\to\infty) \; .
\end{aligned} 
\ee

\section{Conclusions and Outlook}\label{sec:concl}

We have reviewed the main aspects of two theories for long-wavelength, hydrodynamic behaviours that were recently developed: that of soliton gases in integrable dispersive  hydrodynamic equations (or ``integrable turbulence"), and that of generalised hydrodynamics (GHD) for many-body integrable systems. For the former we took the example of the KdV equation, and for the latter, the quantum Lieb-Liniger model, both paradigmatic models in their respective domains.

As noticed in the literature, these theories share a number of similarities, both conceptual and technical. In fact, it is natural to expect that the theory of GHD be able to describe soliton gases, and that the theory of soliton gases be at the basis of GHD. We have shown that this is indeed the case, establishing the precise dictionary between the quantities and concepts studied in soliton gases and in GHD. The crucial observations were the parallels between the soliton gas kinetic equation and the Euler-scale GHD equation, and between the soliton gas dispersion relations and the (classical) thermodynamic Bethe ansatz (TBA) equations. This dictionary has allowed us to clarify aspects of each theories. For instance, we emphasised the geometric interpretation of the soliton gas kinetic equation; and we pointed out that the momentum function, previously seen as a fundamental quantity in GHD, is in fact superfluous and ambiguously defined in the soliton gas interpretation of GHD.

Most importantly, using this dictionary and known results in quantum and classical integrable many-body systems, we went beyond the kinetic equation for the KdV soliton gas, and proposed its exact thermodynamics: free energy, entropy, generalised temperatures. For this purpose, we argued that the classical (and generalised) Yang-Yang equations of TBA correctly describes the soliton gas thermodynamics, if the statistics of the solitons is taken to be that of classical particles, with Maxwell-Boltzmann distribution. The resulting conjectures for susceptibilities of various conserved quantities in the KdV soliton gas were verified numerically, using state-of-the-art numerical methods for generating soliton gases. We compared against other common statistics found in many-body systems (Bose-Eintein, Fermi-Dirac), and confirmed that the Maxwell-Boltzmann statistics gives the most precise predictions.

{One of the pertinent questions arising in relation with soliton gas theory is the issue of ``completeness'' 
of the soliton gas/GHD description of the KdV hydrodynamics. {It is known, by a classification based on the standard inverse scattering method, that the KdV equation admits both solitonic modes (discrete part of the spectrum) and radiative modes (continuous part of the spectrum). Shouldn't a typical KdV gas contain, then, also an ensemble of radiative modes? It is of course possible, naively, to include radiative modes in the hydrodynamic and thermodynamic description of the gas, as the scattering shifts are known.}

But we note that a soliton gas {lying on the full line} does not exhibit  decay as $|x| \to \infty$, and so its spectrum cannot be rigorously classified in the framework of the conventional discrete/continuous dichotomy. The intuitive notion of the purely discrete spectrum of soliton gas comes from the physically natural assumption that, on any sufficiently large interval  $x \in [x_0, x_0+L] \in \mathbb{R}$, a soliton gas can be {\it approximated} by an appropriate $N$-soliton solution with $N \gg 1$.  This assumption is at the heart of the phenomenological construction of soliton gas  outlined in Section~\ref{sec:basic_constr}. The GHD  interpretation  of  soliton gas in Section~\ref{sec:GHDLL}  is consistent with this  ``$N$-soliton paradigm". 

The dispersive hydrodynamic spectral theory of soliton gas developed in \cite{el2003thermodynamic, el2021soliton} and outlined in Section~\ref{sec:nonline_disp_rel}  is based on a more general definition of soliton gas as the thermodynamic limit of finite gap, non-decaying  KdV solutions exhibiting solitons and radiative modes as special cases.  Remarkably, this   ``wave'' approach to soliton gas bridging solitons and radiation leads to the same hydrodynamic description as the ``particle'' GHD construction, which suggests that the GHD in fact captures the ``full" KdV hydrodynamics.

But perhaps the strongest argument in favour of the ``completeness'' of the soliton gas description of random KdV fields (the integrable turbulence \cite{zakharov_turbulence_2009}) is the evidence, at the level of spectral data, that in the limit $N \to \infty$ solitons can provide an approximation to the general solution of the integrable model in question, including the continuous spectrum modes (the latter are approximated by a large number of solitons with small amplitudes), see \cite{faddeev_hamiltonian_2007}, $\S 8$. Although the discussion in \cite{faddeev_hamiltonian_2007} refers to the NLS equation, it is clear that the same arguments are directly applicable to the KdV equation as well. We also point out the relevant recent paper \cite{girotti_rigorous_21} where a soliton gas solution with non-zero reflection coefficient was constructed as the limit as $N \to \infty$ of the $N$-soliton solutions of  the KdV equation.

Of course, {a full understanding} of the completeness of the soliton gas set of ``action-angle variables'' is a significant open problem, {not only in terms of a rigorous proof, but also in terms of its very formulation. One question is as to the existence of consistent subsets of hydrodynamic modes. It is well known that in GHD, one may restrict the initial state to a subset of hydrodynamic modes, which is invariant under time evolution. For instance, in the hard rod gas, one may form an ensemble composed of a finite number of velocities (yet an infinite number of rods). Such restricted ensembles are invariant under evolution (in general, this will be true at least at the Euler scale), and thus have a consistent hydrodynamics. Likewise, a soliton gas certainly gives a consistent hydrodynamics. Another question is as to the definition of the ``right" maximal-entropy ensemble for the KdV field, for instance the thermal states. What is a Boltzmann weight for the KdV field? What Hamiltonian should one use, and what {\em a priori} measure on the field itself? We leave this for future research.}
}

GHD makes a number of far-reaching predictions for the large-scale behaviours of many-body systems, once the statistics of the underlying quasi-particles is known. With Maxwell-Boltzmann statistics for the KdV soliton gas, we made explicit predictions for the matrix of Drude weights, describing ballistic transport of KdV conserved quantity, and for the Euler-scale, dynamical correlation functions.

GHD can be used as the basis of many more predictions, which can immediately be written for soliton gases. As an example, the presence of external potentials coupling to any conserved density in GHD was described in \cite{doyon2017note}, and suggests, for instance, that the kinetic soliton gas equation for evolution under
\beq
    u_t + (\alpha(x) (3 u^2 + u_{xx}))_x = 0
\eeq
with long-wavelength $\alpha(x)$, takes, instead of \eqref{kin_eq0}, the form
\beq
    f_t + (sf)_x + (af)_\eta = 0
\eeq
for an effective acceleration $a$ (whose expression can be inferred from \cite{doyon2017note} using the dictionary described in the present work). More general forms of large-scale space-time dependence in the evolution equation can be introduced and dealt with using GHD \cite{BasGeneralised2019}. Another example is the exact expression for the diffusion operator, a second-derivative additional term in the kinetic equation. This is known in GHD  \cite{de_nardis_hydrodynamic_2018,denardis2019diffusion}, and is conjectured to be applicable to large classes of integrable models, only requiring the knowledge of the scattering shift and the statistics. Thus, the results of the present work can directly be used to make a prediction for diffusion in the KdV soliton gas. A final example is the large-deviation function for fluctuations of soliton transport, which can be obtained from the theory developed in \cite{doyon2020fluctuations,myers2020transport}.

Another aspect is the opportunities that the KdV soliton gases give for a more precise control on its GHD, thanks to the extensive mathematical structure and understanding available. Some of the techniques used in KdV soliton gases may also be transported to other models with a GHD description. For instance, can we adapt Whitham modulation theory to quantum many-body systems?

Finally, it is likely that the arguments and results of the present paper can be extended to soliton gases in other integrable dispersive hydrodynamic systems, particularly the focusing NLS equation, a canonical  model for the description of  modulationally unstable physical systems. The spectral theory of soliton and breather gases for the focusing NLS equation has been developed in \cite{el_spectral_2020} but the counterpart GHD description can provide further insights into the fundamental phenomena of integrable turbulence and rogue wave formation. 

\section*{Acknowledgments}
GE's  work was supported by EPSRC grant EP/W032759/1.  BD's work was supported by EPSRC grant EP/W010194/1.
The authors thank  G. Roberti for sharing  the  numerical code  for the nonlinear spectral synthesis of soliton gas.

\appendix

\section{Solitons as classical particles}\label{app:solclass}

In this Appendix we highlight a few different kinds of soliton gases and compute their GHD (Euler scale) correlation functions and Drude weights. The results of these predictions are then compared to the same correlation functions obtained from direct simulations of a KdV soliton gas in order to confirm our key assumption that KdV solitons in a gas actually behave as classical particles.

\subsection{Soliton condensate}

The notion of soliton condensate as the ``densest possible soliton gas'' for a given spectral support $\Gamma$ was introduced in \cite{el_spectral_2020} in the context of the focusing NLS equation. A detailed theory of  soliton condensates for the KdV equation, adopted here, is  developed in \cite{congy2022dispersive}. 

The spectral description of an equilibrium soliton condensate is obtained by taking the limit $\sigma \to 0$ in  nonlinear dispersion relations \eqref{inta}-\eqref{intat} which yields 
 \be \label{intaC}
\int_\Gamma \log \left|\frac{\eta + \mu}
{\eta-\mu}\right| f(\mu)\rmd\mu
=  \eta\, ,
\ee
\be \label{intatC}
\int_\Gamma \log \left|\frac{\eta + \mu}
{\eta-\mu}\right|v(\mu) \rmd \mu 
=4\eta^3 \, ,
\ee
For $\Gamma=[0,1]$ equations \eqref{intaC}, \eqref{intatC} are explicitly solved by 
\be \label{DOScond}
    f(\eta) = \frac{\eta}{\pi\sqrt{1-\eta^2}} \equiv f_{\rm c}(\eta) \quad , \quad v(\eta)=\frac{6\eta(2\eta^2-1)}{\pi\sqrt{1-\eta^2}} \equiv v_{\rm c}(\eta)\; .
\ee
These solutions describe  the simplest soliton condensate (classified in \cite{congy2022dispersive} as condensate of genus $0$). 
Using \eqref{moments_KdV} we obtain for the condensate $\langle u \rangle=1$, $\langle{u^2}\rangle=1$ so that the variance $\mathcal{A}_{\rm c}=\sqrt{\langle {u^2}\rangle - \langle u \rangle ^2 }= 0$, implying  that with probability $1$ (i.e. almost surely) a typical realisation of KdV soliton condensate is a constant wave field, $u(x,t)=1$. More sophisticated configurations  of soliton condensates exhibiting periodic or quasi-periodic typical realisations arise for non-simply connected spectral supports $\Gamma$ \cite{congy2022dispersive}.

\begin{figure}[ht]
	\includegraphics[width=\textwidth]{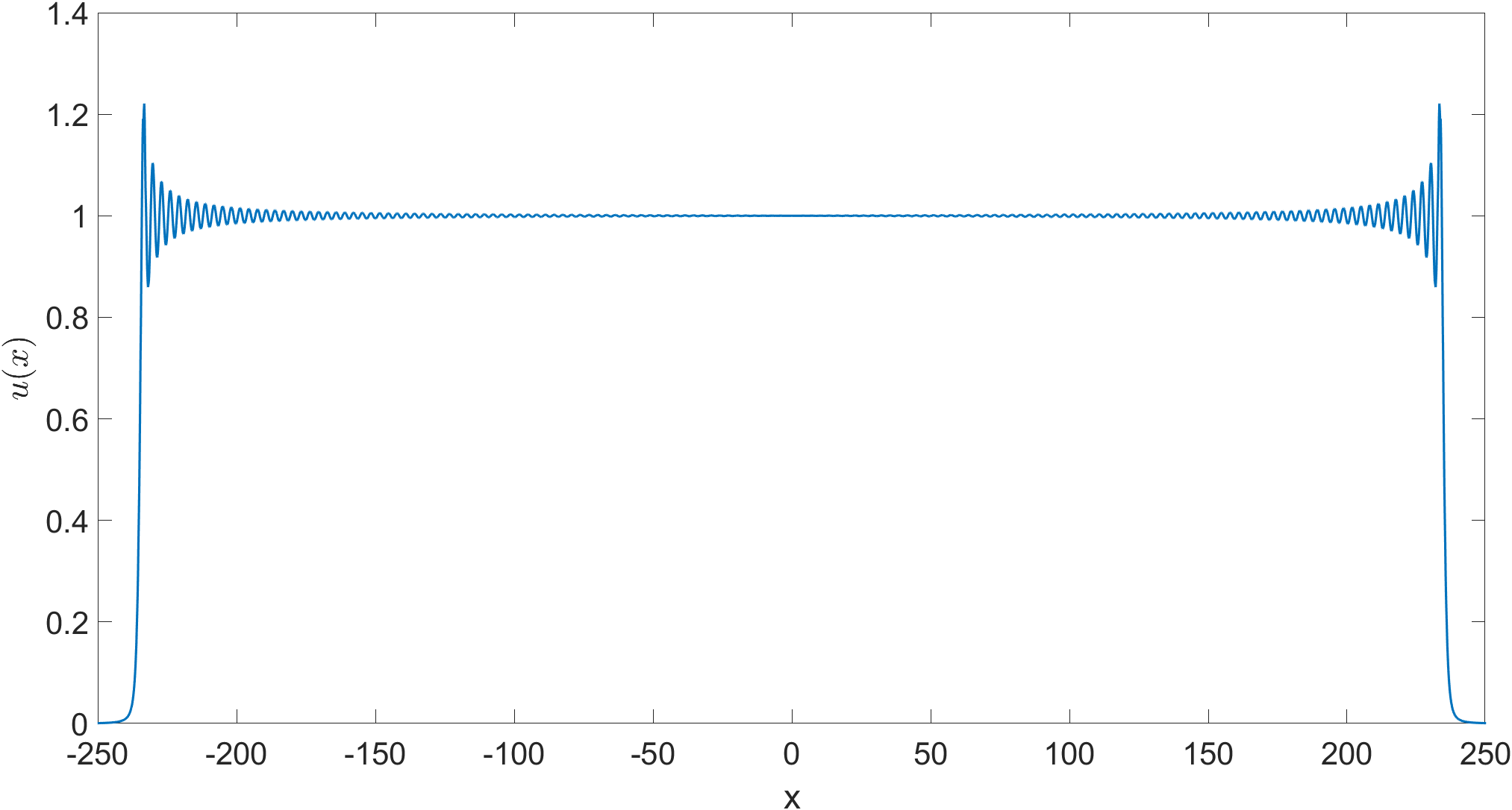}  \caption{Simulated soliton condensate  featuring 150 solitons. Oscillations here are finite size effects: as the number $N$ of solitons in the simulated condensate increases the amplitude of the edge oscillations decreases, and $u(x)\rightarrow 1$ $\forall x \in \mathbb{R}$ as $N \rightarrow \infty$.}
\label{fig:cond}
\end{figure}

A numerical realisation of the genus $0$ soliton condensate as a dense  ensemble of KdV solitons constructed according to the spectral DOS \eqref{DOScond} is presented Figure \ref{fig:cond}. The method for the  numerical realisation of KdV soliton gases used here, based on $N$-soliton solutions with $N$ large, has been developed in \cite{congy2022dispersive}; we present a brief description of it in Appendix \ref{app:gasim}. We should stress that, to achieve the required spectral DOS \eqref{DOScond} in the numerical soliton gas, one has to ``place'' all solitons in a single point, where the  location of a soliton in the gas is defined in terms of the phase of the corresponding norming constant in the $N$-soliton solution. In other words, this means that, in the context of the TBA approach introduced in section \ref{sec:TBAKdV}, for any soliton $i$ of the gas $x_i^-=x_i^\text{right} = x_i^\text{left} = 0$. This implies that, given a condensate supported on an interval of size $L$ at $t=0$, for any two solitons $i$ and $j$ of the condensate $\Delta_i=\Delta_j=L$.

As one can expect, the soliton condensate is not the most interesting case to investigate the statistics of solitons since $\sigma\to 0$ implies that all the dressed quantities, and by extension the correlation functions, vanish. This agrees with the above statement that $u(x,t)=1$ almost everywhere for such a gas, and can also be readily interpreted in GHD terms in the simplest case by looking at equations \eqref{inta}-\eqref{iddr} from which we make the identification
\be\label{iddrident}
    {\rm id}^{\rm dr}(\eta) = \sigma(\eta)f(\eta) = 0  \; ,
\ee
which readily implies
\be
    \mathsf C_{0n} = \mathsf B_{0n} = \mathsf D_{0n} =0
\ee
for any integer $n$, and whatever the statistics.

\subsection{Diluted condensate}

\begin{figure}[ht]
	\includegraphics[width=\textwidth]{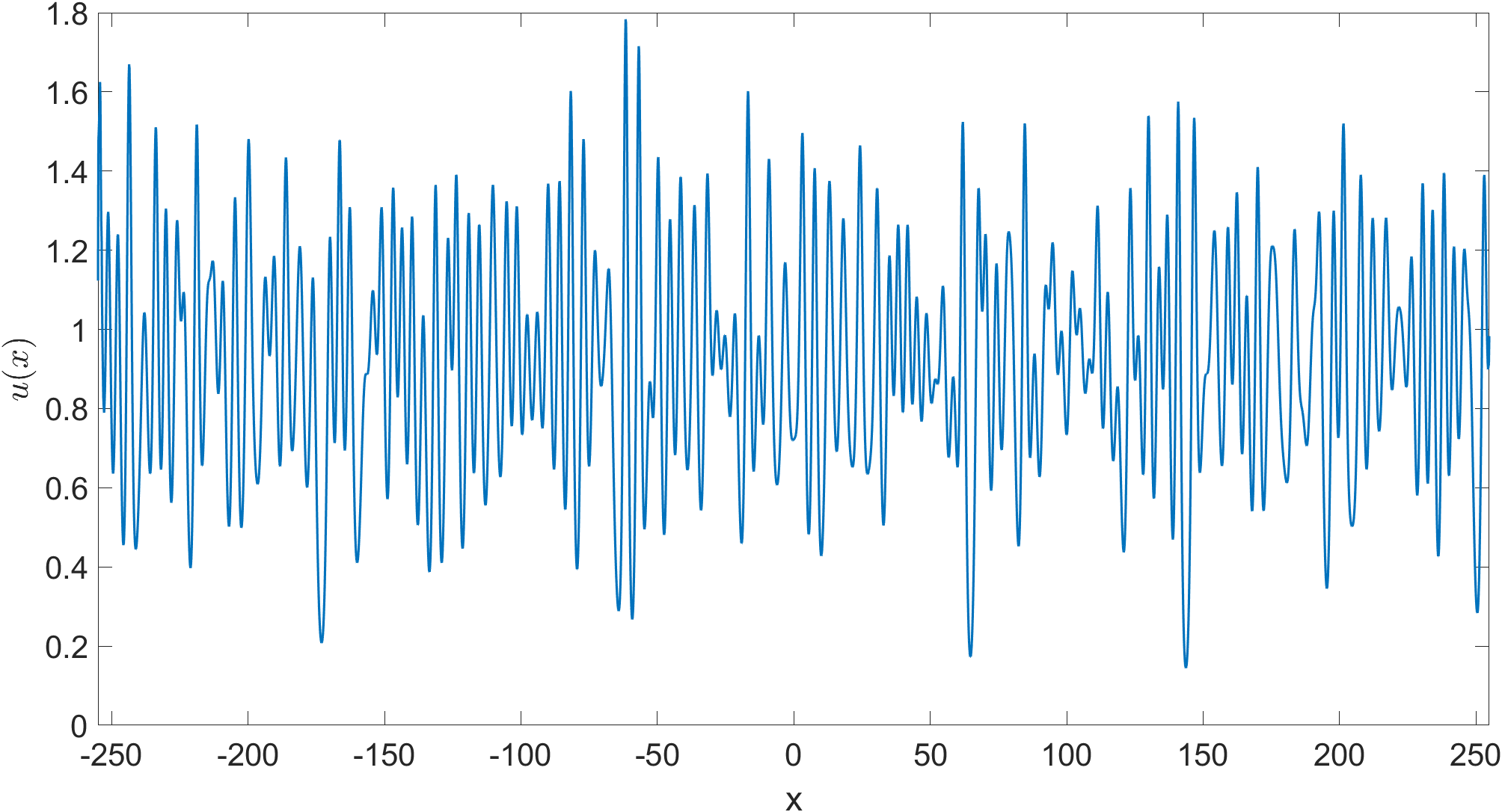}  
\caption{Example of a diluted condensate for $\alpha = 0.91$. One can see that the large and erratic oscillations on display wildly differ from what can be observed when looking at a regular condensate.}
\label{fig:dcreal}
\end{figure}

A more interesting case for the matter at hand would be the ``diluted" condensate \cite{congy2022dispersive} constructed from the spectral DOS
\be\label{DOSdc}
    f(\eta) =  \alpha \frac{\eta}{\pi\sqrt{1-\eta^2}} \; ,
\ee
where $0 < \alpha < 1$.
Notably, even for dilution parameters $\alpha \approx 1$, the resulting soliton gas exhibits spectacularly different qualitative features compared to the ``genuine'' condensate, as exemplified in Figure \ref{fig:dcreal}.

From both the general expression of the dispersion relation \eqref{inta} and the particular case of the condensate \eqref{intaC} we infer
\be\label{sigdc}
    \sigma(\eta) = \frac{\pi(1-\alpha)}{\alpha}\sqrt{1-\eta^2} \; .
\ee
Note that $\sigma>0$ implies $\alpha<1$ which, since $\alpha$ can be directly related to the mean wave field in the gas \eqref{moments_KdV}
\be
    \langle u \rangle = 4 \int^1_0  \eta f(\eta) \dd\eta = \alpha \; ,
\ee
agrees with the notion of  the condensate as the densest possible gas, and that the spectral DOS \eqref{DOSdc} actually corresponds to a diluted version of it. In terms of the TBA approach of section \ref{sec:TBAKdV}, this dilution corresponds to a widening of the asymptotic space in which the impact parameters (i.e. the phases of the norming constants) are distributed.

\begin{figure}[ht]
	\includegraphics[width=\textwidth]{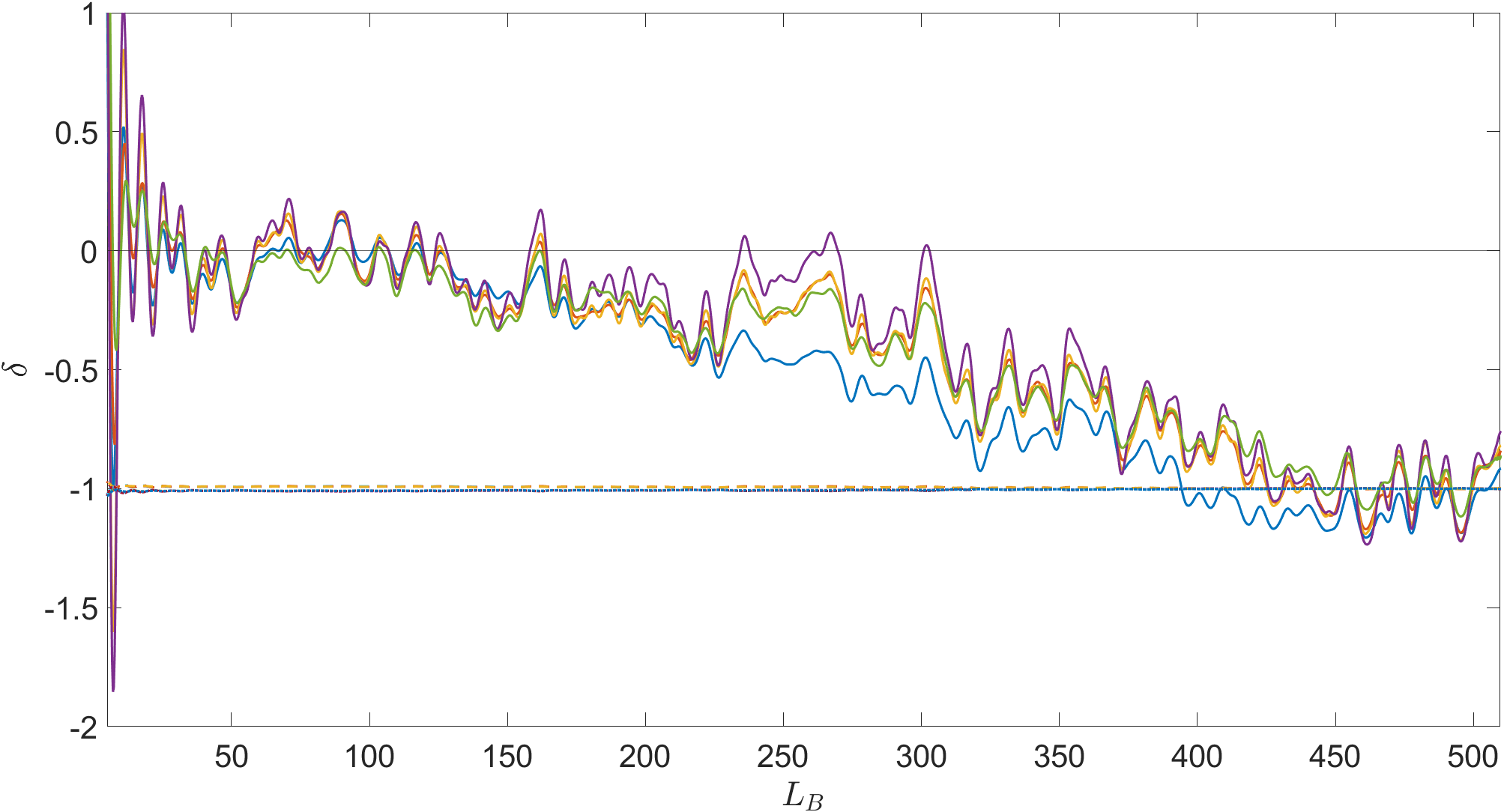}  
\caption{Relative difference $\delta= (\mathsf X_\text{GHD}-\mathsf X_\text{simulations})/\mathsf X_\text{GHD}$, between the GHD predictions and simulations regarding the correlation functions $\mathsf X = \mathsf C_{00}$, $\mathsf C_{01}$, $\mathsf C_{11}$, $\mathsf B_{00}$ and $\mathsf D_{00}$. The horizontal axis represents the size of the box over which correlations are computed; for large boxes the gas cannot be considered homogeneous anymore due to finite size effects. Full curves correspond to Maxwell-Boltzmann statistics, dashed to  Bose-Einstein and dotted to Fermi-Dirac. Results are obtained from the average over 6000 realisations of a diluted condensate of 150 solitons with parameter $\alpha = .91$.}
\label{fig:dc}
\end{figure}

From here it is possible to compute the simplest correlation function, recalling the identification \eqref{iddrident}
\be
\begin{aligned}
    \mathsf C_{00} &= \int_0^1 \dd\eta\,f(\eta)\left(h_0^{\rm dr}(\eta)\right)^2 \\
    & = \int^1_0 \dd\eta\,f(\eta)\left(4\ {\rm id}^{\rm dr}(\eta)\right)^2 \\
    & = \frac{32}{3\pi}\alpha(1-\alpha)^2 \; ,
\end{aligned}
\ee
in the classical case. However, it is typically difficult to obtain explicit analytical expressions, and one usually has to resort to numerical integration (details of which are found in Appendix \ref{app:numint}). Here we start by solving the second integral equation \eqref{intat} of the dispersion relations, with $\sigma(\eta)$ given by \eqref{sigdc}; this is equivalent to computing $\left(4\eta^3\right)^{\rm dr} \equiv 3h_1^{\rm dr}/4$. Recalling that in \eqref{intat} $v(\eta) \equiv s(\eta)f(\eta)$, the previous computation directly yields the effective velocity
\be
    s = \frac{\left(4\eta^3\right)^{\rm dr}}{\sigma(\eta)f(\eta)} = \frac{\left(4\eta^3\right)^{\rm dr}}{{\rm id}^{\rm dr}(\eta)} \; ,
\ee
and we recover the canonical GHD expression \eqref{veffratio} for the choice of momentum function \eqref{solpart}. From there, using expressions \eqref{conncorr} and \eqref{drude}, we can compute, among other quantities, $\mathsf C_{01}$, $\mathsf C_{11}$, $\mathsf B_{00}$ and $\mathsf D_{00}$ for different statistical factors, which we then compare, in both Figure \ref{fig:dc} and Table \ref{tab:dc}, to correlations obtained from a simulated gas. As we can see, even though the data is admittedly noisy, simulations seem to indicate solitons behave as classical particles.

\begin{table}
\centering
\begin{tabular}{ |p{1cm}||p{2cm}|p{2cm}|p{2cm}|p{2.5cm}| }
 \hline
  & MB & FD & BE & Simulations\\
 \hline
 $\mathsf C_{00}$ & 0.0235 & -2.28 & 2.32 & 0.022 $\pm$ 0.003\\
 $\mathsf C_{01}$ & 0.027 & -3.18 & 3.23 & 0.024 $\pm$ 0.004\\
 $\mathsf C_{11}$ & 0.042 & -4.48 & 4.56 & 0.039 $\pm$ 0.005\\
 $\mathsf B_{00}$ & 0.081 & -9.53 & 9.69 & 0.072 $\pm$ 0.01\\
 $\mathsf D_{00}$ & 0.377 & -40.3 & 41.1 &  0.325 $\pm$ 0.035\\
 \hline
\end{tabular}
\caption{GHD predictions regarding a diluted condensate soliton gas for various statistics displayed against the simulated correlation functions (averaged over $L_B\in[50;200]$) with standard deviation.}
\label{tab:dc}
\end{table}

Computing correlations for charges of order $n\ge 2$ would require solving the integral equation defined by the dressing operation for $\left(\eta^{2n+1}\right)^{\rm dr}$, which at this point should be rather straightforward.

\subsection{Uniform spectral distribution}

Another case we may consider is that of a uniform spectral distribution $f(\e) = \alpha$, for which some analytical results are still easy to obtain. Recalling again that $\iddr = \s f(\e)$, we have
\begin{equation}\label{iddrUnif}
	\begin{aligned}
		\iddr(\eta) &= \eta + \alpha\int_{\eta_0}^1 \dd\mu\,\log\left|\frac{\mu-\eta}{\mu+\eta}\right|\\
		&= \eta + \alpha\left[\eta\log\frac{\eta^2-\eta_0^2}{1-\eta^2}+\log\frac{1-\eta}{1+\eta}-\eta_0\log\frac{\eta-\eta_0}{\eta+\eta_0}\right] \; .
	\end{aligned} 
\end{equation}
The cut-off $\eta_0$ is here to ensure that $\sigma$ (as well as $\iddr$) stays non-negative. By noting that $\sigma(\eta)$ reaches its minimal value for $\eta=\eta_\text{min}$
\begin{equation}\label{etamin}
	\eta_\text{min} = \sqrt{\frac{1+\eta_0^2e^\alpha}{1+e^\alpha}} \; ,
\end{equation} 
and imposing that $\sigma(\eta_\text{min}) = 0$, we obtain an implicit relation between $\eta_0$ and $\alpha$
\begin{equation}\label{eta0}
	\left(\sqrt{1+e^\alpha}-\sqrt{1+\eta_0^2e^\alpha}\right)^2 = \left(\sqrt{1+\eta_0^2e^\alpha}-\eta_0\sqrt{1+e^\alpha}\right)^{2\eta_0}(1-\eta_0^2)^{1-\eta_0}e^\alpha \; .
\end{equation}
It may not appear clearly from this last relation \eqref{eta0}, but $\eta_0$ actually grows monotonously with $\alpha$ which, since
\begin{equation}
	\langle u \rangle = 4\int_{\eta_0}^1 \eta\dos(\eta)\dd\eta = 2\alpha(1-\eta_0^2) \; ,
\end{equation}
can be interpreted as a way to ensure the gas remains less dense than the condensate.

\begin{figure}[ht]
	\includegraphics[width=\textwidth]{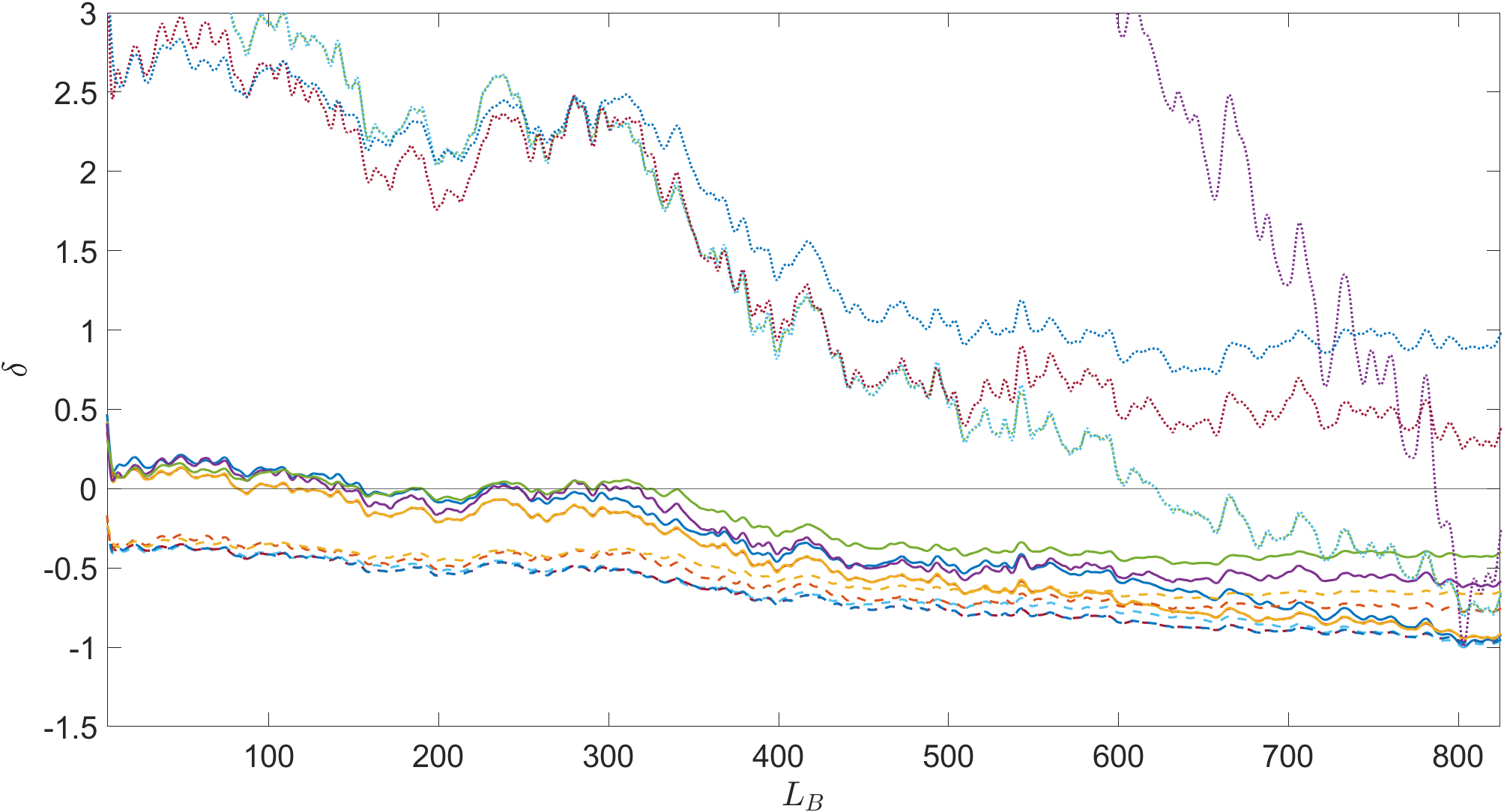}  
\caption{Relative difference $\delta= (\mathsf X_\text{GHD}-\mathsf X_\text{simulations})/\mathsf X_\text{GHD}$, between the GHD predictions and simulations regarding the correlation functions $\mathsf X =\mathsf C_{00}$, $\mathsf C_{01}$, $\mathsf C_{11}$, $\mathsf B_{00}$ and $\mathsf D_{00}$. The horizontal axis represents the size of box over which correlations are computed; for large boxes the gas cannot be considered homogeneous anymore due to finite size effects. Full curves correspond to Maxwell-Boltzmann statistics, dashed to Bose-Einstein and dotted to Fermi-Dirac. Results are obtained from the average over 4000 realisations of a simulated gas of 200 soltions with uniform spectral distribution $\alpha = .36$ and cut-off $\eta_0 = .4$.}
\label{fig:unif}
\end{figure}

Table \ref{tab:unif} compiles the results shown in Figure~\ref{fig:unif}: GHD predictions for different correlation functions, assuming either Maxwell-Boltzmann, Fermi-Dirac or Bose-Einstein statistics, are compared to results of the simulations. As with the diluted condensate, computing correlation functions  require numerical integration and, as before, simulations favour Maxwell-Boltzmann statistics.

\begin{table}
\centering
\begin{tabular}{ |p{1cm}||p{2cm}|p{2cm}|p{2cm}|p{2cm}| }
 \hline
  & MB & FD & BE & Simulations\\
 \hline
 $\mathsf C_{00}$ & 0.22 & 0.028 & 0.41 & 0.2 $\pm$ 0.03\\
 $\mathsf C_{01}$ & 0.28 & 0.072 & 0.49 & 0.23 $\pm$ 0.04\\
 $\mathsf C_{11}$ & 0.39 & 0.12 & 0.66 & 0.36 $\pm$ 0.05\\
 $\mathsf B_{00}$ & 0.84 & 0.22 & 1.46 & 0.69 $\pm$ 0.11\\
 $\mathsf D_{00}$ & 3.48 & 1.05 & 5.49 & 3.39 $\pm$ 0.31\\
 \hline
\end{tabular}
\caption{GHD predictions regarding a soliton gas with uniform DOS for various statistics displayed against the simulated correlation functions (averaged over $L_B\in[100;400]$) with standard deviation.}
\label{tab:unif}
\end{table}

\subsection{Linear spectral distribution}

\begin{figure}[ht]
	\includegraphics[width=\textwidth]{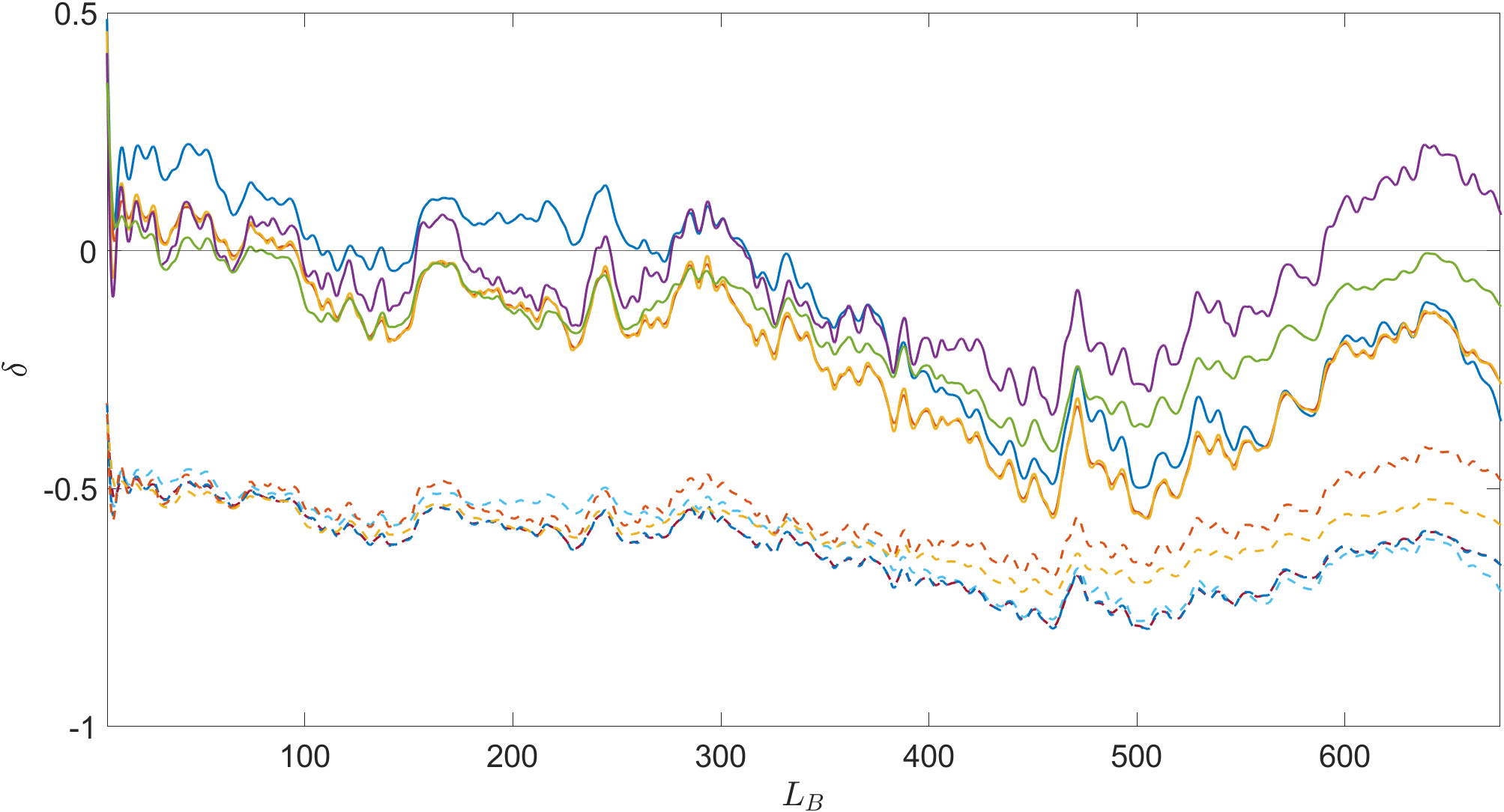}  
\caption{Relative difference $\delta= (\mathsf X_\text{GHD}-\mathsf X_\text{simulations})/\mathsf X_\text{GHD}$, between the GHD predictions and simulations regarding the correlation functions $\mathsf X =\mathsf C_{00}$, $\mathsf C_{01}$, $\mathsf C_{11}$, $\mathsf B_{00}$ and $\mathsf D_{00}$. The horizontal axis represents the size of box over which correlations are computed; for large boxes the gas cannot be considered homogeneous anymore due to finite size effects. Full curves correspond to Maxwell-Boltzmann statistics and dashed to Bose-Einstein; relative difference for Fermi-Dirac statistics are too large for their inclusion to be relevant. Results are obtained from the average over 3000 realisations of a simulated gas of 150 soltions with linear spectral distribution $\alpha = .26$ and cut-off $\eta_0 = .4$.}
\label{fig:lin}
\end{figure}

The last example we shall consider soliton gas realised as result of a long-time KdV evolution of an infinite sequence of broad parabolic pulses randomly distributed on $\mathbb{R}$ \cite{gurevich2000statistical}. The spectral DOS of such a gas grows linearly with the spectral parameter, $f(\eta) = 2\alpha\eta$. As before, we can compute the dressed identity
\begin{equation}\label{iddrLin}
	\begin{aligned}
		\iddr(\eta) = 2\alpha\eta\sigma(\eta) &= \eta + 2\alpha\int_{\eta_0}^1 \mu\,\log\left|\frac{\mu-\eta}{\mu+\eta}\right|\dd\mu\\
		&= \eta + 2\alpha\left\{(\eta_0-1)\eta + (\eta_0^2-\eta^2){\rm Ath}\left[\frac{\eta_0}{\eta}\right] + (\eta^2-1){\rm Ath}\left[\eta\right]\right\} \; ,
	\end{aligned} 
\end{equation}
for which a cut-off needs to be defined as well so that $\sigma>0$ and the average density
\be
    \langle u \rangle =  \frac{8\alpha}{3}(1-\eta_0^3) \; ,
\ee
remains lower than that of the condensate. Comparison between simulations et GHD predictions once again favours Maxwell-Boltzmann statistics (c.f. Figure \ref{fig:lin} and Table \ref{tab:lin}).

\begin{table}
\centering
\begin{tabular}{ |p{1cm}||p{2cm}|p{2cm}|p{2cm}|p{2cm}| }
 \hline
  & MB & FD & BE & Simulations\\
 \hline
 $\mathsf C_{00}$ & 0.2 & -0.05 & 0.45 & 0.2 $\pm$ 0.01\\
 $\mathsf C_{01}$ & 0.25 & -0.03 & 0.54 & 0.23 $\pm$ 0.01\\
 $\mathsf C_{11}$ & 0.36 & -0.03 & 0.75 & 0.34 $\pm$ 0.02\\
 $\mathsf B_{00}$ & 0.76 & -0.09 & 1.62 & 0.68 $\pm$ 0.04\\
 $\mathsf D_{00}$ & 3.26 & -0.27 & 6.79 & 2.9 $\pm$ 0.14\\
 \hline
\end{tabular}
\caption{GHD predictions regarding a soliton gas with linear DOS for various statistics displayed against the simulated correlation functions (averaged over $L_B\in[100;300]$) with standard deviation.}
\label{tab:lin}
\end{table}

\bigskip

While the previous three examples do not constitute a proof that soliton behave as classical particles, we feel they serve as a good indicator. Ultimately, it does not seem outlandish to assume Maxwell-Boltzmann statistics is the correct best fitted to soliton gases.

\section{Computing dressed quantities numerically}
\label{app:numint}

We start by recalling the definition of the dressing operation \eqref{dressing}
\beqa\label{dressing_mom}
	a^{\rm dr}(\eta) &=& a(\eta) +\int^1_{\eta_0} \frc{\dd\mu}{2\pi}\,\varphi(\mu,\eta) n(\mu) a^{\rm dr}(\mu) \\
	&=& a(\eta) +\int^1_{\eta_0} \frc{\dd\mu}{\sigma(\mu)}\,\log\left|\frc{\eta-\mu}{\eta+\mu}\right| a^{\rm dr}(\mu) \; ,
\eeqa
given the momentum function $P(\eta) = 4\eta^2$. Computing any dressed quantity then requires solving a Fredholm integral equation of the second kind, we do it by way of Nyström method as described in \cite{atkinson2009numerical}. Because the logarithmic kernel of the KdV soliton gas is singular for $\eta = \mu$ one has to be careful when approximating the integral by a weighted sum. Simpson's rule for example does not provide accurate results, instead we will deal with the singular part of the kernel separately.

Let $n>1$ an integer so that $h = (1-\eta_0)/n$ and $\eta_j = \eta_0 + jh$ for $j=0, 1, ... n$. We shall approximate the non-singular part of the integrand (recall that $\sigma >0$) by piecewise linear interpolation between nodes $\{\eta_j\}$
\be
    \left[\frac{a^{\rm dr}(\mu)}{\sigma(\mu)}\right]_n = \frac{1}{h}\left[(\eta_j-\mu)\frac{a^{\rm dr}(\eta_{j-1})}{\sigma(\eta_{j-1})} + (\mu - \eta_{j-1})\frac{a^{\rm dr}(\eta_j)}{\sigma(\eta_{j})}\right] \; .
\ee
Under such approximation, the integral equation \eqref{dressing_mom} can be replaced by the linear system
\be\label{linsys}
    a^{\rm dr}(\eta_i) = a(\eta_i) + \sum^n_{j=0} \frac{w_j(\e_i)-h\log|\eta_i + \eta_j|}{\sigma(\eta_{j})}a^{\rm dr}(\eta_j) \; ,
\ee
with weights
\be
    w_0(\eta) = \frac{1}{h}\int_{\eta_0}^{\eta_1} (\eta_1-\mu)\log|\eta-\mu|\dd\mu \; ,
\ee
\be
    w_n(\eta) = \frac{1}{h}\int_{\eta_{n-1}}^{\eta_n} (\mu-\e_{n-1})\log|\eta-\mu|\dd\mu \; ,
\ee
\be
    w_j(\eta) = \frac{1}{h}\left[\int_{\eta_j}^{\eta_{j-1}} (\mu-\e_{j-1})\log|\eta-\mu|\dd\mu + \int_{\eta_{j}}^{\eta_{j+1}} (\eta_{j+1}-\mu)\log|\eta-\mu|\dd\mu \right] \; ,
\ee
easily computed analytically. Once written in matrix form it only takes a few seconds to solve system \eqref{linsys} on a mid-range laptop using Matlab for $n=4000$. Depending on how well-behaved $\sigma$ is, it may be relevant to go for quadratic interpolation with graded mesh, but it usually seems unnecessary.

\section{Numerical generation of a soliton gas}\label{app:gasim}

In this Appendix we briefly outline the scheme used to numerically generate the soliton gases presented in Appendix \ref{app:solclass}. This procedure is based on \cite{huang1992darboux} and was originally used to generate the numerical synthesis of  soliton and breather gases for the focusing NLS equation in \cite{gelash_strongly_2018, roberti_numerical_2021}. The modification of this scheme for the generation of KdV soliton gas adopted here  has been developed in \cite{congy2022dispersive}; we refer the interested reader to the latter reference for an in depth presentation.

Our scheme generates exact $N$-soliton solutions of KdV by algebraic recursive procedures. The Darboux transformation relates the $N$-soliton Jost solution $J_N(\e)$ of the KdV auxiliary IST system to $J_{N-1}(\e)$, so that
\be
    J_N(\e) = D_N(\e)D_{N-1}(\e)...D_1(\e)J_0(\e)\; ,
\ee
where
\be
J_0(\e) = 
    \begin{pmatrix}
\exp[\e x - 4\e^3 t] & -\exp[4\e^3 t-\e x ] \\
0 & -2\e\exp[ 4\e^3 t-\e x ]
\end{pmatrix} \; ,
\ee
and $D_i(\e)$ is the $i$-th Darboux operator which depends only on the spectral parameter $\e_i$, the norming constant $b_i\equiv(-1)^i\exp\left[2\e_ix_i^-\right]$ and on $J_{i-1}(\e_i)$ \cite{huang1992darboux}. As such, the scheme can be broken down into three main steps
\begin{enumerate}
    \item We start by sampling a set of $N$ spectral parameters $\{\e_i\}$ and $N$ impact parameters $\{x_i^-\}$ from two given distributions; in the simulations of Appendix \ref{app:solclass} spectral parameters were sampled from the according DOS, and impact parameters from a uniform distribution over the asymptotic space.
    \item We compute the norming constants $\{b_i\}$ and then successively construct the Jost solutions $\{J_i\}$  as well as the Darboux operators $\{D_i\}$.
    \item We compute the $N$-soliton solution in the form
\be
u_N = 4\sum_{i=1}^N \e_iK_i \; ,
\ee
where the $\{K_i\}$'s, as the $\{D_i\}$'s, depend on the spectral parameter $\e_i$, the norming constant $b_i$ and on $J_{i-1}(\e_i)$ in a non-trivial manner \cite{huang1992darboux}.
\end{enumerate}

We stress once again that $N$-soliton solutions thus produced are exact solutions of KdV, approximating  the soliton gas in the large $N$ limit. This provides precise control over the numerically generated soliton gas, while also being computationally faster than numerically solving KdV through pseudo-spectral methods. However, this scheme necessitates an extremely high precision: though no rigorous stability analysis has been performed a good rule of thumb seems to be that generating $N$-soliton solutions requires $2N$ digits of precision, making the scheme unwieldy for $N$ greater than a few hundreds. That is why, in Appendix \ref{app:solclass}, rather than construct a soliton gas featuring thousands of solitons, we make use of ergodicty and average over several realisations of 150-soliton or 200-soliton solutions.

\bibliographystyle{plain}

\begin{thebibliography}{10}

\bibitem{aizenman1975ergodic}
M. Aizenman, S. Goldstein, and J.~L. Lebowitz.
\newblock Ergodic properties of an infinite one dimensional hard rod system.
\newblock {\em Communications in Mathematical Physics}, 39(4):289--301, 1975.

\bibitem{arutyunov2019factorised}
G. Arutyunov.
\newblock Factorised scattering theory.
\newblock In {\em Elements of Classical and Quantum Integrable Systems}, pages
  239--287. Springer, 2019.

\bibitem{atkinson2009numerical}
K. Atkinson and W. Han.
\newblock Numerical solution of fredholm integral equations of the second kind.
\newblock In {\em Theoretical Numerical Analysis}, pages 473--549. Springer,
  2009.

\bibitem{BasGeneralised2019}
A. Bastianello, V. Alba and J.-S. Caux, \newblock Generalized hydrodynamics with space-time inhomogeneous interaction. \newblock In {\em Physical Review Letters} 123(13):130602, 2019.

\bibitem{bastianello2019Integrability}
A. Bastianello and A. De Luca, 
\newblock Integrability-Protected Adiabatic Reversibility in Quantum Spin Chains.
\newblock In {\em Physical Review Letters} 122(24):240606, 2019.

\bibitem{bastianello2018generalized}
A. Bastianello, B. Doyon, G. Watts, and T. Yoshimura.
\newblock Generalized hydrodynamics of classical integrable field theory: the
  sinh-gordon model.
\newblock In {\em SciPost Physics}, 4(6):045, 2018.

\bibitem{belokolos_algebro-geometric_1994}
E.~D. Belokolos, A.~I. Bobenko, V.~Z. Enolski, A.~R. Its, and V.~B. Matveev.
\newblock {\em Algebro-geometric approach to nonlinear integrable equations}.
\newblock Springer, 1994.

\bibitem{bertini2016transport}
B. Bertini, M. Collura, J. De~Nardis, and M. Fagotti.
\newblock Transport in out-of-equilibrium x x z chains: Exact profiles of
  charges and currents.
\newblock In {\em Physical review letters}, 117(20):207201, 2016.

\bibitem{bettelheim2020whitham}
E. Bettelheim.
\newblock The Whitham approach to the c→ 0 limit of the Lieb--Liniger model and generalized hydrodynamics.
\newblock In {\em Journal of Physics A: Mathematical and Theoretical}, 53(20):205204, 2020.

\bibitem{boldrighini_one-dimensional_1983}
C.~Boldrighini, R.~L. Dobrushin, and Yu~M. Sukhov.
\newblock One-dimensional hard rod caricature of hydrodynamics.
\newblock In {\em Journal of Statistical Physics}, 31(3):577--616, 1983.

\bibitem{bonnes2014light}
L. Bonnes, F. H.~L. Essler, and A.~M. L\"auchli.
\newblock ``Light-cone'' dynamics after quantum quenches in spin chains.
\newblock In {\em Physical review letters}, 113(18):187203, 2014.

\bibitem{bulchandani_classical_2017}
V.~B. Bulchandani.
\newblock On classical integrability of the hydrodynamics of quantum integrable
  systems.
\newblock In {\em Journal of Physics A: Mathematical and Theoretical},
  50(43):435203, 2017.

\bibitem{BCM19} V. B. Bulchandani, X. Cao and J. E. Moore, 
\newblock Kinetic theory of quantum and classical Toda lattices.
\newblock In {\em Journal of Physics A: Mathematical and Theoretical} 52(33):33LT01, 2019.

\bibitem{bulchandani2021quasi-particle}
V.~B. Bulchandani, M. Kulkarni, J.~E. Moore, and X. Cao.
\newblock quasi-particle kinetic theory for calogero models.
\newblock In {\em Journal of Physics A: Mathematical and Theoretical},
  54(47):474001, 2021.

\bibitem{bulchandani2017solvable}
V.~B. Bulchandani, R. Vasseur, C. Karrasch, and J.~E. Moore.
\newblock Solvable hydrodynamics of quantum integrable systems.
\newblock In {\em Physical review letters}, 119(22):220604, 2017.

\bibitem{bulchandani2018bethe}
V.~B. Bulchandani, R. Vasseur, C. Karrasch, and J.~E. Moore.
\newblock Bethe-Boltzmann hydrodynamics and spin transport in the XXZ chain.
\newblock In {\em Physical review B}, 97(4):045407, 2018.

\bibitem{cao2018incomplete}
X. Cao, V.~B. Bulchandani, and J.~E. Moore.
\newblock Incomplete thermalization from trap-induced integrability breaking:
  Lessons from classical hard rods.
\newblock In {\em Physical review letters}, 120(16):164101, 2018.

\bibitem{castro2016emergent}
O.~A. Castro-Alvaredo, B. Doyon, and T. Yoshimura.
\newblock Emergent hydrodynamics in integrable quantum systems out of
  equilibrium.
\newblock In {\em Physical Review X}, 6(4):041065, 2016.

\bibitem{caux2019hydrodynamics}
J.-S. Caux, Benjamin Doyon, J. Dubail, R. Konik, T.
  Yoshimura, et~al.
\newblock Hydrodynamics of the interacting bose gas in the quantum newton
  cradle setup.
\newblock In {\em SciPost Physics}, 6(6):070, 2019.

\bibitem{congy2022dispersive}
T.~Congy, G. A.~El, G.~Roberti, and A.~Tovbis.
\newblock Dispersive hydrodynamics of soliton condensates for the kdv equation.
\newblock {\em To be submitted}, 2022.

\bibitem{congy_soliton_2021}
T. Congy, G. A. El, and G. Roberti.
\newblock Soliton gas in bidirectional dispersive hydrodynamics.
\newblock In {\em Physical Review E}, 103(4):042201, 2021.

\bibitem{costa_soliton_2014}
A. Costa, A.~R. Osborne, D.~T. Resio, S. Alessio, E.
  Chrivì, E. Saggese, K. Bellomo, and C.~E. Long.
\newblock Soliton {Turbulence} in {Shallow} {Water} {Ocean} {Surface} {Waves}.
\newblock In {\em Physical Review Letters}, 113(10):108501, 2014.

\bibitem{cubero2021form}
A.~C. Cubero, T. Yoshimura, and H. Spohn.
\newblock Form factors and generalized hydrodynamics for integrable systems.
\newblock In {\em Journal of Statistical Mechanics: Theory and Experiment},
  2021(11):114002, 2021.
  
\bibitem{davies1990higher}
B.~Davies.
\newblock Higher conservation laws for the quantum non-linear schr{\"o}dinger
  equation.
\newblock In {\em Physica A: Statistical Mechanics and its Applications},
  167(2):433--456, 1990.
  
 \bibitem{davies2011higher}
B. Davies and V.~E. Korepin.
\newblock Higher conservation laws for the quantum non-linear schr{\"o}dinger
  equation.
\newblock {\em arXiv preprint arXiv:1109.6604}, 2011.


\bibitem{de2016equilibration}
A. De~Luca and G. Mussardo.
\newblock Equilibration properties of classical integrable field theories.
\newblock In {\em Journal of Statistical Mechanics: Theory and Experiment},
  2016(6):064011, 2016.

\bibitem{de_nardis_hydrodynamic_2018}
J. De~Nardis, D. Bernard, and B. Doyon.
\newblock Hydrodynamic {Diffusion} in {Integrable} {Systems}.
\newblock In {\em Physical Review Letters}, 121(16):160603, 2018.

\bibitem{denardis2019diffusion}
J.~De Nardis, D. Bernard, and B. Doyon.
\newblock {Diffusion in generalized hydrodynamics and quasi-particle
  scattering}.
\newblock {\em SciPost Physics}, 6(4):049, 2019.

\bibitem{doyon2017thermalization}
B. Doyon.
\newblock Thermalization and pseudolocality in extended quantum systems.
\newblock In {\em Communications in Mathematical Physics}, 351(1):155--200, 2017.

\bibitem{doyon2018exact}
B. Doyon.
\newblock Exact large-scale correlations in integrable systems out of
  equilibrium.
\newblock In {\em SciPost Physics}, 5(5):054, 2018.

\bibitem{doyon2019generalized}
B. Doyon.
\newblock Generalized hydrodynamics of the classical toda system.
\newblock In {\em Journal of Mathematical Physics}, 60(7):073302, 2019.

\bibitem{doyon_lecture_2020}
B. Doyon.
\newblock Lecture notes on {Generalised} {Hydrodynamics}.
\newblock In {\em SciPost Physics Lecture Notes}, page~18, 2020.

\bibitem{doyon2017large}
B. Doyon, J. Dubail, R. Konik, and T. Yoshimura.
\newblock Large-scale description of interacting one-dimensional bose gases:
  generalized hydrodynamics supersedes conventional hydrodynamics.
\newblock In {\em Physical review letters}, 119(19):195301, 2017.

\bibitem{doyon2020fluctuations}
B. Doyon and J. Myers.
\newblock Fluctuations in ballistic transport from euler hydrodynamics.
\newblock In {\em Annales Henri Poincar{\'e}}, volume~21, pages 255--302.
  Springer, 2020.

\bibitem{SciPostPhys.3.6.039}
B.~Doyon and H.~Spohn, Drude Weight for the Lieb-Liniger Bose Gas. In {\em SciPost Physics},~3(6): 039,~2017.

\bibitem{doyon2018geometric}
B. Doyon, H. Spohn, and T. Yoshimura.
\newblock A geometric viewpoint on generalized hydrodynamics.
\newblock In {\em Nuclear Physics B}, 926:570--583, 2018.

\bibitem{doyon2017note}
B. Doyon and T. Yoshimura.
\newblock {A note on generalized hydrodynamics: inhomogeneous fields and other
  concepts}.
\newblock In {\em SciPost Physics}, 2(2):014, 2017.

\bibitem{doyon_soliton_2018}
B. Doyon, T. Yoshimura, and J.-S. Caux.
\newblock Soliton {Gases} and {Generalized} {Hydrodynamics}.
\newblock In {\em Physical Review Letters}, 120(4):045301, 2018.

\bibitem{DurninTherma2020} J. Durnin, M. J. Bhaseen and B. Doyon, 
\newblock Non-equilibrium dynamics and weakly broken integrability
\newblock In {\em Physical Review Letters} 127(13):130601, 2021.

\bibitem{eisert2015quantum}
J. Eisert, M. Friesdorf, and C. Gogolin.
\newblock Quantum many-body systems out of equilibrium.
\newblock In {\em Nature Physics}, 11(2):124--130, 2015.

\bibitem{el_kinetic_2005}
G.~A. El and A.~M. Kamchatnov.
\newblock Kinetic equation for a dense soliton gas.
\newblock In {\em Physical Review Letters}, 95(20):204101, 2005.

\bibitem{el_kinetic_2011}
G.~A. El, A.~M. Kamchatnov, M.~V. Pavlov, and S.~A. Zykov.
\newblock Kinetic {Equation} for a {Soliton} {Gas} and {Its} {Hydrodynamic}
  {Reductions}.
\newblock In {\em Journal of Nonlinear Science}, 21(2):151--191, 2011.

\bibitem{el2003thermodynamic}
G. A. El.
\newblock The thermodynamic limit of the whitham equations.
\newblock In {\em Physics Letters A}, 311(4):374 -- 383, 2003.

\bibitem{el_spectral_2020}
G. A. El and A. Tovbis.
\newblock Spectral theory of soliton and breather gases for the focusing
  nonlinear {Schr{\"o}dinger} equation.
\newblock In {\em Physical Review E}, 101(5):052207, 2020.

\bibitem{el2021soliton}
G. A. El.
\newblock Soliton gas in integrable dispersive hydrodynamics.
\newblock In {\em Journal of Statistical Mechanics: Theory and Experiment},
  2021(11):114001, 2021.

\bibitem{essler2016quench}
F.~H. L.  Essler and M. Fagotti.
\newblock Quench dynamics and relaxation in isolated integrable quantum spin
  chains.
\newblock In {\em Journal of Statistical Mechanics: Theory and Experiment},
  2016(6):064002, 2016.

\bibitem{faddeev_hamiltonian_2007}  
L.D.~Faddeev and L.A.~Takhtajan 
\newblock{\em Hamiltonian Methods in the Theory of Solitons}.
\newblock Springer, 2007


\bibitem{ferapontov2021kinetic}
E.V.~Ferapontov and M.V.~Pavlov.
\newblock Kinetic equation for soliton gas: integrable reductions.
\newblock In {\em Journal of Nonlinear Science}, 32(2):1-22, 2022.

\bibitem{flaschka_multiphase_1980}
H.~Flaschka, M.~G. Forest, and D.~W. McLaughlin.
\newblock Multiphase averaging and the inverse spectral solution of the
  {Korteweg}-de {Vries} equation.
\newblock In {\em Communications on Pure and Applied Mathematics}, 33:739--784, 1980.

\bibitem{franchini2017introduction}
F. Franchini.
\newblock {\em An introduction to integrable techniques for one-dimensional
  quantum systems}.
\newblock Springer, 2017.

\bibitem{friedman2020diffusive}
A.~J. Friedman, S. Gopalakrishnan, and R. Vasseur.
\newblock Diffusive hydrodynamics from integrability breaking.
\newblock In {\em Physical Review B}, 101(18):180302, 2020.

\bibitem{gardner_method_1967}
C.~S. Gardner, J.~M. Greene, M.~D. Kruskal, and R.~M. Miura.
\newblock Method for {Solving} the {Korteweg}-{deVries} {Equation}.
\newblock In {\em Physical Review Letters}, 19(19):1095--1097, 1967.

\bibitem{gelash_strongly_2018}
A.~A. Gelash and D.~S. Agafontsev.
\newblock Strongly interacting soliton gas and formation of rogue waves.
\newblock In {\em Physical Review E}, 98(4):042210, 2018.

\bibitem{girotti_rigorous_21}
M. Girotti, T. Grava, R. Jenkins, and K. D. T.- R. McLaughlin 
\newblock Rigorous Asymptotics of a KdV Soliton Gas.
\newblock In {\em Communications in Mathematical Physics}, 384(2):733-84, 2021.

\bibitem{gopalakrishnan2020hydrodynamics}
S. Gopalakrishnan, D.~A. Huse, V. Khemani, and R. Vasseur.
\newblock Hydrodynamics of operator spreading and quasi-particle diffusion in
  interacting integrable systems.
\newblock In {\em Physical Review B}, 98(22):220303, 2018.

\bibitem{gurevich_development_1999}
A.~V. Gurevich, K.~P. Zybkin, and G.~A. El{\rq}.
\newblock Development of stochastic oscillations in a one-dimensional dynamical
  system described by the {Korteweg-de Vries} equation.
\newblock In {\em Journal of Experimental and Theoretical Physics},
  88(1):182--195, 1999.

\bibitem{gurevich2000statistical}
A. V.~Gurevich, N. G.~Mazur, and K. P.~Zybin.
\newblock Statistical limit in a completely integrable system with
  deterministic initial conditions.
\newblock In {\em Journal of Experimental and Theoretical Physics},
  90(4):695--713, 2000.

\bibitem{huang1992darboux}
N.-N. Huang.
\newblock Darboux transformations for the korteweg-de-vries equation.
\newblock In {\em Journal of Physics A: Mathematical and General}, 25(2):469,
  1992.

\bibitem{ilievski2017microscopic}
E. Ilievski and J. De~Nardis.
\newblock Microscopic origin of ideal conductivity in integrable quantum
  models.
\newblock In {\em Physical review letters}, 119(2):020602, 2017.

\bibitem{ilievski2015complete}
E. Ilievski, J. De~Nardis, B. Wouters, J.-S. Caux, F.~H. L. Essler, and
  T. Prosen.
\newblock Complete generalized gibbs ensembles in an interacting theory.
\newblock In {\em Physical review letters}, 115(15):157201, 2015.

\bibitem{ilievski2016quasilocal}
E. Ilievski, M. Medenjak, T. Prosen, and L. Zadnik.
\newblock Quasilocal charges in integrable lattice systems.
\newblock In {\em Journal of Statistical Mechanics: Theory and Experiment},
  2016(6):064008, 2016.

\bibitem{langen2015experimental}
T. Langen, S. Erne, R. Geiger, B. Rauer, T. Schweigler,
  Ma. Kuhnert, W. Rohringer, I.~E. Mazets, T. Gasenzer, and
  J. Schmiedmayer.
\newblock Experimental observation of a generalized gibbs ensemble.
\newblock In {\em Science}, 348(6231):207--211, 2015.

\bibitem{lax_zero_1991}
P.~D. Lax.
\newblock The zero dispersion limit, a deterministic analogue of turbulence.
\newblock In {\em Communications on Pure and Applied Mathematics},
  44(8-9):1047--1056, 1991.

\bibitem{lax_small_1983}
P.~D. Lax and C.~D. Levermore.
\newblock The small dispersion limit of the \{{Korteweg}-de {Vries}\} equation:
  2.
\newblock In {\em Communications on Pure and Applied Mathematics}, 36(5):571--593, 1983.

\bibitem{lax_integrals_1968}
P. D. Lax.
\newblock Integrals of nonlinear equations of evolution and solitary waves.
\newblock In {\em Communications on Pure and Applied Mathematics}, 21:467--490, 1968.

\bibitem{levkovich2016bethe}
F. Levkovich-Maslyuk.
\newblock The bethe ansatz.
\newblock In {\em Journal of Physics A: Mathematical and Theoretical},
  49(32):323004, 2016.

\bibitem{lieb1963exact}
E.~H. Lieb and W. Liniger.
\newblock Exact analysis of an interacting bose gas. i. the general solution
  and the ground state.
\newblock In {\em Physical Review}, 130(4):1605, 1963.

\bibitem{malvania2020ghd}
N. Malvania, Y. Zhang, Y. Le, J. Dubail, M. Rigol and D. S. Weiss, Generalized hydrodynamics in strongly interacting 1D Bose gases. In {\em Science} 373(6559):1129, 2021.

\bibitem{moller2020extension}
F. M\o{}ller, C. Li, I. Mazets, H.-P. Stimming, T. Zhou, Z. Zhu, X. Chen, J. Schmiedmayer,
Extension of the Generalized Hydrodynamics to Dimensional Crossover Regime. In {\em Physical Review Letters} 126(09):090602, 2020.

\bibitem{mossel2012generalized}
J. Mossel and J.-S. Caux.
\newblock Generalized tba and generalized gibbs.
\newblock In {\em Journal of Physics A: Mathematical and Theoretical},
  45(25):255001, 2012.

\bibitem{myers2020transport}
J. Myers, M.~J. Bhaseen, R.~J. Harris, and B. Doyon.
\newblock {Transport fluctuations in integrable models out of equilibrium}.
\newblock In {\em SciPost Physics}, 8(1):007, 2020.

\bibitem{piroli2017transport}
L. Piroli, J. De~Nardis, M. Collura, B. Bertini, and M.
  Fagotti.
\newblock Transport in out-of-equilibrium xxz chains: Nonballistic behavior and
  correlation functions.
\newblock In {\em Physical Review B}, 96(11):115124, 2017.

\bibitem{pozsgay2020algebraic}
B. Pozsgay.
\newblock Algebraic construction of current operators in integrable spin
  chains.
\newblock In {\em Physical Review Letters}, 125(7):070602, 2020.

\bibitem{redor_experimental_2019}
I. Redor, E. Barth{\'e}lemy, H. Michallet, M. Onorato, and
  N. Mordant.
\newblock Experimental {Evidence} of a {Hydrodynamic} {Soliton} {Gas}.
\newblock In {\em Physical Review Letters}, 122(21):214502, 2019.

\bibitem{roberti_numerical_2021}
G.~Roberti, G.~El, A.~Tovbis, F.~Copie, P.~Suret, and S.~Randoux.
\newblock Numerical spectral synthesis of breather gas for the focusing
  nonlinear {Schrödinger} equation.
\newblock In {\em Physical Review E}, 103(4):042205, 2021.

\bibitem{rosenzweig2019mean}
M. Rosenzweig.
\newblock The mean-field limit of the lieb-liniger model.
\newblock {\em arXiv preprint arXiv:1912.07585}, 2019.

\bibitem{schemmer2019generalised}
M.~Schemmer, I.~Bouchoule, B.~Doyon, and J.~Dubail.
\newblock Generalized hydrodynamics on an atom chip.
\newblock In {\em Physical Review Letters}, 122(09):090601, 2019.

\bibitem{spohn2012large}
H. Spohn.
\newblock {\em Large scale dynamics of interacting particles}.
\newblock Springer Science \& Business Media, 2012.

\bibitem{SpoToda} H. Spohn
\newblock Generalized Gibbs ensembles of the classical Toda chain.
\newblock In {\em Journal of Statistical Physics} 180(1):4--22, 2020.

\bibitem{spohn2020collision}
H. Spohn.
\newblock Collision rate ansatz for the classical toda lattice.
\newblock In {\em Physical Review E}, 101(6):060103, 2020.

\bibitem{suret_nonlinear_2020}
P. Suret, A. Tikan, F. Bonnefoy, F. Copie, G.
  Ducrozet, A. A. Gelash, G. Prabhudesai, G. Michel, A.
  Cazaubiel, E. Falcon, G. A. El, and S. Randoux.
\newblock Nonlinear {Spectral} {Synthesis} of {Soliton} {Gas} in {Deep}-{Water}
  {Surface} {Gravity} {Waves}.
\newblock In {\em Physical Review Letters}, 125(26):264101, 2020.

\bibitem{takayama1985extended}
H. Takayama and M. Ishikawa.
\newblock On the extended ideal gas phenomenological and the bethe ansatz
  approaches to the thermodynamics of integrable soliton-bearing systems.
\newblock In {\em Progress of Theoretical Physics}, 74(3):479--489, 1985.

\bibitem{timonen1986exact}
J.~Timonen, R. K.~Bullough, and D. J.~Pilling.
\newblock Exact bethe-ansatz thermodynamics for the sine-gordon model in the
  classical limit: Effect of long strings.
\newblock In {\em Physical Review B}, 34(9):6525, 1986.

\bibitem{van2016introduction}
S.~J. van Tongeren.
\newblock Introduction to the thermodynamic bethe ansatz.
\newblock In {\em Journal of Physics A: Mathematical and Theoretical},
  49(32):323005, 2016.
  
    \bibitem{tovbis2022recent}
  {A.~Tovbis and F.~Wang, Recent developments in spectral theory of the focusing NLS soliton and breather gases: the thermodynamic limit of average densities, fluxes and certain meromorphic differentials; periodic gases.
 \newblock {\em arXiv preprint arXiv:2203.03566}, 2022}. 

\bibitem{whitham_linear_1974}
G.~B. Whitham.
\newblock {\em Linear and Nonlinear Waves}.
\newblock John Wiley \& Sons, Inc., 1974.

\bibitem{yang1969thermodynamics}
C.-N. Yang and C.~P. Yang.
\newblock Thermodynamics of a one-dimensional system of bosons with repulsive
  delta-function interaction.
\newblock In {\em Journal of Mathematical Physics}, 10(7):1115--1122, 1969.

\bibitem{yoshimura2020collision}
T. Yoshimura and H. Spohn.
\newblock Collision rate ansatz for quantum integrable systems.
\newblock In {\em SciPost Physics}, 9(3):040, 2020.

\bibitem{zakharov_kinetic_1971}
V.E. Zakharov.
\newblock Kinetic equation for solitons.
\newblock In {\em Journal of Experimental and Theoretical Physics},
  33(3):538--541, 1971.

\bibitem{zakharov_turbulence_2009}
V.~E. Zakharov.
\newblock Turbulence in integrable systems.
\newblock In {\em Studies in Applied Mathematics}, 122(3):219--234, 2009.

\bibitem{zamolodchikov1990thermodynamic}
A.~B. Zamolodchikov.
\newblock Thermodynamic bethe ansatz in relativistic models: Scaling 3-state
  potts and lee-yang models.
\newblock In {\em Nuclear Physics B}, 342(3):695--720, 1990.


\end{thebibliography}

\end{document}